\documentclass[longauth,twocolumn]{aa}

\usepackage{graphicx}
\usepackage{natbib}
\usepackage{scalerel}
\usepackage{lscape}
\usepackage{ulem}
\usepackage{soul}
\usepackage{siunitx}
\usepackage[table]{xcolor}

\usepackage{txfonts}
\usepackage[pdfencoding=auto,psdextra]{hyperref}
\hypersetup{
    colorlinks=true,
    linkcolor=blue,
    filecolor=magenta,      
    urlcolor=blue,
    citecolor=blue
}
\makeatletter
\renewcommand*\aa@pageof{, page \thepage{} of \pageref*{LastPage}}
\makeatother

\usepackage[utf8]{inputenc}

\usepackage[switch, modulo]{lineno}

\usepackage{euclid}
\usepackage{hhline}

\begin{document}
%
%

\title{\Euclid: Quick Data Release (Q1) -- Photometric studies of known transients\thanks{This paper is published on
       behalf of the Euclid Consortium}}

   
\newcommand{\orcid}[1]{\href{https://orcid.org/#1}{\includegraphics[scale=0.0037]{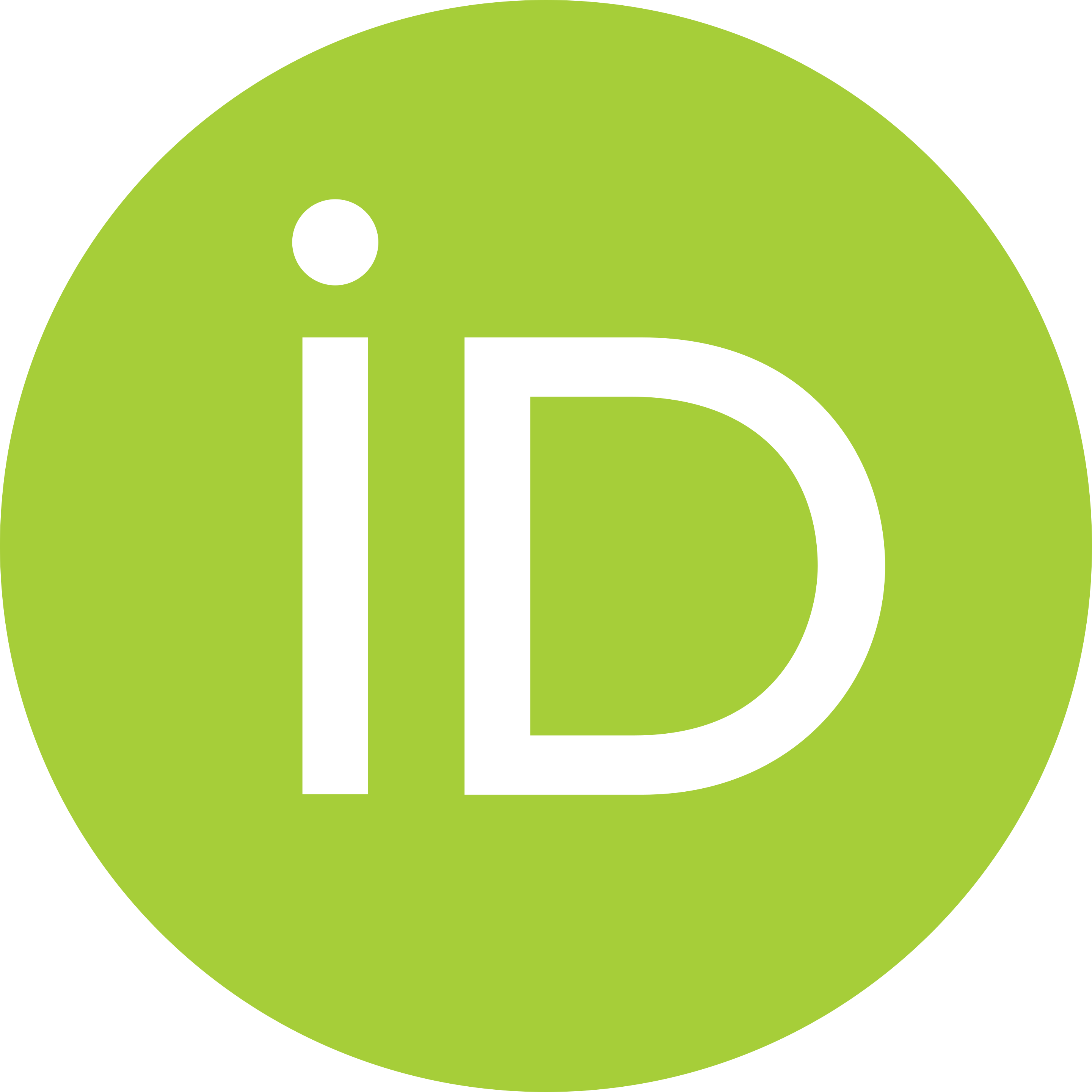}}{}} 
\author{C.~Duffy\orcid{0000-0001-6662-0200}\thanks{\email{c.j.duffy@lancaster.ac.uk}}\inst{\ref{aff1}}
\and E.~Cappellaro\orcid{0000-0001-5008-8619}\inst{\ref{aff2}}
\and M.~T.~Botticella\orcid{0000-0002-3938-692X}\inst{\ref{aff3}}
\and I.~M.~Hook\orcid{0000-0002-2960-978X}\inst{\ref{aff1}}
\and F.~Poidevin\orcid{0000-0002-5391-5568}\inst{\ref{aff4},\ref{aff5}}
\and T.~J.~Moriya\orcid{0000-0003-1169-1954}\inst{\ref{aff6},\ref{aff7},\ref{aff8}}
\and A.~A.~Chrimes\orcid{0000-0001-9842-6808}\inst{\ref{aff9},\ref{aff10}}
\and V.~Petrecca\orcid{0000-0002-3078-856X}\inst{\ref{aff3},\ref{aff11}}
\and K.~Paterson\orcid{0000-0001-8340-3486}\inst{\ref{aff12}}
\and A.~Goobar\orcid{0000-0002-4163-4996}\inst{\ref{aff13}}
\and L.~Galbany\orcid{0000-0002-1296-6887}\inst{\ref{aff14},\ref{aff15}}
\and R.~Kotak\orcid{0000-0001-5455-3653}\inst{\ref{aff16}}
\and C.~Gall\orcid{0000-0002-8526-3963}\inst{\ref{aff17}}
\and C.~M.~Gutierrez\orcid{0000-0001-7854-783X}\inst{\ref{aff18}}
\and C.~Tao\orcid{0000-0001-7961-8177}\inst{\ref{aff19}}
\and L.~Izzo\orcid{0000-0001-9695-8472}\inst{\ref{aff3}}
\and N.~Aghanim\orcid{0000-0002-6688-8992}\inst{\ref{aff20}}
\and B.~Altieri\orcid{0000-0003-3936-0284}\inst{\ref{aff21}}
\and A.~Amara\inst{\ref{aff22}}
\and S.~Andreon\orcid{0000-0002-2041-8784}\inst{\ref{aff23}}
\and N.~Auricchio\orcid{0000-0003-4444-8651}\inst{\ref{aff24}}
\and C.~Baccigalupi\orcid{0000-0002-8211-1630}\inst{\ref{aff25},\ref{aff26},\ref{aff27},\ref{aff28}}
\and M.~Baldi\orcid{0000-0003-4145-1943}\inst{\ref{aff29},\ref{aff24},\ref{aff30}}
\and A.~Balestra\orcid{0000-0002-6967-261X}\inst{\ref{aff2}}
\and S.~Bardelli\orcid{0000-0002-8900-0298}\inst{\ref{aff24}}
\and A.~Basset\inst{\ref{aff31}}
\and P.~Battaglia\orcid{0000-0002-7337-5909}\inst{\ref{aff24}}
\and A.~Biviano\orcid{0000-0002-0857-0732}\inst{\ref{aff26},\ref{aff25}}
\and A.~Bonchi\orcid{0000-0002-2667-5482}\inst{\ref{aff32}}
\and E.~Branchini\orcid{0000-0002-0808-6908}\inst{\ref{aff33},\ref{aff34},\ref{aff23}}
\and M.~Brescia\orcid{0000-0001-9506-5680}\inst{\ref{aff35},\ref{aff3}}
\and J.~Brinchmann\orcid{0000-0003-4359-8797}\inst{\ref{aff36},\ref{aff37}}
\and S.~Camera\orcid{0000-0003-3399-3574}\inst{\ref{aff38},\ref{aff39},\ref{aff40}}
\and V.~Capobianco\orcid{0000-0002-3309-7692}\inst{\ref{aff40}}
\and C.~Carbone\orcid{0000-0003-0125-3563}\inst{\ref{aff41}}
\and J.~Carretero\orcid{0000-0002-3130-0204}\inst{\ref{aff42},\ref{aff43}}
\and R.~Casas\orcid{0000-0002-8165-5601}\inst{\ref{aff15},\ref{aff44}}
\and S.~Casas\orcid{0000-0002-4751-5138}\inst{\ref{aff45}}
\and M.~Castellano\orcid{0000-0001-9875-8263}\inst{\ref{aff46}}
\and G.~Castignani\orcid{0000-0001-6831-0687}\inst{\ref{aff24}}
\and S.~Cavuoti\orcid{0000-0002-3787-4196}\inst{\ref{aff3},\ref{aff47}}
\and A.~Cimatti\inst{\ref{aff48}}
\and C.~Colodro-Conde\inst{\ref{aff4}}
\and G.~Congedo\orcid{0000-0003-2508-0046}\inst{\ref{aff49}}
\and C.~J.~Conselice\orcid{0000-0003-1949-7638}\inst{\ref{aff50}}
\and L.~Conversi\orcid{0000-0002-6710-8476}\inst{\ref{aff51},\ref{aff21}}
\and Y.~Copin\orcid{0000-0002-5317-7518}\inst{\ref{aff52}}
\and F.~Courbin\orcid{0000-0003-0758-6510}\inst{\ref{aff53},\ref{aff54}}
\and H.~M.~Courtois\orcid{0000-0003-0509-1776}\inst{\ref{aff55}}
\and M.~Cropper\orcid{0000-0003-4571-9468}\inst{\ref{aff56}}
\and A.~Da~Silva\orcid{0000-0002-6385-1609}\inst{\ref{aff57},\ref{aff58}}
\and H.~Degaudenzi\orcid{0000-0002-5887-6799}\inst{\ref{aff59}}
\and G.~De~Lucia\orcid{0000-0002-6220-9104}\inst{\ref{aff26}}
\and A.~M.~Di~Giorgio\orcid{0000-0002-4767-2360}\inst{\ref{aff60}}
\and C.~Dolding\orcid{0009-0003-7199-6108}\inst{\ref{aff56}}
\and H.~Dole\orcid{0000-0002-9767-3839}\inst{\ref{aff20}}
\and F.~Dubath\orcid{0000-0002-6533-2810}\inst{\ref{aff59}}
\and C.~A.~J.~Duncan\orcid{0009-0003-3573-0791}\inst{\ref{aff50}}
\and X.~Dupac\inst{\ref{aff21}}
\and S.~Dusini\orcid{0000-0002-1128-0664}\inst{\ref{aff61}}
\and A.~Ealet\orcid{0000-0003-3070-014X}\inst{\ref{aff52}}
\and S.~Escoffier\orcid{0000-0002-2847-7498}\inst{\ref{aff19}}
\and M.~Farina\orcid{0000-0002-3089-7846}\inst{\ref{aff60}}
\and F.~Faustini\orcid{0000-0001-6274-5145}\inst{\ref{aff32},\ref{aff46}}
\and S.~Ferriol\inst{\ref{aff52}}
\and S.~Fotopoulou\orcid{0000-0002-9686-254X}\inst{\ref{aff62}}
\and M.~Frailis\orcid{0000-0002-7400-2135}\inst{\ref{aff26}}
\and P.~Franzetti\inst{\ref{aff41}}
\and S.~Galeotta\orcid{0000-0002-3748-5115}\inst{\ref{aff26}}
\and K.~George\orcid{0000-0002-1734-8455}\inst{\ref{aff63}}
\and B.~Gillis\orcid{0000-0002-4478-1270}\inst{\ref{aff49}}
\and C.~Giocoli\orcid{0000-0002-9590-7961}\inst{\ref{aff24},\ref{aff30}}
\and P.~G\'omez-Alvarez\orcid{0000-0002-8594-5358}\inst{\ref{aff64},\ref{aff21}}
\and A.~Grazian\orcid{0000-0002-5688-0663}\inst{\ref{aff2}}
\and F.~Grupp\inst{\ref{aff65},\ref{aff63}}
\and S.~Gwyn\orcid{0000-0001-8221-8406}\inst{\ref{aff66}}
\and S.~V.~H.~Haugan\orcid{0000-0001-9648-7260}\inst{\ref{aff67}}
\and J.~Hoar\inst{\ref{aff21}}
\and H.~Hoekstra\orcid{0000-0002-0641-3231}\inst{\ref{aff68}}
\and W.~Holmes\inst{\ref{aff69}}
\and F.~Hormuth\inst{\ref{aff70}}
\and A.~Hornstrup\orcid{0000-0002-3363-0936}\inst{\ref{aff71},\ref{aff72}}
\and P.~Hudelot\inst{\ref{aff73}}
\and K.~Jahnke\orcid{0000-0003-3804-2137}\inst{\ref{aff12}}
\and M.~Jhabvala\inst{\ref{aff74}}
\and E.~Keih\"anen\orcid{0000-0003-1804-7715}\inst{\ref{aff75}}
\and S.~Kermiche\orcid{0000-0002-0302-5735}\inst{\ref{aff19}}
\and B.~Kubik\orcid{0009-0006-5823-4880}\inst{\ref{aff52}}
\and K.~Kuijken\orcid{0000-0002-3827-0175}\inst{\ref{aff68}}
\and M.~K\"ummel\orcid{0000-0003-2791-2117}\inst{\ref{aff63}}
\and M.~Kunz\orcid{0000-0002-3052-7394}\inst{\ref{aff76}}
\and H.~Kurki-Suonio\orcid{0000-0002-4618-3063}\inst{\ref{aff77},\ref{aff78}}
\and Q.~Le~Boulc'h\inst{\ref{aff79}}
\and A.~M.~C.~Le~Brun\orcid{0000-0002-0936-4594}\inst{\ref{aff80}}
\and D.~Le~Mignant\orcid{0000-0002-5339-5515}\inst{\ref{aff81}}
\and P.~Liebing\inst{\ref{aff56}}
\and S.~Ligori\orcid{0000-0003-4172-4606}\inst{\ref{aff40}}
\and P.~B.~Lilje\orcid{0000-0003-4324-7794}\inst{\ref{aff67}}
\and V.~Lindholm\orcid{0000-0003-2317-5471}\inst{\ref{aff77},\ref{aff78}}
\and I.~Lloro\orcid{0000-0001-5966-1434}\inst{\ref{aff82}}
\and D.~Maino\inst{\ref{aff83},\ref{aff41},\ref{aff84}}
\and E.~Maiorano\orcid{0000-0003-2593-4355}\inst{\ref{aff24}}
\and O.~Mansutti\orcid{0000-0001-5758-4658}\inst{\ref{aff26}}
\and S.~Marcin\inst{\ref{aff85}}
\and O.~Marggraf\orcid{0000-0001-7242-3852}\inst{\ref{aff86}}
\and M.~Martinelli\orcid{0000-0002-6943-7732}\inst{\ref{aff46},\ref{aff87}}
\and N.~Martinet\orcid{0000-0003-2786-7790}\inst{\ref{aff81}}
\and F.~Marulli\orcid{0000-0002-8850-0303}\inst{\ref{aff88},\ref{aff24},\ref{aff30}}
\and R.~Massey\orcid{0000-0002-6085-3780}\inst{\ref{aff89}}
\and E.~Medinaceli\orcid{0000-0002-4040-7783}\inst{\ref{aff24}}
\and M.~Melchior\inst{\ref{aff90}}
\and Y.~Mellier\inst{\ref{aff91},\ref{aff73}}
\and M.~Meneghetti\orcid{0000-0003-1225-7084}\inst{\ref{aff24},\ref{aff30}}
\and E.~Merlin\orcid{0000-0001-6870-8900}\inst{\ref{aff46}}
\and G.~Meylan\inst{\ref{aff92}}
\and M.~Moresco\orcid{0000-0002-7616-7136}\inst{\ref{aff88},\ref{aff24}}
\and P.~W.~Morris\orcid{0000-0002-5186-4381}\inst{\ref{aff93}}
\and L.~Moscardini\orcid{0000-0002-3473-6716}\inst{\ref{aff88},\ref{aff24},\ref{aff30}}
\and C.~Neissner\orcid{0000-0001-8524-4968}\inst{\ref{aff94},\ref{aff43}}
\and R.~C.~Nichol\orcid{0000-0003-0939-6518}\inst{\ref{aff22}}
\and S.-M.~Niemi\inst{\ref{aff9}}
\and J.~W.~Nightingale\orcid{0000-0002-8987-7401}\inst{\ref{aff95}}
\and C.~Padilla\orcid{0000-0001-7951-0166}\inst{\ref{aff94}}
\and S.~Paltani\orcid{0000-0002-8108-9179}\inst{\ref{aff59}}
\and F.~Pasian\orcid{0000-0002-4869-3227}\inst{\ref{aff26}}
\and K.~Pedersen\inst{\ref{aff17}}
\and W.~J.~Percival\orcid{0000-0002-0644-5727}\inst{\ref{aff96},\ref{aff97},\ref{aff98}}
\and V.~Pettorino\inst{\ref{aff9}}
\and S.~Pires\orcid{0000-0002-0249-2104}\inst{\ref{aff99}}
\and G.~Polenta\orcid{0000-0003-4067-9196}\inst{\ref{aff32}}
\and M.~Poncet\inst{\ref{aff31}}
\and L.~A.~Popa\inst{\ref{aff100}}
\and F.~Raison\orcid{0000-0002-7819-6918}\inst{\ref{aff65}}
\and R.~Rebolo\inst{\ref{aff101},\ref{aff5},\ref{aff4}}
\and A.~Renzi\orcid{0000-0001-9856-1970}\inst{\ref{aff102},\ref{aff61}}
\and J.~Rhodes\orcid{0000-0002-4485-8549}\inst{\ref{aff69}}
\and G.~Riccio\inst{\ref{aff3}}
\and E.~Romelli\orcid{0000-0003-3069-9222}\inst{\ref{aff26}}
\and M.~Roncarelli\orcid{0000-0001-9587-7822}\inst{\ref{aff24}}
\and R.~Saglia\orcid{0000-0003-0378-7032}\inst{\ref{aff63},\ref{aff65}}
\and Z.~Sakr\orcid{0000-0002-4823-3757}\inst{\ref{aff103},\ref{aff104},\ref{aff105}}
\and D.~Sapone\orcid{0000-0001-7089-4503}\inst{\ref{aff106}}
\and B.~Sartoris\orcid{0000-0003-1337-5269}\inst{\ref{aff63},\ref{aff26}}
\and J.~A.~Schewtschenko\orcid{0000-0002-4913-6393}\inst{\ref{aff49}}
\and M.~Schirmer\orcid{0000-0003-2568-9994}\inst{\ref{aff12}}
\and P.~Schneider\orcid{0000-0001-8561-2679}\inst{\ref{aff86}}
\and T.~Schrabback\orcid{0000-0002-6987-7834}\inst{\ref{aff107}}
\and A.~Secroun\orcid{0000-0003-0505-3710}\inst{\ref{aff19}}
\and G.~Seidel\orcid{0000-0003-2907-353X}\inst{\ref{aff12}}
\and S.~Serrano\orcid{0000-0002-0211-2861}\inst{\ref{aff15},\ref{aff108},\ref{aff44}}
\and P.~Simon\inst{\ref{aff86}}
\and C.~Sirignano\orcid{0000-0002-0995-7146}\inst{\ref{aff102},\ref{aff61}}
\and G.~Sirri\orcid{0000-0003-2626-2853}\inst{\ref{aff30}}
\and J.~Skottfelt\orcid{0000-0003-1310-8283}\inst{\ref{aff109}}
\and L.~Stanco\orcid{0000-0002-9706-5104}\inst{\ref{aff61}}
\and J.~Steinwagner\orcid{0000-0001-7443-1047}\inst{\ref{aff65}}
\and P.~Tallada-Cresp\'{i}\orcid{0000-0002-1336-8328}\inst{\ref{aff42},\ref{aff43}}
\and A.~N.~Taylor\inst{\ref{aff49}}
\and I.~Tereno\inst{\ref{aff57},\ref{aff110}}
\and R.~Toledo-Moreo\orcid{0000-0002-2997-4859}\inst{\ref{aff111}}
\and F.~Torradeflot\orcid{0000-0003-1160-1517}\inst{\ref{aff43},\ref{aff42}}
\and I.~Tutusaus\orcid{0000-0002-3199-0399}\inst{\ref{aff104}}
\and L.~Valenziano\orcid{0000-0002-1170-0104}\inst{\ref{aff24},\ref{aff112}}
\and T.~Vassallo\orcid{0000-0001-6512-6358}\inst{\ref{aff63},\ref{aff26}}
\and G.~Verdoes~Kleijn\orcid{0000-0001-5803-2580}\inst{\ref{aff113}}
\and Y.~Wang\orcid{0000-0002-4749-2984}\inst{\ref{aff114}}
\and J.~Weller\orcid{0000-0002-8282-2010}\inst{\ref{aff63},\ref{aff65}}
\and A.~Zacchei\orcid{0000-0003-0396-1192}\inst{\ref{aff26},\ref{aff25}}
\and G.~Zamorani\orcid{0000-0002-2318-301X}\inst{\ref{aff24}}
\and F.~M.~Zerbi\inst{\ref{aff23}}
\and E.~Zucca\orcid{0000-0002-5845-8132}\inst{\ref{aff24}}
\and C.~Burigana\orcid{0000-0002-3005-5796}\inst{\ref{aff115},\ref{aff112}}
\and R.~Cabanac\orcid{0000-0001-6679-2600}\inst{\ref{aff104}}
\and L.~Gabarra\orcid{0000-0002-8486-8856}\inst{\ref{aff116}}
\and V.~Scottez\inst{\ref{aff91},\ref{aff117}}
\and D.~Scott\orcid{0000-0002-6878-9840}\inst{\ref{aff118}}
\and M.~Sullivan\orcid{0000-0001-9053-4820}\inst{\ref{aff119}}}
										   
\institute{Department of Physics, Lancaster University, Lancaster, LA1 4YB, UK\label{aff1}
\and
INAF-Osservatorio Astronomico di Padova, Via dell'Osservatorio 5, 35122 Padova, Italy\label{aff2}
\and
INAF-Osservatorio Astronomico di Capodimonte, Via Moiariello 16, 80131 Napoli, Italy\label{aff3}
\and
Instituto de Astrof\'{\i}sica de Canarias, V\'{\i}a L\'actea, 38205 La Laguna, Tenerife, Spain\label{aff4}
\and
Universidad de La Laguna, Departamento de Astrof\'{\i}sica, 38206 La Laguna, Tenerife, Spain\label{aff5}
\and
National Astronomical Observatory of Japan, National Institutes of Natural Sciences, 2-21-1 Osawa, Mitaka, Tokyo 181-8588, Japan\label{aff6}
\and
Graduate Institute for Advanced Studies, SOKENDAI, 2-21-1 Osawa, Mitaka, Tokyo 181-8588, Japan\label{aff7}
\and
School of Physics and Astronomy, Monash University, Clayton, VIC 3800, Australia\label{aff8}
\and
European Space Agency/ESTEC, Keplerlaan 1, 2201 AZ Noordwijk, The Netherlands\label{aff9}
\and
Department of Astrophysics/IMAPP, Radboud University, PO Box 9010, 6500 GL Nijmegen, The Netherlands\label{aff10}
\and
Dipartimento di Fisica "E. Pancini", Universita degli Studi di Napoli Federico II, Via Cinthia 6, 80126, Napoli, Italy\label{aff11}
\and
Max-Planck-Institut f\"ur Astronomie, K\"onigstuhl 17, 69117 Heidelberg, Germany\label{aff12}
\and
Oskar Klein Centre for Cosmoparticle Physics, Department of Physics, Stockholm University, Stockholm, SE-106 91, Sweden\label{aff13}
\and
Institut de Ciencies de l'Espai (IEEC-CSIC), Campus UAB, Carrer de Can Magrans, s/n Cerdanyola del Vall\'es, 08193 Barcelona, Spain\label{aff14}
\and
Institut d'Estudis Espacials de Catalunya (IEEC),  Edifici RDIT, Campus UPC, 08860 Castelldefels, Barcelona, Spain\label{aff15}
\and
Department of Physics and Astronomy, Vesilinnantie 5, 20014 University of Turku, Finland\label{aff16}
\and
DARK, Niels Bohr Institute, University of Copenhagen, Jagtvej 155, 2200 Copenhagen, Denmark\label{aff17}
\and
Instituto de Astrof\'\i sica de Canarias, c/ Via Lactea s/n, La Laguna 38200, Spain. Departamento de Astrof\'\i sica de la Universidad de La Laguna, Avda. Francisco Sanchez, La Laguna, 38200, Spain\label{aff18}
\and
Aix-Marseille Universit\'e, CNRS/IN2P3, CPPM, Marseille, France\label{aff19}
\and
Universit\'e Paris-Saclay, CNRS, Institut d'astrophysique spatiale, 91405, Orsay, France\label{aff20}
\and
ESAC/ESA, Camino Bajo del Castillo, s/n., Urb. Villafranca del Castillo, 28692 Villanueva de la Ca\~nada, Madrid, Spain\label{aff21}
\and
School of Mathematics and Physics, University of Surrey, Guildford, Surrey, GU2 7XH, UK\label{aff22}
\and
INAF-Osservatorio Astronomico di Brera, Via Brera 28, 20122 Milano, Italy\label{aff23}
\and
INAF-Osservatorio di Astrofisica e Scienza dello Spazio di Bologna, Via Piero Gobetti 93/3, 40129 Bologna, Italy\label{aff24}
\and
IFPU, Institute for Fundamental Physics of the Universe, via Beirut 2, 34151 Trieste, Italy\label{aff25}
\and
INAF-Osservatorio Astronomico di Trieste, Via G. B. Tiepolo 11, 34143 Trieste, Italy\label{aff26}
\and
INFN, Sezione di Trieste, Via Valerio 2, 34127 Trieste TS, Italy\label{aff27}
\and
SISSA, International School for Advanced Studies, Via Bonomea 265, 34136 Trieste TS, Italy\label{aff28}
\and
Dipartimento di Fisica e Astronomia, Universit\`a di Bologna, Via Gobetti 93/2, 40129 Bologna, Italy\label{aff29}
\and
INFN-Sezione di Bologna, Viale Berti Pichat 6/2, 40127 Bologna, Italy\label{aff30}
\and
Centre National d'Etudes Spatiales -- Centre spatial de Toulouse, 18 avenue Edouard Belin, 31401 Toulouse Cedex 9, France\label{aff31}
\and
Space Science Data Center, Italian Space Agency, via del Politecnico snc, 00133 Roma, Italy\label{aff32}
\and
Dipartimento di Fisica, Universit\`a di Genova, Via Dodecaneso 33, 16146, Genova, Italy\label{aff33}
\and
INFN-Sezione di Genova, Via Dodecaneso 33, 16146, Genova, Italy\label{aff34}
\and
Department of Physics "E. Pancini", University Federico II, Via Cinthia 6, 80126, Napoli, Italy\label{aff35}
\and
Instituto de Astrof\'isica e Ci\^encias do Espa\c{c}o, Universidade do Porto, CAUP, Rua das Estrelas, PT4150-762 Porto, Portugal\label{aff36}
\and
Faculdade de Ci\^encias da Universidade do Porto, Rua do Campo de Alegre, 4150-007 Porto, Portugal\label{aff37}
\and
Dipartimento di Fisica, Universit\`a degli Studi di Torino, Via P. Giuria 1, 10125 Torino, Italy\label{aff38}
\and
INFN-Sezione di Torino, Via P. Giuria 1, 10125 Torino, Italy\label{aff39}
\and
INAF-Osservatorio Astrofisico di Torino, Via Osservatorio 20, 10025 Pino Torinese (TO), Italy\label{aff40}
\and
INAF-IASF Milano, Via Alfonso Corti 12, 20133 Milano, Italy\label{aff41}
\and
Centro de Investigaciones Energ\'eticas, Medioambientales y Tecnol\'ogicas (CIEMAT), Avenida Complutense 40, 28040 Madrid, Spain\label{aff42}
\and
Port d'Informaci\'{o} Cient\'{i}fica, Campus UAB, C. Albareda s/n, 08193 Bellaterra (Barcelona), Spain\label{aff43}
\and
Institute of Space Sciences (ICE, CSIC), Campus UAB, Carrer de Can Magrans, s/n, 08193 Barcelona, Spain\label{aff44}
\and
Institute for Theoretical Particle Physics and Cosmology (TTK), RWTH Aachen University, 52056 Aachen, Germany\label{aff45}
\and
INAF-Osservatorio Astronomico di Roma, Via Frascati 33, 00078 Monteporzio Catone, Italy\label{aff46}
\and
INFN section of Naples, Via Cinthia 6, 80126, Napoli, Italy\label{aff47}
\and
Dipartimento di Fisica e Astronomia "Augusto Righi" - Alma Mater Studiorum Universit\`a di Bologna, Viale Berti Pichat 6/2, 40127 Bologna, Italy\label{aff48}
\and
Institute for Astronomy, University of Edinburgh, Royal Observatory, Blackford Hill, Edinburgh EH9 3HJ, UK\label{aff49}
\and
Jodrell Bank Centre for Astrophysics, Department of Physics and Astronomy, University of Manchester, Oxford Road, Manchester M13 9PL, UK\label{aff50}
\and
European Space Agency/ESRIN, Largo Galileo Galilei 1, 00044 Frascati, Roma, Italy\label{aff51}
\and
Universit\'e Claude Bernard Lyon 1, CNRS/IN2P3, IP2I Lyon, UMR 5822, Villeurbanne, F-69100, France\label{aff52}
\and
Institut de Ci\`{e}ncies del Cosmos (ICCUB), Universitat de Barcelona (IEEC-UB), Mart\'{i} i Franqu\`{e}s 1, 08028 Barcelona, Spain\label{aff53}
\and
Instituci\'o Catalana de Recerca i Estudis Avan\c{c}ats (ICREA), Passeig de Llu\'{\i}s Companys 23, 08010 Barcelona, Spain\label{aff54}
\and
UCB Lyon 1, CNRS/IN2P3, IUF, IP2I Lyon, 4 rue Enrico Fermi, 69622 Villeurbanne, France\label{aff55}
\and
Mullard Space Science Laboratory, University College London, Holmbury St Mary, Dorking, Surrey RH5 6NT, UK\label{aff56}
\and
Departamento de F\'isica, Faculdade de Ci\^encias, Universidade de Lisboa, Edif\'icio C8, Campo Grande, PT1749-016 Lisboa, Portugal\label{aff57}
\and
Instituto de Astrof\'isica e Ci\^encias do Espa\c{c}o, Faculdade de Ci\^encias, Universidade de Lisboa, Campo Grande, 1749-016 Lisboa, Portugal\label{aff58}
\and
Department of Astronomy, University of Geneva, ch. d'Ecogia 16, 1290 Versoix, Switzerland\label{aff59}
\and
INAF-Istituto di Astrofisica e Planetologia Spaziali, via del Fosso del Cavaliere, 100, 00100 Roma, Italy\label{aff60}
\and
INFN-Padova, Via Marzolo 8, 35131 Padova, Italy\label{aff61}
\and
School of Physics, HH Wills Physics Laboratory, University of Bristol, Tyndall Avenue, Bristol, BS8 1TL, UK\label{aff62}
\and
Universit\"ats-Sternwarte M\"unchen, Fakult\"at f\"ur Physik, Ludwig-Maximilians-Universit\"at M\"unchen, Scheinerstrasse 1, 81679 M\"unchen, Germany\label{aff63}
\and
FRACTAL S.L.N.E., calle Tulip\'an 2, Portal 13 1A, 28231, Las Rozas de Madrid, Spain\label{aff64}
\and
Max Planck Institute for Extraterrestrial Physics, Giessenbachstr. 1, 85748 Garching, Germany\label{aff65}
\and
NRC Herzberg, 5071 West Saanich Rd, Victoria, BC V9E 2E7, Canada\label{aff66}
\and
Institute of Theoretical Astrophysics, University of Oslo, P.O. Box 1029 Blindern, 0315 Oslo, Norway\label{aff67}
\and
Leiden Observatory, Leiden University, Einsteinweg 55, 2333 CC Leiden, The Netherlands\label{aff68}
\and
Jet Propulsion Laboratory, California Institute of Technology, 4800 Oak Grove Drive, Pasadena, CA, 91109, USA\label{aff69}
\and
Felix Hormuth Engineering, Goethestr. 17, 69181 Leimen, Germany\label{aff70}
\and
Technical University of Denmark, Elektrovej 327, 2800 Kgs. Lyngby, Denmark\label{aff71}
\and
Cosmic Dawn Center (DAWN), Denmark\label{aff72}
\and
Institut d'Astrophysique de Paris, UMR 7095, CNRS, and Sorbonne Universit\'e, 98 bis boulevard Arago, 75014 Paris, France\label{aff73}
\and
NASA Goddard Space Flight Center, Greenbelt, MD 20771, USA\label{aff74}
\and
Department of Physics and Helsinki Institute of Physics, Gustaf H\"allstr\"omin katu 2, 00014 University of Helsinki, Finland\label{aff75}
\and
Universit\'e de Gen\`eve, D\'epartement de Physique Th\'eorique and Centre for Astroparticle Physics, 24 quai Ernest-Ansermet, CH-1211 Gen\`eve 4, Switzerland\label{aff76}
\and
Department of Physics, P.O. Box 64, 00014 University of Helsinki, Finland\label{aff77}
\and
Helsinki Institute of Physics, Gustaf H{\"a}llstr{\"o}min katu 2, University of Helsinki, Helsinki, Finland\label{aff78}
\and
Centre de Calcul de l'IN2P3/CNRS, 21 avenue Pierre de Coubertin 69627 Villeurbanne Cedex, France\label{aff79}
\and
Laboratoire d'etude de l'Univers et des phenomenes eXtremes, Observatoire de Paris, Universit\'e PSL, Sorbonne Universit\'e, CNRS, 92190 Meudon, France\label{aff80}
\and
Aix-Marseille Universit\'e, CNRS, CNES, LAM, Marseille, France\label{aff81}
\and
NOVA optical infrared instrumentation group at ASTRON, Oude Hoogeveensedijk 4, 7991PD, Dwingeloo, The Netherlands\label{aff82}
\and
Dipartimento di Fisica "Aldo Pontremoli", Universit\`a degli Studi di Milano, Via Celoria 16, 20133 Milano, Italy\label{aff83}
\and
INFN-Sezione di Milano, Via Celoria 16, 20133 Milano, Italy\label{aff84}
\and
University of Applied Sciences and Arts of Northwestern Switzerland, School of Computer Science, 5210 Windisch, Switzerland\label{aff85}
\and
Universit\"at Bonn, Argelander-Institut f\"ur Astronomie, Auf dem H\"ugel 71, 53121 Bonn, Germany\label{aff86}
\and
INFN-Sezione di Roma, Piazzale Aldo Moro, 2 - c/o Dipartimento di Fisica, Edificio G. Marconi, 00185 Roma, Italy\label{aff87}
\and
Dipartimento di Fisica e Astronomia "Augusto Righi" - Alma Mater Studiorum Universit\`a di Bologna, via Piero Gobetti 93/2, 40129 Bologna, Italy\label{aff88}
\and
Department of Physics, Institute for Computational Cosmology, Durham University, South Road, Durham, DH1 3LE, UK\label{aff89}
\and
University of Applied Sciences and Arts of Northwestern Switzerland, School of Engineering, 5210 Windisch, Switzerland\label{aff90}
\and
Institut d'Astrophysique de Paris, 98bis Boulevard Arago, 75014, Paris, France\label{aff91}
\and
Institute of Physics, Laboratory of Astrophysics, Ecole Polytechnique F\'ed\'erale de Lausanne (EPFL), Observatoire de Sauverny, 1290 Versoix, Switzerland\label{aff92}
\and
California Institute of Technology, 1200 E California Blvd, Pasadena, CA 91125, USA\label{aff93}
\and
Institut de F\'{i}sica d'Altes Energies (IFAE), The Barcelona Institute of Science and Technology, Campus UAB, 08193 Bellaterra (Barcelona), Spain\label{aff94}
\and
School of Mathematics, Statistics and Physics, Newcastle University, Herschel Building, Newcastle-upon-Tyne, NE1 7RU, UK\label{aff95}
\and
Waterloo Centre for Astrophysics, University of Waterloo, Waterloo, Ontario N2L 3G1, Canada\label{aff96}
\and
Department of Physics and Astronomy, University of Waterloo, Waterloo, Ontario N2L 3G1, Canada\label{aff97}
\and
Perimeter Institute for Theoretical Physics, Waterloo, Ontario N2L 2Y5, Canada\label{aff98}
\and
Universit\'e Paris-Saclay, Universit\'e Paris Cit\'e, CEA, CNRS, AIM, 91191, Gif-sur-Yvette, France\label{aff99}
\and
Institute of Space Science, Str. Atomistilor, nr. 409 M\u{a}gurele, Ilfov, 077125, Romania\label{aff100}
\and
Consejo Superior de Investigaciones Cientificas, Calle Serrano 117, 28006 Madrid, Spain\label{aff101}
\and
Dipartimento di Fisica e Astronomia "G. Galilei", Universit\`a di Padova, Via Marzolo 8, 35131 Padova, Italy\label{aff102}
\and
Institut f\"ur Theoretische Physik, University of Heidelberg, Philosophenweg 16, 69120 Heidelberg, Germany\label{aff103}
\and
Institut de Recherche en Astrophysique et Plan\'etologie (IRAP), Universit\'e de Toulouse, CNRS, UPS, CNES, 14 Av. Edouard Belin, 31400 Toulouse, France\label{aff104}
\and
Universit\'e St Joseph; Faculty of Sciences, Beirut, Lebanon\label{aff105}
\and
Departamento de F\'isica, FCFM, Universidad de Chile, Blanco Encalada 2008, Santiago, Chile\label{aff106}
\and
Universit\"at Innsbruck, Institut f\"ur Astro- und Teilchenphysik, Technikerstr. 25/8, 6020 Innsbruck, Austria\label{aff107}
\and
Satlantis, University Science Park, Sede Bld 48940, Leioa-Bilbao, Spain\label{aff108}
\and
Centre for Electronic Imaging, Open University, Walton Hall, Milton Keynes, MK7~6AA, UK\label{aff109}
\and
Instituto de Astrof\'isica e Ci\^encias do Espa\c{c}o, Faculdade de Ci\^encias, Universidade de Lisboa, Tapada da Ajuda, 1349-018 Lisboa, Portugal\label{aff110}
\and
Universidad Polit\'ecnica de Cartagena, Departamento de Electr\'onica y Tecnolog\'ia de Computadoras,  Plaza del Hospital 1, 30202 Cartagena, Spain\label{aff111}
\and
INFN-Bologna, Via Irnerio 46, 40126 Bologna, Italy\label{aff112}
\and
Kapteyn Astronomical Institute, University of Groningen, PO Box 800, 9700 AV Groningen, The Netherlands\label{aff113}
\and
Infrared Processing and Analysis Center, California Institute of Technology, Pasadena, CA 91125, USA\label{aff114}
\and
INAF, Istituto di Radioastronomia, Via Piero Gobetti 101, 40129 Bologna, Italy\label{aff115}
\and
Department of Physics, Oxford University, Keble Road, Oxford OX1 3RH, UK\label{aff116}
\and
ICL, Junia, Universit\'e Catholique de Lille, LITL, 59000 Lille, France\label{aff117}
\and
Department of Physics and Astronomy, University of British Columbia, Vancouver, BC V6T 1Z1, Canada\label{aff118}
\and
Department of Physics and Astronomy, University of Southampton, Southampton, SO17 1BJ, UK\label{aff119}}         
%
%
\abstract{We report on serendipitous \Euclid observations of previously known transients, using the \Euclid Q1 data release. By cross-matching with the Transient Name Server (TNS) we identify 164 transients that coincide with the data release. Although the \Euclid Q1 release only includes single-epoch data, we are able to make \Euclid photometric measurements at the location of 161 of these transients. \Euclid obtained deep photometric measurements or upper limits of these transients in the \IE, \YE, \JE, and \HE bands at various phases of the transient light-curves, including before, during, and after the observations of ground-based transient surveys. Approximately 70\% of known transients reported in the six months before the \Euclid observation date and with discovery magnitude brighter than 24 were detected in \Euclid $\IE$ images. Our observations include one of the earliest near-infrared detections of a Type~Ia supernova (SN~2024pvw) 15~days prior to its peak brightness, and the late-phase (435.9~days post peak) observations of the enigmatic core-collapse SN~2023aew. \Euclid deep photometry provides valuable information on the nature of these transients such as their progenitor systems and power sources{, with late time observations being a uniquely powerful contribution}. In addition, \Euclid is able to detect the host galaxies of some transients that were previously classed as hostless. The Q1 data demonstrate the power of the \Euclid data even with only single-epoch observations available, as will be the case for much larger areas of sky in the Euclid Wide Survey.}
%
%
%
%
   \titlerunning{\Euclid: Q1. Photometric studies of known transients}
   \authorrunning{C.~Duffy et. al}
   \maketitle
%
%
%
%
\section{\label{sec:Intro}Introduction}
The primary objective of the \Euclid mission is to constrain cosmological parameters by observing distant galaxies \citep{EuclidSkyOverview}. Although \Euclid does not focus on time-domain astronomy, the potential exploration of the transient sky with \Euclid has long been considered \citep{Laureijs11,Astier14}. \Euclid's high sensitivity and high resolution using the VIS and NISP instruments \citep{EuclidSkyVIS,EuclidSkyNISP} make it ideal for the detection of (faint) point sources, like transients. The discovery of AT~2023adqt by \Euclid clearly demonstrated its capability to find transients by conducting image subtractions of its own images \citep{EuclidSkyOverview}. The high near-infrared (NIR) sensitivity in principle allows us to discover intrinsically red, dust enshrouded, and/or high-redshift transients that often escape detection by optical transient surveys. The recent discovery of high-redshift supernovae (SNe) from \textit{The James Webb Space Telescope} (JWST) showed the potential of exploring NIR time-domain astronomy from space \citep{DeCoursey24}.

Within the \Euclid survey, Euclid Deep Fields (EDFs) are planned to be observed repeatedly throughout the survey period \citep{Scaramella-EP1}. Although the cadence of the EDF observations is sparse, many Type~Ia and core-collapse SNe are expected to be detected, providing valuable information on their progenitor stars, rates, and intrinsic physical properties, as well as their dust formation. The \Euclid observations of Type~Ia SNe, combined with ground-based survey data, can improve their use as cosmological distance indicators, and hence improve constraints on the cosmological parameters \citep{Bailey23}. In addition, a large number of high-redshift superluminous and pair-instability SNe are expected to be discovered in the EDFs, depending on their uncertain event rates at high redshifts \citep{Inserra18,Moriya22,Tanikawa23,briel2024}. 

Most of the \Euclid survey time is devoted to the Euclid Wide Survey (EWS) in which each survey field is only visited once \citep{Scaramella-EP1}. In such a field, it is impossible to search for transients by only using the single-epoch \Euclid data. However, live transients can serendipitously appear in \Euclid exposures, providing photometry in the optical ($\IE$) and NIR ($\YE$, $\JE$, and $\HE$) bands. 

Depending on the timing of the \Euclid observation relative to the transients' evolution, these \Euclid detections can be very early (i.e., shortly after explosion in the case of explosive transients), very late, or at any phase in between. Early detections, or even upper limits close to the time of explosion, are particularly valuable, since they constrain the date of explosion, which is essential to estimate the properties of their progenitors. For example, the early photometric observations of Type~II SNe have been used to infer the progenitor radii and circumstellar environments \citep[e.g.,][]{Gall15,Gonzalez-Gaitan15,Forster18}. The early observations of Type~Ia SNe contain information on their progenitor systems, as well as their explosion mechanisms \citep[e.g.,][]{Maoz14,Deckers22}. 

Even if the \Euclid observation is long before the appearance of transients, the deep \Euclid data will allow us to identify the progenitors and/or host galaxies of the transients. The late-phase observations of transients can provide information on the amount of the radioactive nuclei \citep[e.g.,][]{Leibundgut03,Tucker22}, possible dust formation \citep[e.g.,][]{Lucy89,Kotak09,Fox11,Gall11,Gall14,Szalai19}, and circumstellar environments \citep[e.g.,][]{Fox13} of the transients.

The \Euclid Q1 \citep{Q1cite} data release provides data corresponding to a single visit of the EDFs, and therefore mimics what we expect to obtain in the EWS. In this paper, we report \Euclid photometry of the transients that were identified by other transient surveys before and after the \Euclid observations of the Q1 fields. {We show that in the absence of multiple epochs for difference imaging, using PSF-fitting photometry \Euclid can provide essential information on the nature of these transients and their environments through its deep photometry from single-epoch data. This serves to underline the prospect for transient science with \Euclid in the EWS. Furthermore, it highlights the future potential for transient detection that can be achieved through image differencing in \Euclid fields with multi-epoch observations, e.g., the EDFs.}

Throughout this work, the photometric measurements are reported in the AB system, unless otherwise specified. We use the \textit{Planck} 2018 flat $\Lambda $CDM cosmology model \citep[$\Omega_{\rm m}=0.31, H_{0}=67.7\,\rm kms^{-1}$;][]{planck2018i}.

\section{Data and target selection}
\subsection{Q1 data products}

The Q1 data release \citep{Q1-TP001} contains several different data products. For the purpose of this work we make use of the background-subtracted mosaic image files. \Euclid imaging observations are composed of four dithered exposures taken in sequence for each filter. These have been median-stacked to create an intermediate data product (not released as part of Q1, but anticipated for future releases), which eliminates chip gaps and reduces the effects of artefacts such as cosmic ray strikes. The stacked images have then been trimmed and combined with other images to produce the mosaic image files that correspond to predefined sky tiles. This data were accessed using the ESA Datalabs service \citep{Navarro2024}.

Since mosaic images are composites of multiple \Euclid observations, they are not supplied with time information. Nevertheless, using the \Euclid science archive we were able to identify which observations made up a given part of a mosaic coterminous with a location of interest. This allowed us to add the necessary time information for each of the objects we consider in this work. In future data releases we anticipate that we would make use of the intermediate stacked image product since they are to be provided with time information.

\subsection{Target selection}\label{sec:selection}
\begin{figure*}
    \centering
    \includegraphics[width=0.9\linewidth]{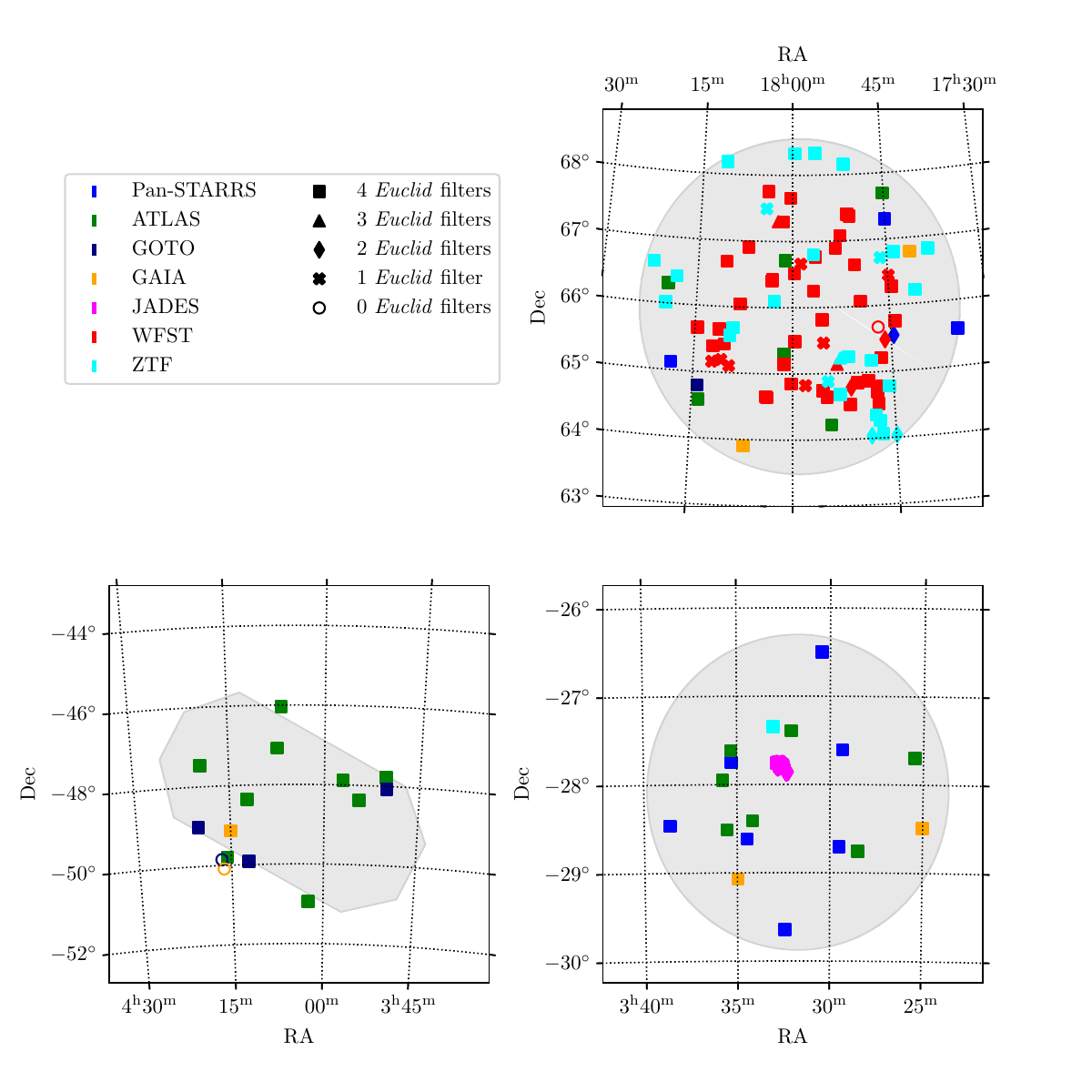}
    \caption{Sky regions of the three Euclid Deep fields. \textit{Top right}: the EDF-North. \textit{Bottom left}: the EDF-South. \textit{Bottom right}: the EDF-Fornax. In each panel the footprints of the EDFs have been overlaid with the positions of transients considered in this work. Colour coding indicates the discovery survey. Marker shape denotes the number of \Euclid filters in which we were able to make a measurement (including upper limits). Those with zero \Euclid filters had exposures available, but  were not suitable for performing the photometric procedure, so no photometric measurement (or upper limit) could be derived.}
    \label{fig:deep-field-area}
\end{figure*}

\begin{table}[hb]
    \caption{Euclid Deep Field spatial information and approximate time of observation in Q1. }
    \smallskip
    \label{tab:Q1Area}
    \smallskip
    \begin{tabular}{llll}
    \hline
    \hline
        EDF & Field Centre& Area &  Date\\
            &  (RA Dec) & (deg$^{2}$) & (2024)\\
        \hline
       Fornax & {0}\ra{03;31;43.6},   \ang{-28;05;18.6} & 10\, & {05--08 Aug} \\
       North & \ra{17;58;55.9},  +\ang{66;01;03.7} & 20\, & {17--19 Jul} \\
       South & {0}\ra{04;04;57.84},   \ang{-48;25;22.8} & 23\, & {05--08 Sep} \\
        \hline
    \end{tabular}
\end{table}

The Q1 data release nominally covers the EDFs (see \autoref{tab:Q1Area}), though in reality it is composed of the mosaic sky tiles that overlap with the EDFs. As such, the data release covers an area slightly larger than that defined by the EDFs. As a consequence of this, some known transients, which we identified outside the EDFs, have been included in this work since data is available at their reported location (see Fig. \ref{fig:deep-field-area} for examples of this).

We selected known transients that overlap both spatially and temporally with the Q1 data products. In order to select a manageable number of transients for consideration, we limited the selection to relatively recent or bright transients. 
The selection criteria were applied to each field separately because the \Euclid observation dates of the fields differ (see \autoref{tab:Q1Area}). The selection criteria used were:
\begin{itemize}
    \item any transient reported after the \Euclid observations up to {5 December 2024}\footnote{Date on which target list was finalised.};
    \item any transient reported up to 1 year prior to \Euclid observations;
    \item any transient reported from between 1--3 years prior to \Euclid observations with a discovery brightness of $18$ or brighter in the band quoted by the relevant discovery survey. 
    \end{itemize}

\begin{figure}
    \centering
    \includegraphics[width=\linewidth]{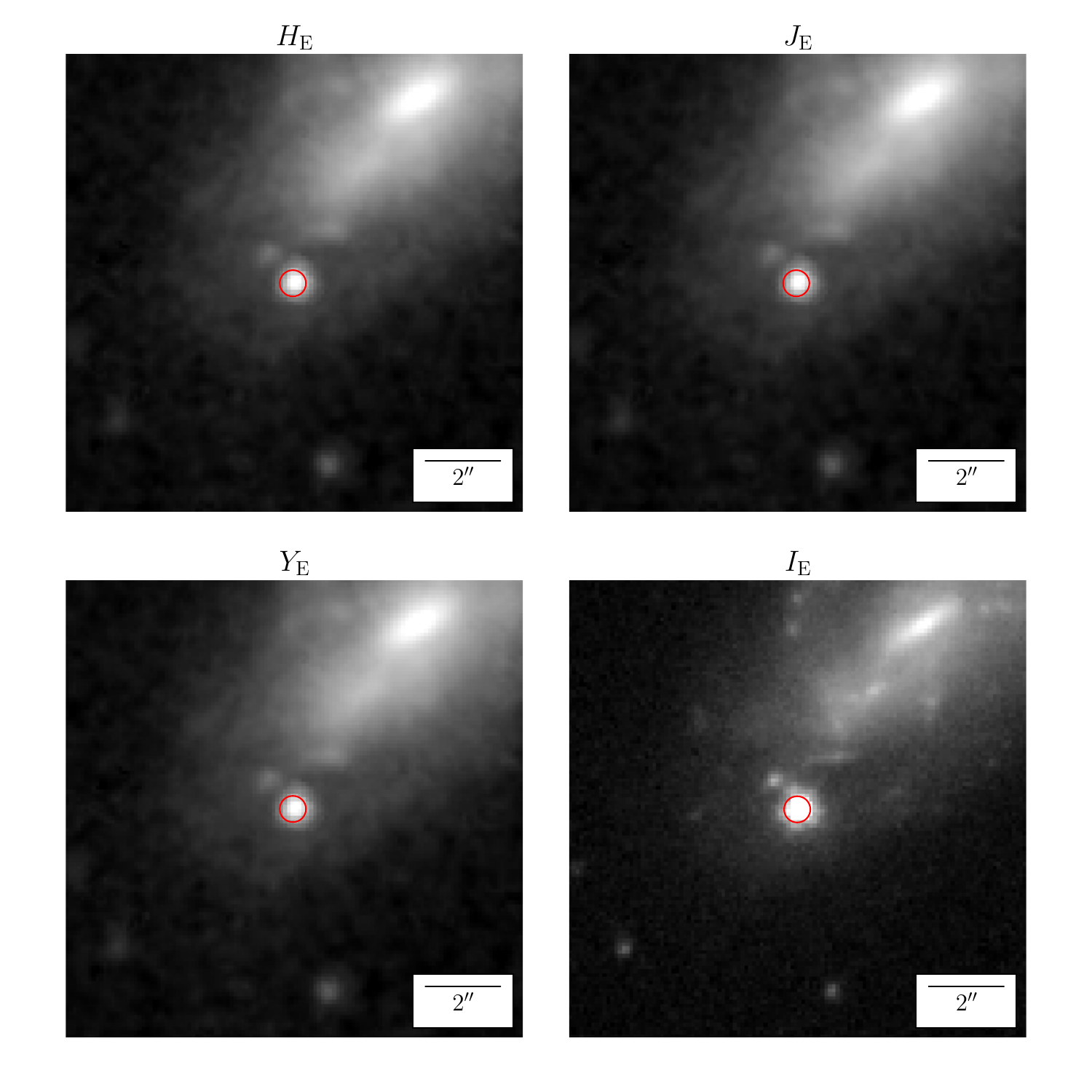}
    \caption{Images of AT 2024pnv in each of the \Euclid bands. The transient is highlighted with a red circle centred on the reported coordinates from TNS.} 
    \label{fig:example-cutouts}
\end{figure}

A total of 164 transients reported to the Transient Name Server (TNS)\footnote{\href{https://sandbox.wis-tns.org}{https://sandbox.wis-tns.org}} meet these criteria. Their positions are shown in Fig. \ref{fig:deep-field-area} and overlaid on the footprints of the EDFs. In \autoref{tab:completeList} we show the complete catalogue of (161) transients with recovered \Euclid photometry that are included in this figure. The discrepancy in source number arises from there being three sources within the Q1 area that it was not possible to make any form of measurement on. This is because they lie at the edge of the field, or an image, making them unsuitable for the photometric procedure detailed above in any filter, whilst still making up part of the Q1 release. \autoref{tab:surveys_statistics} details the distribution of the discovery surveys of the transients that we follow up. In Fig. \ref{fig:example-cutouts} we show the \Euclid images in each of the four \Euclid bands of one example transient, AT~2024pnv. In \autoref{apx:cutouts} we show the $\IE$ cutouts of each of the targets.

\begin{table}[t]
    \caption{Discovery survey statistics of the 161 sources followed up with \Euclid, as listed in \autoref{tab:completeList}.}
    \smallskip
    \label{tab:surveys_statistics}
    \smallskip
    \centering
    \begin{tabular}{lc}
    \hline
    \hline
        Survey & Number of soures\\
        \hline
     Pan-STARRS$^{a}$ & 11 \\     
       ATLAS$^{b}$ & 22 \\
     GOTO$^{c}$ & 4  \\     
     \textit{Gaia}$^{d}$ & 5 \\   
      JADES$^{e}$ & 49 \\
     WFST$^{f}$ & 49  \\
       ZTF$^{g}$ & 28  \\
       \hline
    \end{tabular}
    {\begin{flushleft}
\footnotesize{$^{a}$ Panoramic Survey Telescope and Rapid Response System, \citet{2018AAS...23143601F}. $^{b}$ Asteroid Terrestrial-impact Last Alert System, \citet{Tonry2018b}. $^{c}$ Gravitational-wave Optical Transient Observer, \citet{2024SPIE13094E..1XD}. $^{d}$\citet{2016A&A...595A...1G}, $^{e}$ JWST Advanced Deep Extragalactic Survey. \citet{2023arXiv230602465E,DeCoursey24}. $^{f}$ Wide Fast Survey Telescope, \citet{2023SCPMA..6609512W}. $^{g}$ Zwicky Transient Facility, \citet{bellm2019,graham2019}.}
\end{flushleft}}
\end{table}

\section{Euclid photometry}

\subsection{PSF photometry}\label{sec:euclid_photomtry}

We measured the photometry of each selected transient by fitting the point spread-function (PSF) using the \texttt{ecsnoopy} package\footnote{\texttt{ecsnoopy} is a Python package developed by E. Cappellaro for transient photometry through PSF-fitting and/or template subtraction. It extensively utilises \texttt{astropy} \citep{2022ApJ...935..167A} and, in particular, \texttt{photutils} \citep{larry_bradley_2024_12585239}. A detailed description of the package is available at \href{http://sngroup.oapd.inaf.it/ecsnoopy.html}{http://sngroup.oapd.inaf.it/ecsnoopy.html}.}. The PSF fitting procedure consists of several steps: modelling the PSF on the images; subtracting the background; fitting the source using the PSF model; and generating both a model image of the fitted source and a residual image (the difference between the observed data and the fitted model) to evaluate the quality of the fit. We construct the PSF model by selecting a sample of several dozen isolated, bright stars, using the background-subtracted images for both VIS and the three filters of NISP.

{Accurate background subtraction is crucial to account for potential contamination to the PSF fitting process from the `local background', i.e., host galaxy light and any small-scale residuals that may remain after background subtraction during image processing. With multi-epoch observation this can be addressed using template subtraction, particularly when the transient is located within a bright galaxy, near a galaxy nucleus, or embedded in a compact galaxy. However, since we only have a single epoch of observations, template image subtraction cannot be applied here, and we must estimate  the `local background' around the transient position. In order to do this we interpolate the counts measured outside a circular region centred on the source, typically with a radius of twice the Full Width at Half Max (FWHM; depending on the specific case). This value was then subtracted from the counts at the transient’s position. The `local background' noise was quantified as the standard deviation of the local background counts.}

{In the fitting process, the PSF centre is allowed to shift to account for uncertainties in the transient coordinates reported by the discovery survey. The offset between the coordinates of the source detected on \Euclid image and the coordinates of the transient reported by TNS is then measured (see \autoref{tab:completeList}). }

We note that a source detected in the single-epoch \Euclid image, without a template image as reference, may not correspond to the transient reported by the TNS. It could instead be a nearby contaminating stellar source, a compact \ion{H}{ii} region, or, for more distant sources, an unresolved host galaxy. On the other hand, some transients labelled by Lasair as an {\it orphan} (see Sect.~\ref{sec:ztf_survey}), i.e., those lacking a detected host galaxy, are detected in the \Euclid images near a faint extended source -- likely the host galaxy -- that is only visible due to the superior depth of the \Euclid exposures. This highlights the potential of the \Euclid images for various applications, such as identifying the hosts of apparent orphan transient sources (see Sect.~\ref{sec:hostless}) or pinpointing the progenitor stars of nearby transients.

\begin{figure*}
    \centering
    \begin{minipage}{.33\textwidth}
    \fbox{\includegraphics[width=0.9\linewidth]{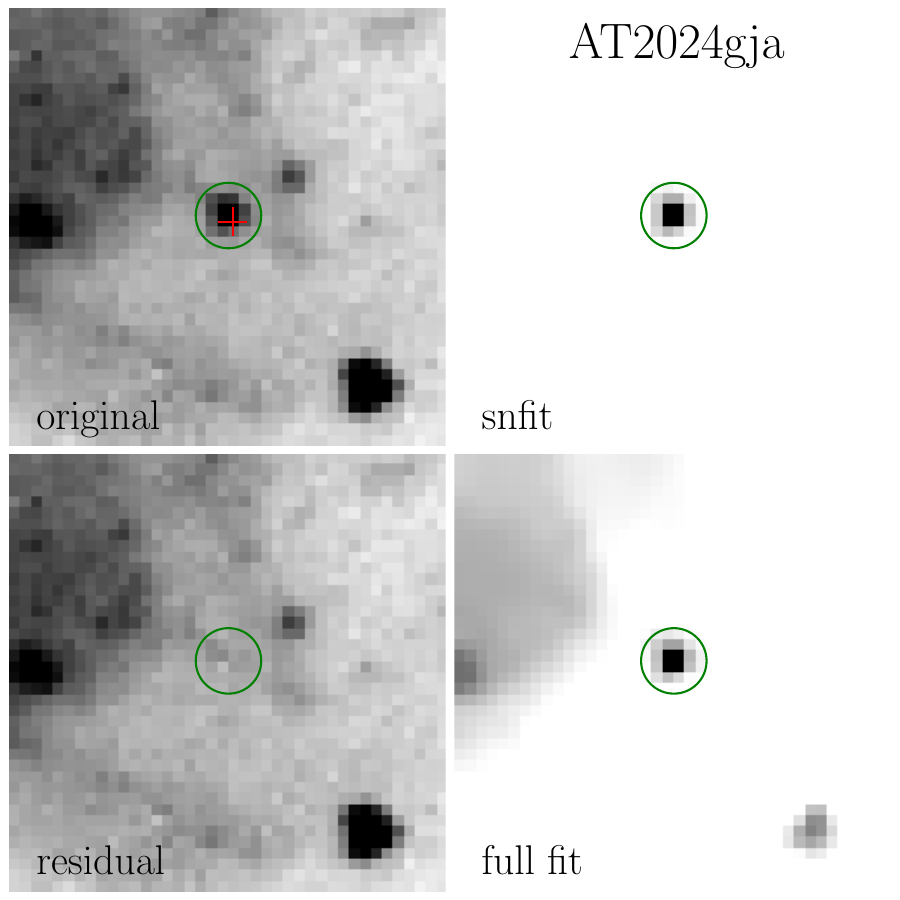}}
\end{minipage}%
\begin{minipage}{.33\textwidth}
    \centering
    \fbox{\includegraphics[width=0.9\linewidth]{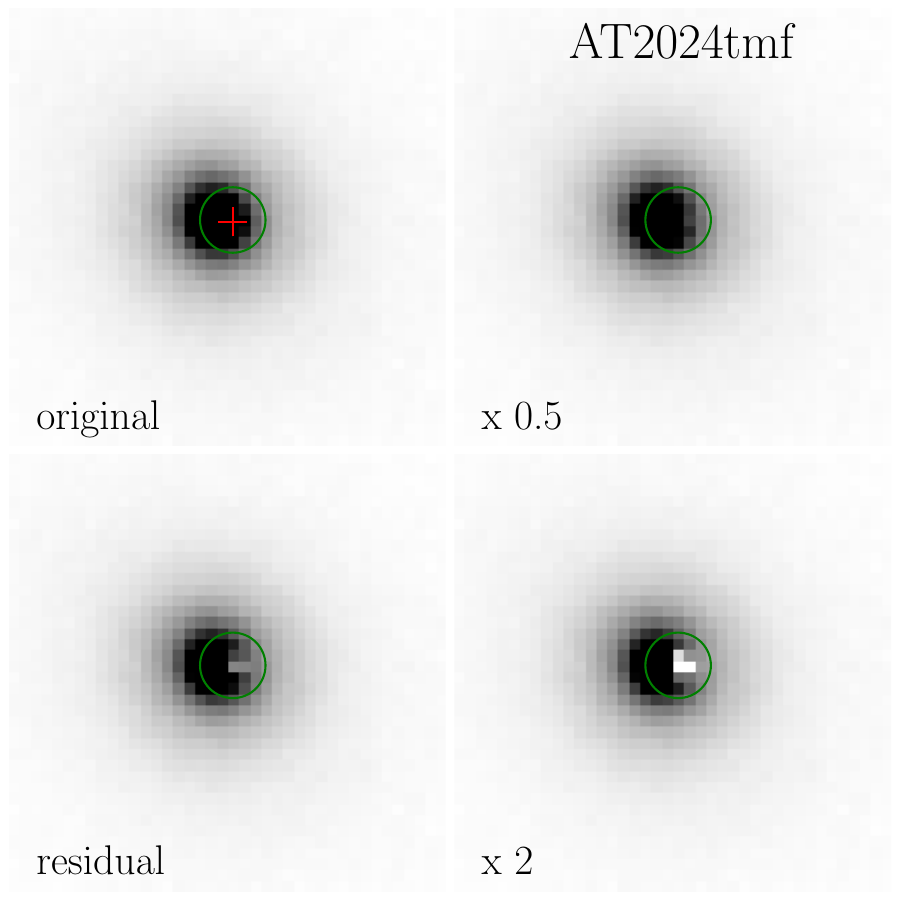}}
\end{minipage}%
\begin{minipage}{.33\textwidth}
    \centering
    \fbox{\includegraphics[width=0.9\linewidth]{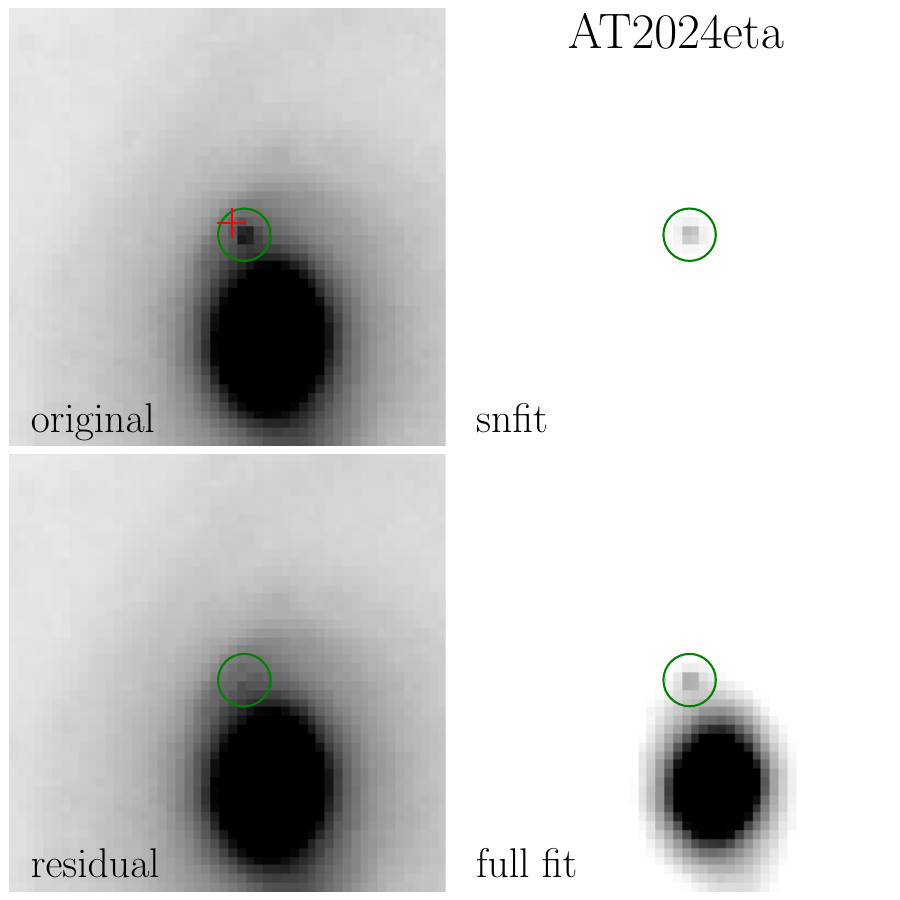}}
\end{minipage}%
\caption{
\textit{Left}: Example of the PSF fitting for AT 2024gja in $\IE$. The figure shows clockwise from top-left; a cutout of the original frame, the fitted PSF, the full fit including PSF and background, and the residual after subtraction of the fitted source. The green circle is centred at the fitted position, whereas the red cross marks the transient position reported in TNS.
\textit{Centre}: Example of the $\IE$ magnitude limit estimate for AT 2024tmf. The limit is intended as the magnitude of the faintest source that can remain hidden in the background noise. The lower left panel shows the residual after subtraction of a PSF with $\rm S/N=2.5$. To highlight the significance of this limit, the two right panels show the residual for a PSF scaled by 0.5 and 2.0 of the adopted limit, corresponding to a star 0.75 mag fainter and brighter, respectively, than the adopted limit.
\textit{Right}: Example of the $\IE$ PSF fitting for AT 2024eta. As explained in Sect.~\ref{sec:euclid_photomtry}, because of the low S/N and relatively large offset from the discovery position, for this transient we conservatively report an upper limit of $23.6$. Each cutout shown is $5\arcsec\times5\arcsec$.
}
\label{fig:psf_fitting}
\end{figure*}

With these considerations in mind, the significance of a detection using PSF-fitting can be defined in terms of several factors: the S/N of the detection; the positional uncertainties of the source; and the potential presence of a host galaxy, particularly for orphan transients. A detection is considered significant if the fitted source is clearly distinguishable from background noise (i.e., the ratio between the peak of the fitted PSF and the `local background' noise in the residual image exceeds 2.5\footnote{{The detection threshold and S/N over the full stellar profile are related for a given profile width \citep{1990PASP..102..949H}. For \Euclid, where stars have FWHM of 2 pixels, a detection threshold of 2.5 corresponds to an S/N of approximately 3.}).}. Additionally, the source's position must align within a reasonable uncertainty radius with the transient position reported by TNS (ranging from \ang{;;0.1} to \ang{;;0.3}, depending on the accuracy of the astrometric calibration of the survey that discovered the transient), and there must be a clear positional offset between the transient and the host galaxy centre (indicating that the transient is distinct from the host in the case of orphan transients).

If the detection exceeds the S/N threshold, the positional uncertainty is small, and there is a clear offset between the transient's position and the centre of the host galaxy, the detection is considered significant. However, if the S/N is low or the offset from the TNS reported position is large, the detection is flagged as uncertain, and an upper limit on the transient’s flux may be reported instead.

Examples of (i) a robust detection, (ii) an upper limit, and (iii) an inconclusive case are outlined in the following and in Fig.~\ref{fig:psf_fitting}. The three examples are as follows:
\begin{enumerate}[(i)]
    \item The robust detection of AT~2024gja (Fig.~\ref{fig:psf_fitting}, left). The detected source has brightness $24.108\pm0.026$, $\rm S/N \simeq30$ and a position consistent with that reported by TNS (with an offset of \ang{;;0.037}).
    \item The upper limit for AT~2024tmf (Fig.~\ref{fig:psf_fitting}, centre). Because the transient is located close to the galaxy centre we could report only a relatively bright upper limit of $21.3$ in the absence of a template image to remove the host galaxy background.
    \item The inconclusive detection of SN~2024eta (Fig.~\ref{fig:psf_fitting}, right). A faint source with $\IE \simeq23.6$ and $\rm S/N=1.5$ was detected \ang{;;0.22} away from the position reported by TNS. Conservatively, we report this as an upper limit in the Q1 release.
\end{enumerate}

The magnitudes measured are calibrated using the photometric zero point provided in the data release {\citep[see][for details]{Q1-TP002,Q1-TP003}. In addition to PSF-fitting magnitudes, we also measured aperture magnitudes. The two measurements show no significant systematic offset. We prefer PSF photometry becasuse it isolates the stellar-like source and is less sensitive to deviant pixels.}
As previously mentioned, and as shown in Fig.~\ref{fig:deep-field-area}, there are three sources contained within the Q1 release that were unsuitable for the photometric procedure outlined above because they are located near the edge of an image. Similarly, there are several sources in \autoref{tab:completeList} where a detection or upper limit is reported in one or more \Euclid filters, while other filters have no data.

\subsection{Overview of photometric measurements}

The photometric measurements for all sources, both detections and upper limits, are provided in \autoref{tab:completeList}, with more detail in the supplementary online table. For each source, we list the magnitude with its associated error, a flag indicating whether it is a robust source detection or an upper limit, the detection threshold, and, for each source detection, the distance (in arcseconds) between the PSF fit centre and the transient position reported by TNS. Additional notes are included where necessary.

We report detections for 59 sources in \IE and 40, 36, and 30 in \YE, \JE, and \HE, respectively.  Their magnitudes range from 17 to 26 in both VIS and NIR. In general, VIS frames are deeper and, due to the sharper PSF, allow measurement of fainter magnitudes, even in brighter and more crowded backgrounds. Indeed, the average magnitude of the detected sources is $23.0$ in \IE and $21.7$ in \JE. Because of the superior depth of the VIS images, there were no NIR-only detections (i.e., all objects that were detected in the NIR were also detected in the VIS image). 

We also measured upper limits for 93 sources in \IE and 91, 88, and 93 in \YE \JE, and \HE, respectively. In favourable conditions, such as an isolated transient, the upper limiting magnitude for \IE reaches as faint as 28. In total, we were able to obtain a measurement or an upper limit for 163 transients reported by TNS. This number differs from the total reported in Sect. \ref{sec:selection} due to sources found at the edges of images, where it was not possible to recover a suitable area for performing the PSF-fitting procedure. From this point onward, we will consider only those transients for which we were able to obtain photometric measurements.

\begin{figure}
    \centering
    \includegraphics[width=\linewidth]{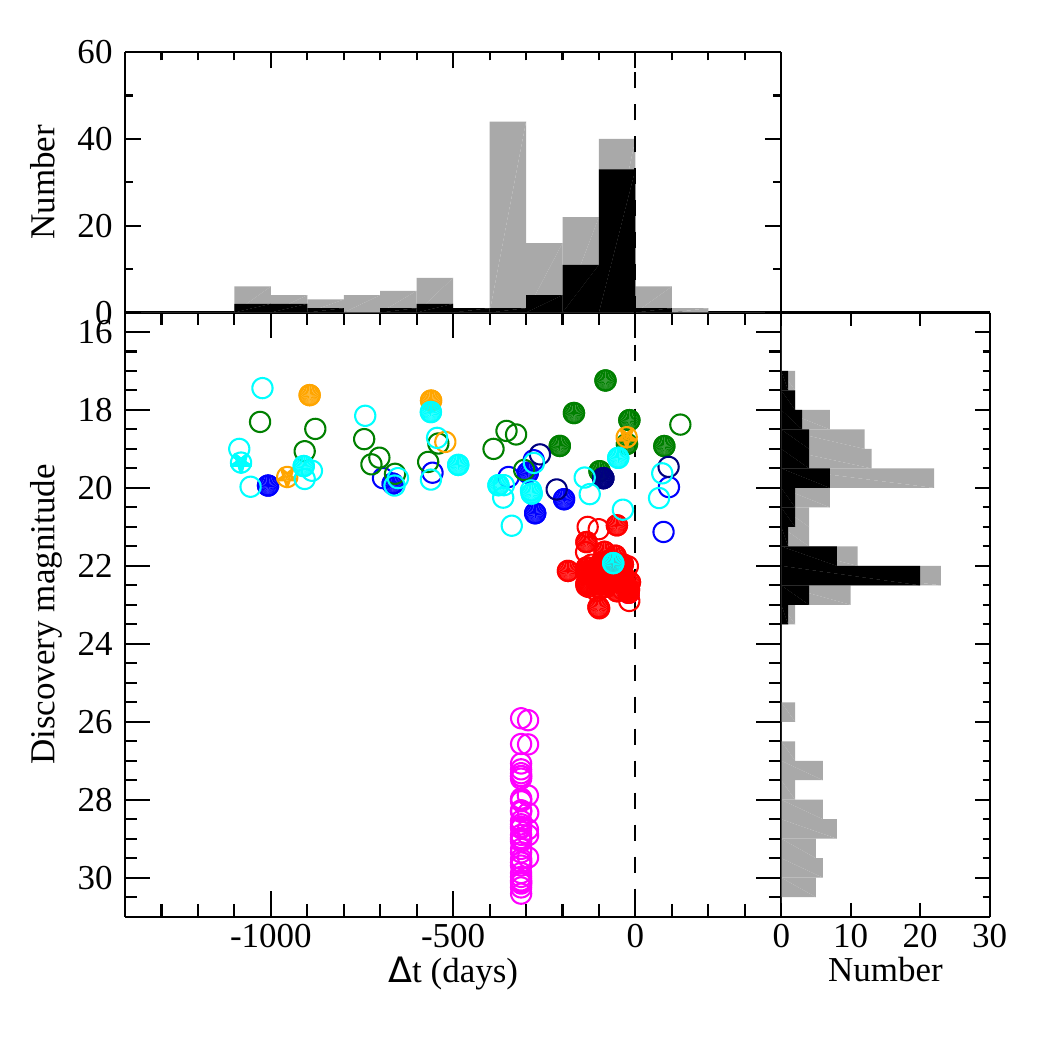}
    \caption{Distribution of the targets in parameter space of discovery magnitude and $\Delta t$, the discovery date minus the \Euclid observation date (in days). Those detected in the \Euclid VIS image are shown with filled symbols, while upper limits are shown with open symbols. Likely supernovae are shown as circles and likely AGN or variable stars are shown with star symbols. The colour of the symbols represents the discovery survey, as in Fig.~\ref{fig:deep-field-area}. The group of sources with discovery magnitudes of about 26 and fainter are all from the JADES survey that was undertaken with JWST \citep[over a small region of sky,][]{2023arXiv230602465E,DeCoursey24}, and are expected to be beyond the limit of \Euclid detection.}
    \label{fig:magdate}
\end{figure}

\autoref{fig:magdate} shows the distribution of our input target list in terms of discovery magnitude and the date difference between the transient's discovery and the \Euclid observation. From this, we can see that the fraction of objects detected in \IE increases as the size of the time difference decreases (i.e., when \Euclid observes the transient closer to its discovery date). Excluding sources with a discovery magnitude fainter than $24$ (this is exclusively JADES sources which we would expected to be too faint for \Euclid) we recover 61\% of transients discovered in the year before \Euclid \IE observation; this rises to 69\% for those discovered in the 6 months before observation, and 82\% for those discovered 100~d before observation.

\begin{figure}
    \centering
    \includegraphics[width=\linewidth]{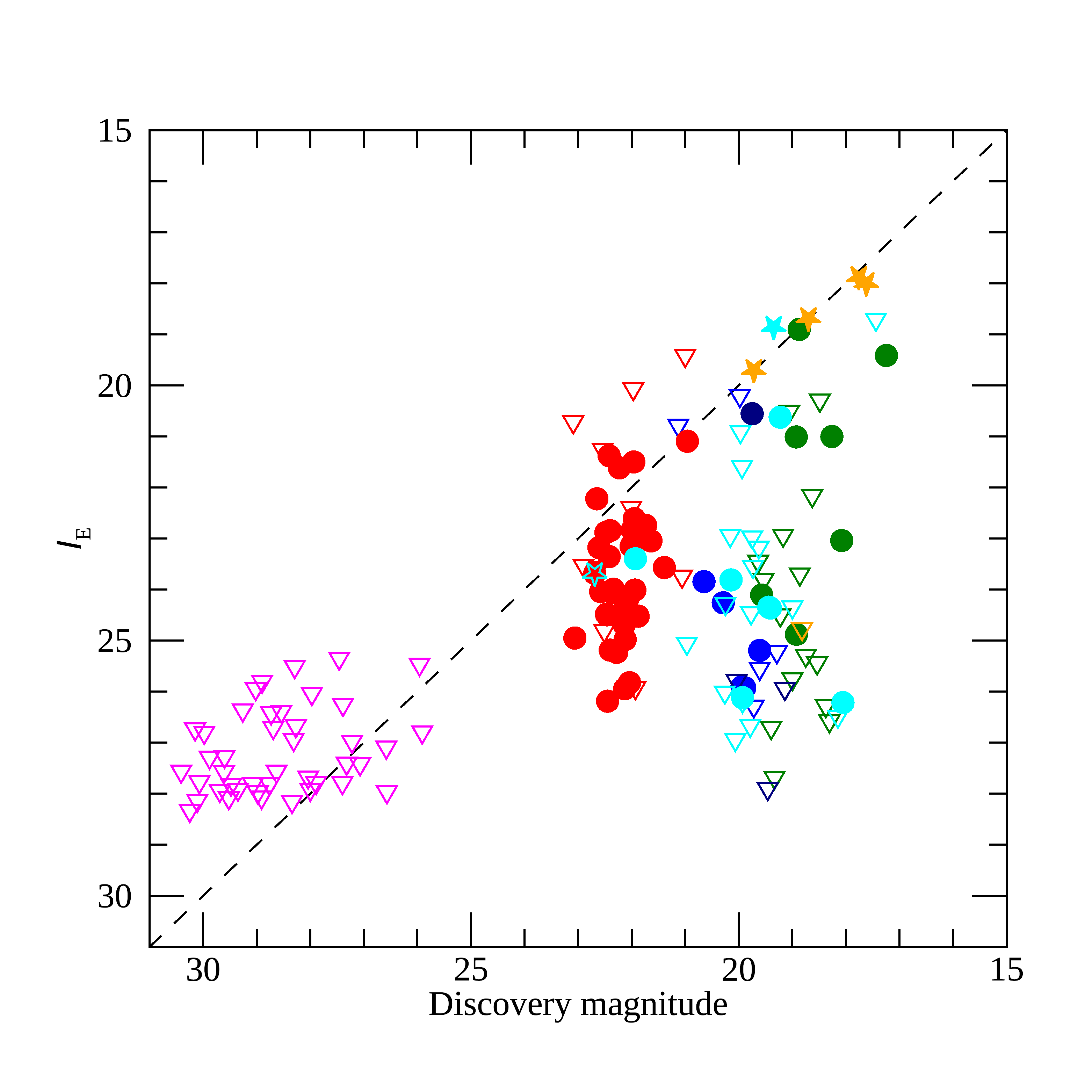}
    \caption{Distribution of the targets in discovery magnitude (in the relevant photometric band of the discovery survey) and \Euclid \IE magnitude. Likely AGN or variable stars are shown with star symbols. Sources detected in the \Euclid VIS image are shown with filled symbols, while upper limits are shown with open triangles for likely SNe or with open star symbols for likely AGN and variable stars. The line of equal magnitude is shown by the dashed line. The colour of the symbols represents the discovery survey as in Fig.~\ref{fig:deep-field-area}.}
    \label{fig:magmag}
\end{figure}

\autoref{fig:magmag} shows the distribution of our input target list in terms of discovery magnitude and \Euclid magnitude (or, if not detected, then the upper limit). The pre-discovery detections allow reliable deep early-time photometry to be provided, see Sect.\,\ref{sec:earlytime} for discussion and examples. The majority of the \Euclid observations presented here are post-discovery, post-peak observations. We see that \Euclid is capable of recovering useful photometry on a transient several magnitudes into the decay of the light curve, see (Sect.\,\ref{sec:latetime} for further discussion and examples.)

\begin{figure*}
    \centering
    \includegraphics[width=\linewidth]{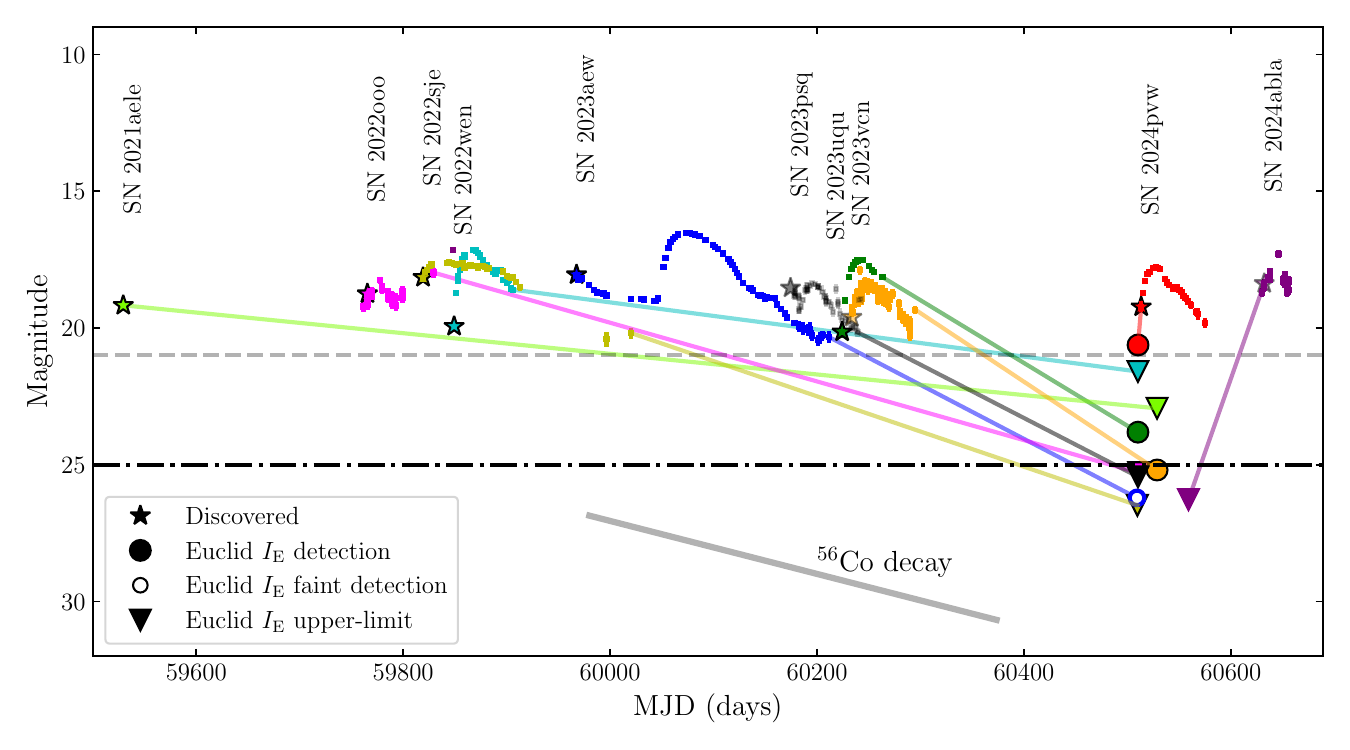}
    \caption{Optical ($r$ and/or $o$-band) light-curves for the 10 transients meeting the selection criteria outlined in Sect. \ref{sec:selection}, which also have a spectroscopic classification. The discovery time is indicated by a star symbol, and each transient is represented by a separate colour. {Thin lines connect the last ground-based photometric points with the \Euclid VIS measurements, as listed in \autoref{tab:completeList}. If the \Euclid epoch was pre-discovery (the case for SN\,2024pvw and SN\,2024abla), we instead connect the \Euclid measurement with the first ground-based point. The $^{56}$Co decay timescale is plotted in grey for reference.} The approximate limiting magnitude of \Euclid VIS imaging for an isolated point source is shown by the black dashed line \citep{EuclidSkyVIS}, and a representative limiting magnitude for ground-based surveys such as ZTF and ATLAS is indicated by the dashed grey line. Detailed light-curves for these transients are given in Figs. \ref{fig:sn2024pvw}, \ref{fig:sn2023uqu}, and \ref{fig:sne_lcs}.}
    \label{fig:lightcurves}
\end{figure*}

\section{\label{sec:groundbased_xmatch} Sources detected in other surveys}

The sources that we follow up in this work have, in addition to \Euclid, been detected in several other surveys. These surveys include optical, high-cadence (typically 2--3 days), large-scale surveys that provide publicly accessible photometry. {Though these surveys may not have originally discovered a source, by their `all sky' nature will have made several observations of a source. }It is by combining the \Euclid photometry, which we present, here with data such as this that we can enhance the scientific impact of \Euclid data for transient science. In the following sections we briefly present some of these surveys and some analysis of those targets as seen by these surveys.

\subsection{\label{sec:atlas_survey} The ATLAS experiment}
{ATLAS is a network of $0.5\,\rm m$ telescopes (two units in Hawaii, one unit in Chile, and one unit in South Africa, with an additional one currently in preparation in Spain), which observing strategy provides identification and orbit constraints for near-earth objects (NEOs) and other potentially hazardous objects.} This full-sky survey naturally provides information about many other types of transients including SNe up to limiting magnitudes of about 20.5--21 in the orange ($o$ band from 560--820~nm) and cyan ($c$ band from 420--650~nm) broad-band filters. Some of the TNS objects included in the list of transients discussed in Sect.~\ref{sec:selection} are ATLAS-discovered transients \citep[][]{Smith2020}.

The ATLAS `forced-photometry' server\footnote{\href{https://fallingstar-data.com/forcedphot/}{https://fallingstar-data.com/forcedphot/}} \citep[][]{Tonry2018a} provides full public access to the photometric measurements over the full history of the survey which started in 2017. Forced-photometry (i.e., photometry measured at a fixed position) was obtained for all the sources listed in \autoref{tab:completeList} by using their coordinates as inputs in the ATLAS photometry server. All measurements with fluxes of S/N ratio lower than 3 were considered as upper limits and filtered out{, as suggested on the photometry server output description page, and the same S/N as the \Euclid detections we report.} The remaining measurements were visually inspected to isolate high-cadence, well-sampled light curves. 

\subsection{\label{sec:ztf_survey} The ZTF experiment}
{ZTF is an experiment based at the Palomar telescope. It has been observing the full northern sky at a cadence of about two days to limiting magnitudes of up to 20.5--21, from beginning of 2018.}
The SDSS \textit{g}- and \textit{r}-band photometry is publicly available and provided by several brokers \citep[e.g., Lasair, ALeRCE, Fink;][]{Young_sherlock_2023,2021AJ....161..141S,2021MNRAS.501.3272M}.
ZTF allows the community to explore a broad range of time-domain science. It can be considered as a pathfinder to the Vera C. Rubin Observatory Legacy Survey of Space and Time (LSST)\footnote{\href{https://https://rubinobservatory.org}{https://rubinobservatory.org}} whose operations are planned to start during 2025. 

Some of the TNS objects discussed in Sect.~\ref{sec:selection} are ZTF-discovered transients. However, not all ZTF transients are reported to TNS. To get the complete list of ZTF transients potentially associated with the transients listed in \autoref{tab:completeList} we cross-match the coordinates of the \Euclid-measured transients with the full ZTF database. To do so, we use the Lasair \citep{2024RASTI...3..362W} watchlist tool\footnote{\href{https://lasair-ztf.lsst.ac.uk/watchlists/}{https://lasair-ztf.lsst.ac.uk/watchlists/}} with an input cross-match radius of \ang{;;0.5}. We find that a total of 42 TNS sources listed in \autoref{tab:completeList} have ZTF data available at their coordinates' location. 

Embedded in Lasair is \texttt{Sherlock} \citep{Young_sherlock_2023} a spatially contextual classifier, which is used to determine the likely object classification. At its most basic \texttt{Sherlock} cross-matches ZTF transients with all major astronomical catalogues that it uses to assign the classification, e.g., AGN, SNe, and variable stars. For sources classified as some type of transient, \texttt{Sherlock} also assigns a likely host or failing that classifies it as an orphan. 

\subsection{\label{sec:groundbased_xmatch_ztf_atlas} Targets detected in ZTF and ATLAS surveys}

To obtain an overview of the full data sample, we produced and inspected individual light curve plots for each of the 161 sources detected with \Euclid (see \autoref{tab:completeList}). Each light curve includes the \Euclid epochs and photometry or upper limits in the \Euclid filters available, the discovery photometry from TNS and, if available, ZTF and ATLAS photometry. The sample contains a total of ten classified SNe while the remaining sources are unclassified, meaning their nature is subject to light curve and contextual interpretations. \autoref{fig:lightcurves} illustrates how the Q1 \Euclid 1 epoch snapshots provide late or early-time information on a pool of classified SNe over a timescale of about three years. {In this figure, for transients discovered before the \Euclid VIS observation, we connect the last ground-based detection with the \Euclid measurement by a thin line. Whether or not the transient was recovered by \Euclid is indicated. If the \Euclid observation was before the first transient report, we instead connect it with the first reported ground-based detection.} Detailed fits have also been made to the light-curves of selected transients (see Sect.~\ref{notable_transients} and Figs.~\ref{fig:sn2024pvw} and \ref{fig:sn2023uqu}).

A total of 42 sources are detected by ZTF, and about 60 sources are well sampled with ATLAS.
From the ten classified SNe eight are detected by ZTF and ATLAS, while two are detected with ATLAS only (see plots in Figs.~\ref{fig:sn2024pvw}, \ref{fig:at2024pcm}, \ref{fig:sn2023uqu}, and \ref{fig:sne_lcs}). Some of these sources are discussed in more detail in Sect. \ref{notable_transients}. A few of the well-sampled light curves are likely from AGN or galactic activities (e.g., AT\,2021vje, Fig.~\ref{fig:other_lcs}), or from stellar activity (e.g., AT\,2022bpn, Fig.~\ref{fig:other_lcs}), as independently corroborated by our \Euclid image analysis and by \texttt{Sherlock}. Finally, a few more sources show unclear SNe-like light curves for which the peak may not be properly sampled (e.g., AT\,2024bkj, Fig.~\ref{fig:other_lcs}).

\subsection{Poorly sampled targets}

The remaining sources are generally poorly sampled and their nature remains unclear in the absence of public spectroscopic classification information. Plots of the 43 faint, JADES-discovered sources show the JWST discovery detections and the \Euclid upper limits, and no high-cadence ATLAS sampling (e.g., AT\,2023adsv and AT\,2023adts, Fig.~\ref{fig:deep_lcs}). The sample of sources detected with the Wide Field Survey Telescope (WFST) also show poorly-sampled light curves, but \Euclid provides deeper imaging than WFST and complementary photometry data points for some of them (e.g., AT\,2024tgn, Fig.~\ref{fig:deep_lcs}). Some statistics on the number of sources detected in the JADES, WFST, and the other surveys (see Fig.~\ref{fig:deep-field-area}) are provided in \autoref{tab:surveys_statistics}.

\section{Notable transients} \label{notable_transients}
\subsection{Transients observed before ground-based transient survey discovery}\label{sec:earlytime}
In this section, we show some examples of transients observed by \Euclid before the ground-based transient survey discovery date.

\subsubsection{SN 2024pvw}\label{sec:SN2024pvw}
\begin{figure}
    \centering 
    \includegraphics[width=\linewidth]{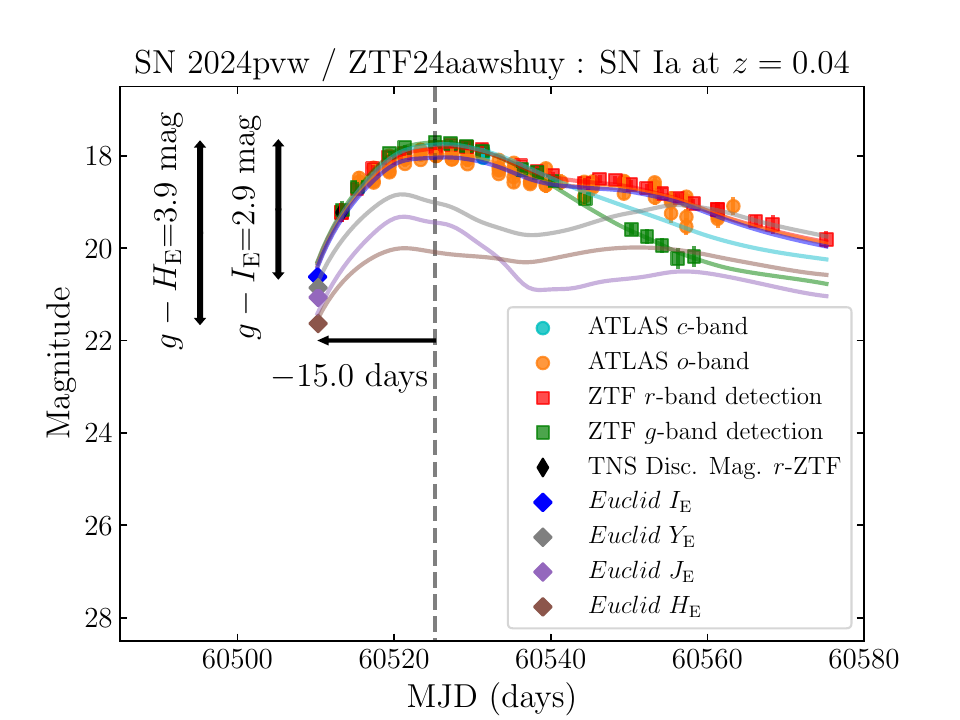}
        \includegraphics[width=\linewidth]
    {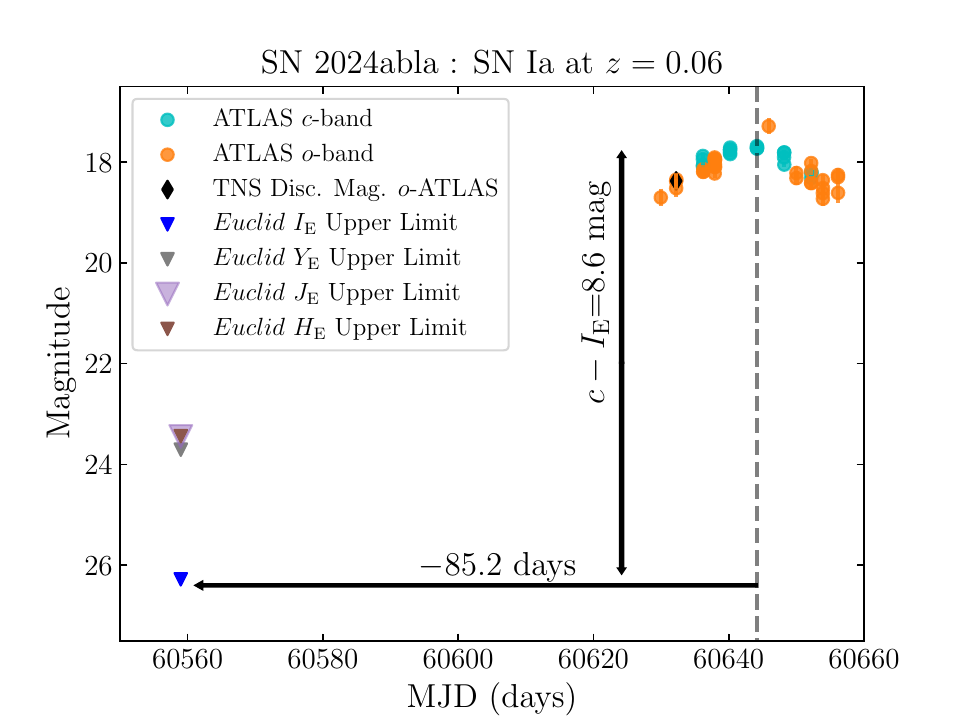}
    \caption{\textit{Top}: Light curve of SN~2024pvw. The \Euclid observations occurred before the ground-based survey discovery, and the resulting \Euclid detections provide the earliest photometry on the light curve. The curves show a Type~Ia SN light curve template in corresponding bands. \textit{Bottom}: Light curve of SN~2024abla. The \Euclid observations occurred almost three months before the peak.}
    \label{fig:sn2024pvw}
\end{figure}

This transient is a known Type~Ia SN \cite[classification reported in TNS by][]{2024TNSCR2607....1S} that was detected by \Euclid before ground-based transient surveys. \autoref{fig:sn2024pvw} (upper panel) shows the \Euclid and ground-based photometry on the light curve. The \Euclid detections were made in all four \Euclid bands. Such early NIR detections are very rare. These detections are 2--3 days earlier than for any Type Ia SNe in the CSP-II Hubble flow cosmological sample \citep{2019PASP..131a4001P}, and about the same phase as for the earliest NIR observations for the iconic nearby Type Ia SN~2011fe \citep{2013ApJ...766...72H}. 

Understanding the {nature} of Type Ia SNe is arguably among the most pressing questions in SN science, especially when considering their use as precision distance estimators in cosmology. While it is considered to be well established that these are explosions of white dwarfs (WDs), it remains unclear whether the binary companion triggering the explosion is another WD, or a non-degenerate star. Another issue of debate is whether the explosion happens at the Chandrasekhar mass or if the main Type Ia SN population used in cosmology arises from sub-Chandrasekhar mass explosions. Early multi-wavelength observations, from hours to days after the explosion, are key for addressing these pressing questions (e.g., \citealt{Maoz14}). For example, early flux excess in Type~Ia SNe is observed in 20--30\% of Type~Ia SNe \citep{Deckers22,Magee22} and they are linked to the existence of a red-giant companion \citep[e.g.,][]{Kasen10}, a dense circumstellar medium (CSM, e.g., \citealt{Piro16,Moriya23}), or a helium detonation \citep[e.g.,][]{Jiang17,Ni23}. Hence, cases like SN 2024pvw, with unusually wide wavelength range early coverage including NIR, can be of substantial interest.  {Specifically, deviation from $t^2$ in the early light curve rise, even in the redder bands observed by \Euclid,  could come from heated material of the white dwarf,  or a companion star. Other instances where a signal could be expected is for a pre-existing accretion disc. Structures in the light curves during the first days after explosion could also originate from radioactive material in the outer parts of the progenitor white dwarf, as discussed by \citet{2015ApJ...799..106G} in connection with early observations of SN~2014J  \citep{2014ApJ...784L..12G}.}

In order to quantify the significance of the early \Euclid photometry of SN~2024pvw, we fit, using \texttt{SNCosmo} \citep{barbary_2024_14025775}, the ZTF and ATLAS light curves by using the SALT3-NIR Type~Ia SN light curve evolution model \citep{Pierel22}. The best fit model is presented in Fig.~\ref{fig:sn2024pvw}. SALT3-NIR covers the wavelength range of 2,500 -- 20,000\,\AA\ and can provide the expected \Euclid photometry at early times. 
The early flux measurements by \Euclid slightly differ from the expected NIR light curves from the best fit model. However, given that the \Euclid NIR flux measurements are performed without conducting image subtractions and hence the measured flux may be contaminated by the host galaxy light, we conclude that no signatures of possible flux anomalies in the early \Euclid light curve were identified in SN~2024pvw. 

\subsubsection{SN 2024abla}
This transient is a known Type~Ia SN at a redshift $z=0.06$ \citep[see classification reported in TNS by][]{2024TNSCR4573....1G} that was observed by \Euclid well before ground-based transient surveys detected the object (see Fig.~\ref{fig:sn2024pvw} lower panel). Deep upper limits in the \Euclid bands were obtained about 70~days before the ATLAS discovery.
{Although we report only upper limits, this demonstrates \Euclid's capacity to make very early time observations. Whilst no precursor would be expected to be observed on this timescale for this system,} some precursor activity before core-collapse SNe is often reported \citep[e.g.,][]{Pastorello07,Ofek13}. The precursors have been observed at a range of timescales, and in many different kinds of core-collapse SNe including typical Type~II SNe \citep[e.g.,][]{Jacobson-Galan22}, interacting SNe \citep[e.g.,][]{Strotjohann21,Brennan2024}, and superluminous SNe \citep[e.g.,][]{nicholl16,Angus19}. The origins of such precursor activities are not well understood, but are likely linked to enhanced mass loss prior to explosion. During the \Euclid mission, it is possible that we will have similar {early time} observations for core-collapse SNe that will allow us to constrain the nature of such precursor activity.
\subsection{Transients observed close to the peak of the light curve}

\subsubsection{AT 2024pcm}
\begin{figure}
    \centering
    \includegraphics[width=\linewidth]{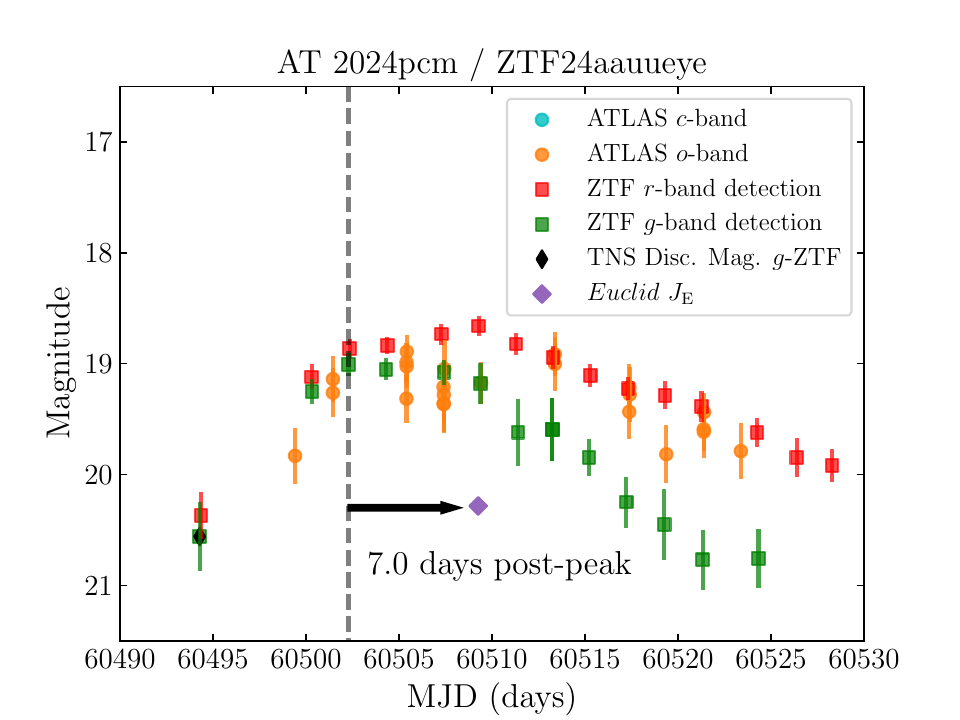}
    \caption{Light curve of AT~2024pcm. The \Euclid observations occurred during the ground-based survey observations, and the resulting \Euclid detections provide NIR photometry 7 days after the \textit{g}-band light curve peak. {Note it was only possible to make a measurement in \JE, see discussion in Sect.~\ref{sec:euclid_photomtry}.}}
    \label{fig:at2024pcm}
\end{figure}

AT~2024pcm was identified as a SN candidate by ZTF \citep{Sollerman24}. Probably due to its relatively faint discovery magnitude ($g_{\mathrm{ZTF}}=20.6$), no spectroscopic classification is available. The light curve is consistent with both type Ia and core-collapse SNe (Fig.~\ref{fig:at2024pcm}). We consider each in turn. If the host galaxy is the irregular spiral, GALEXASC J180413.45+672932.1 ($z=0.0583$), then the peak $g$-band magnitude is $-18.1$ i.e., underluminous by about $1\,\rm mag$ Cf. normal SNe Ia. The lack of a secondary maximum in the ATLAS \textit{o}-band would be consistent with this. Although the ATLAS \textit{o}-band lightcurve is sparsely sampled, it encompasses redder wavelengths (than ZTF$_r$) where the secondary maximum is more prominent. The single \Euclid observation in the \JE band, taken a week after the ZTF \textit{g}-band peak, is $M_{\scriptscriptstyle\rm{J}}= -16.8$. This implies a decline of $1.5$\,mag in $7\,\rm days$, assuming a peak $\JE$-band magnitude for a normal SNe Ia of $-18.27$ \citep{WoodVasey08}. Such a rapid decline would be unexpected. Taken together, the optical and near-IR photometry suggest that the rise to optical peak is too fast (about $9\,\rm days$ taking the last ZTF non-detection into account), and the decline from presumed near-IR peak is also too rapid for AT~2024pcm to be a normal type Ia SN. 
No correction for extinction has been applied, but we note that SN~2024pcm occurred in a relatively isolated region.  There appears to be an extended source closer to the location of the SN than GALEXASC J180413.45+672932.1, but further information on this other source is unfortunately not available. While the rapid rise to $g$-band peak, the shape of the $r$-band light curve, and the absolute peak magnitudes (as above) could all be consistent with certain core-collapse SNe, its large (around $13\,\rm kpc$) projected offset from GALEXASC J180413.45+672932.1 is noteworthy, but not unprecedented. Further imaging data may shed more light on this SN.

\subsection{Transients observed long after ground based transient survey discovery}\label{sec:latetime}
Here we show examples of transients observed by \Euclid long after the ground-based transient survey discovery date. Because of its sensitivity, \Euclid can detect these transients even after their light curves have faded below the detection limit of ground-based transient surveys. 
The two selected examples, shown in Fig.~\ref{fig:sn2023uqu}, were both detected by \Euclid hundreds of days after the peak of their respective light curves. In both cases, detections were made in the \IE, \YE, and \HE bands, while we obtained upper limits in the \JE band.

\subsubsection{SN 2023uqu}
\begin{figure}
    \centering
    \includegraphics[width=\linewidth]{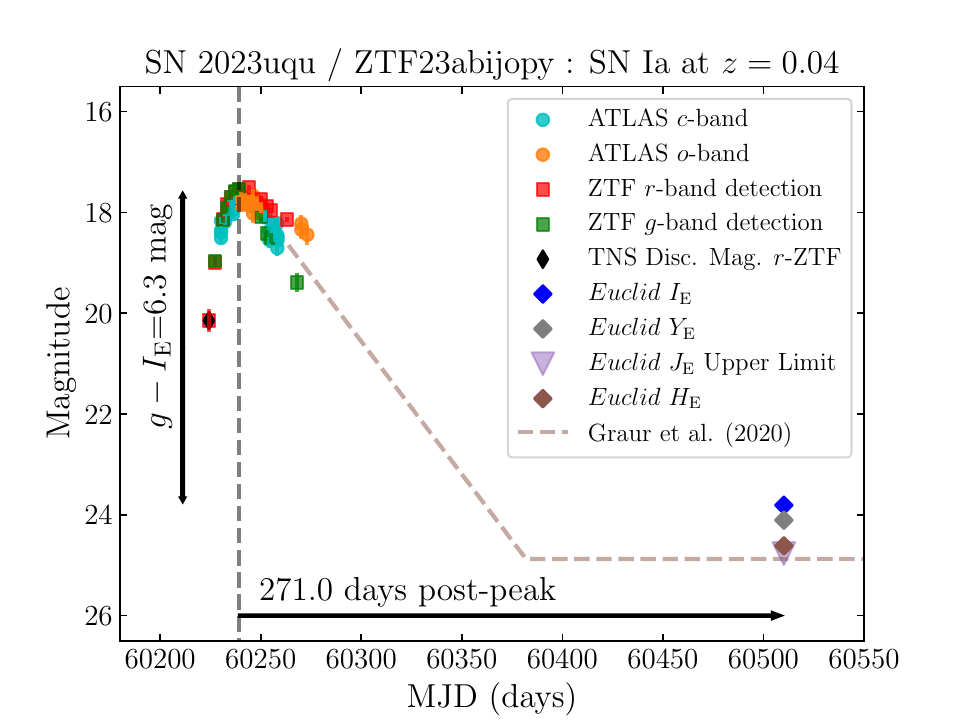}
    \includegraphics[width=\linewidth]{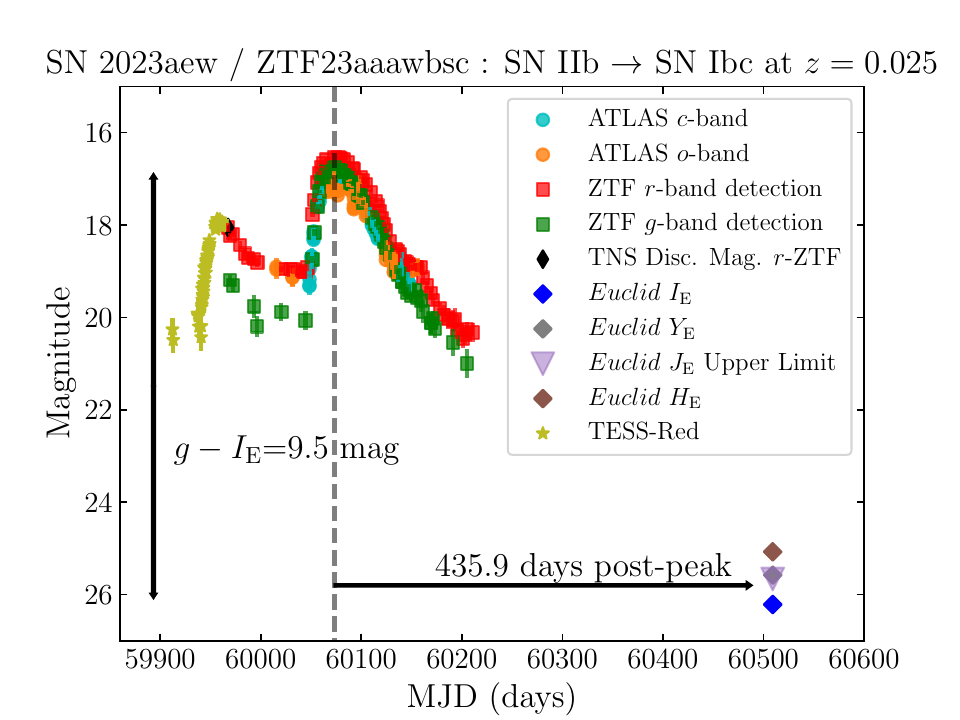}
    \caption{Light curves of SN~2023uqu (top) and SN~2023aew (bottom). These two SNe were observed and detected by \Euclid long after they faded beyond the limit of the ground-based surveys. The dashed line in the top panel is the faint-branch \textit{H}-band light curve template of Type~Ia SNe from \citet{Graur20}.}
    \label{fig:sn2023uqu}
\end{figure}

SN~2023uqu is a Type~Ia SN at $z=0.04$ \citep{2023TNSCR2734....1B}. The \Euclid photometry of this SN is presented in the top panel of Fig.~\ref{fig:sn2023uqu}. 
NIR light curves of Type~Ia SNe are often observed to have a plateau phase at around 150--500~days \citep{Graur20,2023MNRAS.521.4414D}. The \Euclid NIR photometry is consistent with the brightness of a faint NIR plateau (Fig.~\ref{fig:sn2023uqu}), although we cannot discuss the existence of the NIR plateau with only a single-epoch observation. Understanding the diversity in the late-phase NIR light curves in Type~Ia SNe would be essential in understanding the radiative transfer processes in Type~Ia SN ejecta \citep[e.g.,][]{axelrod1980,Fransson96}.

\subsubsection{SN 2023aew}
SN~2023aew was observed by \Euclid about 300~days after the final detection by ZTF (Fig.~\ref{fig:sn2023uqu}, bottom). SN~2023aew is a peculiar SN, which changed its spectroscopic type from Type~IIb to Type~Ibc \citep{Sharma24,Kangas24}. The light curve shows multiple peaks. The initial rise was serendipitously observed by Transiting Exoplanet Survey Satellite \citep[TESS;][]{2015JATIS...1a4003R} and the second peak was observed by ground-based surveys. The \Euclid detection corresponds to 436~days after the second peak. The nature of SN~2023aew is not clear. In particular, the luminosity source causing the multiple peaks has been debated. The final detection in the ZTF \textit{r} band is $20.3\pm 0.2$ at $\mathrm{MJD}=60211.2\,\rm days$ and the \Euclid detection in the $\IE$ band is $26.21\pm 0.06$ at $\mathrm{MJD}=60509.2\,\rm days$. Applying the redshift of $0.025$, the decline rate in the optical band during this time is $0.02~\mathrm{mag~day^{-1}}$. Although this decline rate is faster than that expected from the nuclear decay of ${}^{56}\mathrm{Co}\rightarrow{}^{56}\mathrm{Fe}$ ($0.0098~\mathrm{mag~day^{-1}}$), which is a major late-phase luminosity source of Type~Ibc SNe, the decline rates in the $R$ and $I$ bands in Type~Ibc SNe can be faster in the late phases (e.g., $0.017~\mathrm{mag~day^{-1}}$ in the Type~Ibc sample in \citealt{Hunter2009}) because of the $\gamma$-ray leakage. Thus, the measured decline rate is consistent with the canonical nuclear decay energy input.

Besides the nuclear decay energy, magnetar spin-down, black-hole accretion, and CSM interaction are suggested to be possible power sources forming the second peak \citep{Sharma24,Kangas24}. The light curve decay rate is expected to become slow in late phases in the cases of magnetar spin-down \citep[e.g.,][]{Nicholl2018} and black-hole accretion \citep[e.g.,][]{Dexter2013,Moriya2018}. However, 435.9 days after the peak is still not late enough to observe the predicted decrease in the light curve decay rate \citep[e.g.,][]{Nicholl2018}. In the case of the CSM interaction explanation, the observed light curve decay rate may be consistent, depending on the CSM configuration. It is also possible that the dust formation affects the decline rate in the optical, but we do not observe a significant flux excess in the \Euclid NIR bands. Thus, these power sources could not be excluded by the \Euclid observation presented here. A later phase observation of SN 2023aew by JWST at around 1000 days, combined with this \Euclid observation, would be able to provide a strong constraint on its mysterious power source.

\section{\label{sec:hostless} Transients previously classified as hostless}

The depth of \Euclid photometry, even in single epoch images such as those released in Q1, enables the detection of faint host galaxies. In some cases, \Euclid can detect the host of transients that were previously classed as `hostless' or `orphans'.

In our input target list, 19 objects were classified as orphans by the Sherlock system of Lasair. Eight of these were in fact close to a large galaxy that had presumably been missed by the Sherlock system. The rest showed no host galaxy in the SDSS or in the Legacy survey DR10 images, but showed a faint detection close to the transient position in the \Euclid VIS images. Based on the detection within the \Euclid images, the MER catalogue \citep[\Euclid source catalogue containing photometric and morphological information, see][for a full description]{Q1-TP004} provides a point-like probability for all sources that are detected in VIS. Furthermore, during the PHZ classification \citep[see][for the full details]{Q1-TP005}, \Euclid also provides galaxy, QSO, and star probabilities for the source. The probabilities for the other 11 sources are shown in \autoref{tab:orphan}.

\begin{figure}
    \centering
    \includegraphics[width=\linewidth]{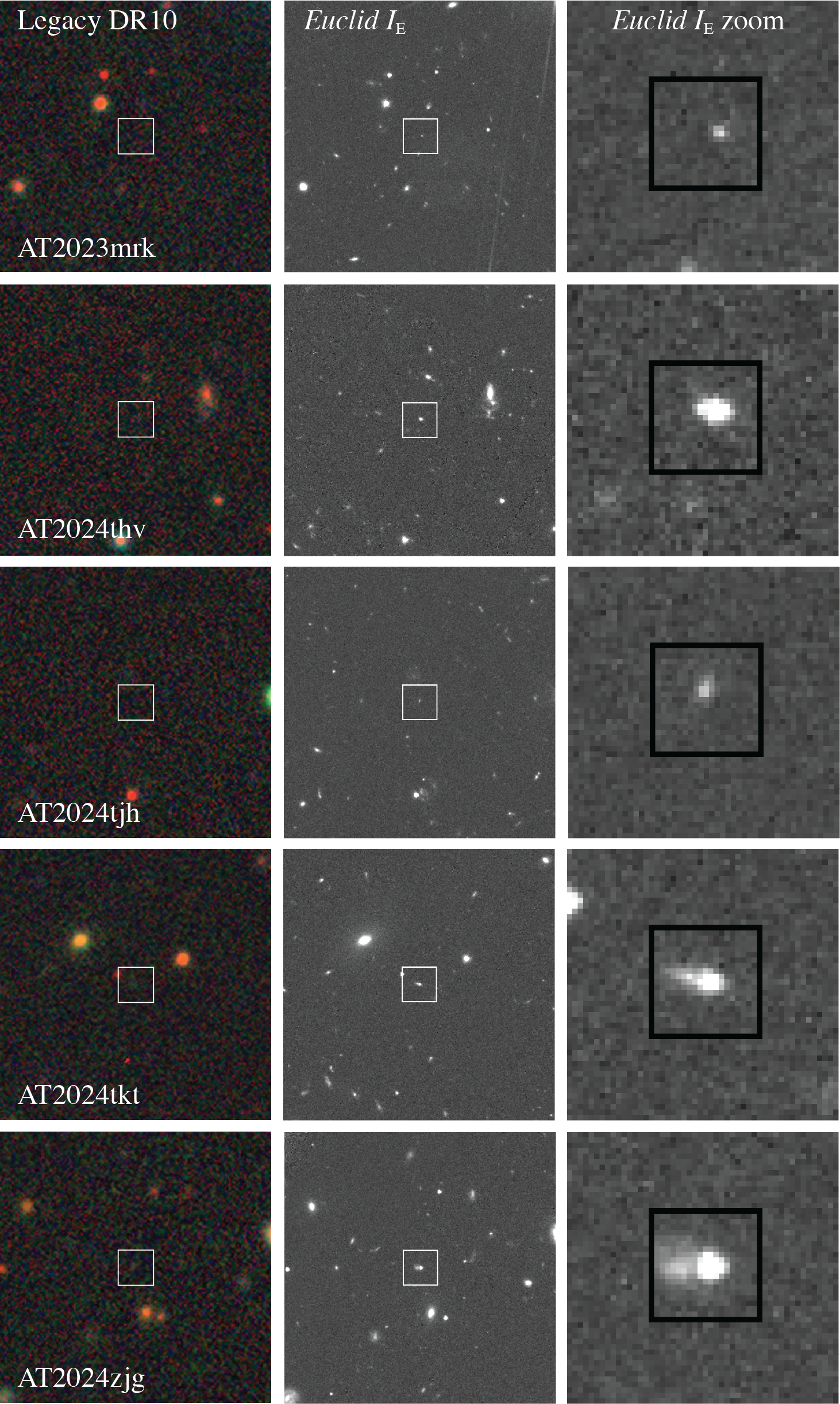} 
    \caption{Example `orphans', i.e., transients that previously had no host galaxy detection. \textit{Top row}: the field around AT~2023mrk, showing the colour Legacy Survey DR10 image \citep{Dey2019} and the deeper \Euclid $\IE$ image, both $40\arcsec$ on a side, and a zoomed-in section of the $\IE$ image $5\arcsec$ on a side, corresponding to the white box in the previous two images. The following rows show the same information for AT~2024thv, AT~2024tjh, AT~2024tkt, and AT2024zjg. The black box marked on the zoomed-in $\IE$ images is $2\arcsec$ across, centred on the transient position reported in TNS. In these cases, \Euclid has detected an extended object close to the transient position (in addition to the transient itself in some cases), which is likely to be the host galaxy.}
    \label{fig:orphans}
\end{figure}

Based on this, the first five sources (AT 2023mrk, AT 2024thv, AT 2024tjh, AT 2024tkt, and AT 2024zjg) have high probabilities of being galaxies. In fact, the clearest case for the detection of an underlaying host galaxy is for the late-time observation of AT 2023mrk, which showed clear photometric evolution akin to a SN around peak (see Fig. \ref{fig:at_lcs}). For AT 2024tkt and AT 2024zjg, we clearly still detect the transient as a point-like source, on top of a faint extended source which is most likely to be the underlying host galaxy (see Fig.~\ref{fig:orphans}).

The remaining six sources (AT 2024tld, AT 2024tkg, AT 2024tsi, AT 2024tsv, AT 2024tny, and AT 2024zdi) have point-like aspects and could either be late-time detections of the transient event in compact dwarf galaxy, variable stars, or QSOs. Each of these only have one detection reported in TNS. While for AT 2024tld and AT 2024tkg we found a magnitude difference of about $2$, indicating either a true transient event or a variable source, AT 2024tsi and AT 2024tny also show changes in magnitude, and AT 2024tsv seems to have a constant magnitude.

\Euclid VIS images showing the five new probable host galaxy detections, along with the previous Legacy survey DR10 images, are shown in Fig. \ref{fig:orphans}. With detections down to depths of \IE $\approx$ 26, these examples highlight the power of \Euclid to uncover potential faint underlying hosts of orphan transients. Without template images, however, we cannot separate the contributions from the host and the transient that may still be present at the time of observation. Therefore, although these sources have photometric redshifts and physical parameter estimates from the \Euclid SED fitting, we refrain from quoting these until we have templates for image subtraction. 

In general, identification of a previously unknown host galaxy may aid in the classification of a transient, particularly if a redshift can be determined. 
{Aside from finding potential (faint) host galaxies of previous apparently `hostless' transients, the superior spatial resolution afforded by \Euclid will allow for firmer host associations.
In addition, correct identification of the host and measurement of its properties are important in the context of cosmological measurements using Type Ia SNe \citep[e.g.,][]{2024ApJ...964..134Q}. Correlations have been found between the Hubble residual and host galaxy properties such as host galaxy mass, size and specific star formation rate \citep{Kelly2010, Lampeitl2010, Sullivan2010}, where the Hubble residual is the difference between the inferred distance modulus to the Type Ia SN, calculated from its apparent luminosity with corrections based on light curve width and colour, and  the expected value at the SN redshift based on the best-fit cosmological model. Hence, in modern Type Ia SN cosmology analyses, corrections are often made to individual SN distance measurements based on the host galaxy properties.  When studying the host galaxies of high redshift Type Ia SN from the SDSS-II and DES-5yr SN Ia samples, \citet{Lampeitl2010} and \citet{2024ApJ...964..134Q} found that 7\% and 6\% of Type Ia SNe were `hostless' to the limits of those surveys (approx $r\approx 23$ and $r\approx 25$, respectively). Based on our measurements from single-epoch Q1 images, we see that \Euclid offers the possibility to detect fainter host galaxies for cosmological samples of Type Ia SNe to $\IE\approx26$ in the EWS, and significantly fainter in the EDFs as their depth is built up over the duration of the mission.}

\begin{table}[ht!]
    \caption{Properties of the sources detected by \Euclid associated with the eleven previously hostless transients.}
    \smallskip
    \label{tab:orphan}
    \smallskip
    \centering
    \begin{tabular}{lcccc}
    \hline
    \hline
Name     &    Point-like & Galaxy & QSO & Star    \\
 & prob. & prob. & prob. & prob.\\
\hline
AT 2023mrk  &   0.116 &  0.839 & 0.067 & 0.175  \\
AT 2024thv  &   0.030 &  0.927 & 0.056 & 0.004  \\
AT 2024tjh  &   0.002 &  0.854 & 0.223 & 0.014  \\
AT 2024tkt  &   0.038 &  0.738 & 0.418 & 0.108  \\
AT 2024zjg  &   0.521 &  0.741 & 0.518 & 0.259  \\
AT 2024tld  &   0.585 &  0.400 & 0.431 & 0.541  \\
AT 2024tkg  &   0.568 &  0.409 & 0.166 & 0.569  \\
AT 2024tsi  &   0.972 &  0.353 & 0.558 & 0.443  \\
AT 2024tsv  &   0.867 &  0.236 & 0.840 & 0.484  \\
AT 2024tny  &   0.680 &  0.211 & 0.505 & 0.564  \\
AT 2024zdi  &   0.796 &  0.365 & 0.583 & 0.386  \\
       \hline
    \end{tabular}
\end{table}

\section{\label{sec:Concl}Conclusions}

We have followed up 161 known transients that were discovered up to 3 years prior to the \Euclid Q1 data release observations. Of these targets we were able to detect 59 sources in \IE and 40, 36, and 30 in \YE, \JE, and \HE respectively with a further 93 sources in \IE and 91,
88, and 93 in \YE, \JE, and \HE, respectively giving upper limits. The relatively high number of upper limits, as opposed to detections arises from using a single epoch of observation, since it is not possible to unambiguously measure the magnitude of transients without subtracting a template image. In cases where the transient appears to be isolated or on a smooth background, we can be more confident of the photometric measurement than in cases where the transient is close to another bright source, e.g., the galaxy nucleus.

The detections give (when disregarding JADES sources) a recovery fraction of 61\% of transients discovered in the year before \Euclid \IE observation, rising to 69\% for those discovered in the 6 months before observation, and 82\% for those discovered 100~days before observation.

In this work we followed up on transients within a sky area of $63.1\,{\rm deg}^{2}$, while the sky area to be released in \Euclid DR1 is anticipated to be $1900\,{\rm deg}^{2}$. Applying the same temporal constraints to the transient follow up we used here, it would suggest that in excess of $5500$ transients will be available for follow up. However, following the start of LSST science operations (mid 2025), we anticipate that this number will increase substantially, particularly into future data releases.

\Euclid DR1 will overcome the ambiguity between detections and upper limits that we have found in this work, because the EDFs will have multiple epochs available, allowing image subtractions to be performed. Beyond follow-up observations we further anticipate that the EDF survey data will allow us to discover new transients using \Euclid alone which are beyond the reach of ground-based surveys. The ambiguity raised here between detections and upper limits will persist into the EWS and as such it may be necessary to employ strategies such as scene modelling using high-resolution ground imaging from surveys like LSST in order to overcome this challenge.

Future work in the field of transients with \Euclid will include the generation of joint derived data products with the Vera C. Rubin telescope \citep{2022zndo...5836022G}. These will include joint cutouts and light curves of transients using photometry in all of the filters provided by both telescopes. This will enhance SNe rates and cosmology, as well as improved study of extragalactic transient environments. Early examples of such products can be seen in Duffy et al. (in prep.)

\begin{acknowledgements}
\AckEC
\AckQone
\AckDatalabs
L.G. acknowledges financial support from AGAUR, CSIC, MCIN and AEI 10.13039/501100011033 under projects PID2023-151307NB-I00, PIE 20215AT016, CEX2020-001058-M, ILINK23001, COOPB2304, and 2021-SGR-01270. FP and CGC acknowledge support from the Spanish Ministerio de Ciencia, Innovación y Universidades (MICINN) under grant numbers PID2022-141915NB-C21. TJM is supported by the Grants-in-Aid for Scientific Research of the Japan Society for the Promotion of Science (JP24K00682, JP24H01824, JP21H04997, JP24H00002, JP24H00027, JP24K00668) and by the Australian Research Council (ARC) through the ARC's Discovery Projects funding scheme (project DP240101786). AG acknowledges support from the Swedish Research Council (Dnr 2020-03444) and the Swedish National Space Agency (Dnr 2023-00226). KP acknowledges funding by the German Space Agency DLR under grant number 50~QE~2303. CD And IH acknowledge support from the Science and Technologies Facilities Council [grant ST/Y005511/1]; IH futher acknowledges [grants ST/V000713/1 and ST/Y001230/1] and support from the Leverhulme Trust [International Fellowship IF-2023-027]. 
RK acknowledges support from the Research Council of Finland (340613).
AAC acknowledges support through the European Space Agency (ESA) research fellowship programme. EC acknowledges support from MIUR, PRIN 2020 (grant 2020KB33TP) ``Multimessenger astronomy in the Einstein Telescope Era (METE)''. V.P. acknowledges the financial contribution
from PRIN-MIUR 2022 funded by the European Union –
Next Generation EU, and from the Timedomes grant within the “INAF 2023 Finanziamento della Ricerca Fondamentale”. L.I. acknowledges financial support from the YES Data Grant Program (PI: Izzo) {\it Multi-wavelength and multi messenger analysis of relativistic supernovae}.
CG acknowledges support from a Villum Young Investigator grant (project number 25501) and a Villum Experiment grant (VIL69896) from VILLUM FONDEN. 
MTB  acknowledges financial support from  the Mini Grant Program 1.05.23.04.02 (PI: Maria Teresa Botticella) {\it Euclid Transient Search}.
The Legacy Surveys consist of three individual and complementary projects: the Dark Energy Camera Legacy Survey (DECaLS; Proposal ID \#~2014B-0404; PIs: David Schlegel and Arjun Dey), the Beijing-Arizona Sky Survey (BASS; NOAO Prop. ID \#~2015A-0801; PIs: Zhou Xu and Xiaohui Fan), and the Mayall z-band Legacy Survey (MzLS; Prop. ID \#~2016A-0453; PI: Arjun Dey). DECaLS, BASS and MzLS together include data obtained, respectively, at the Blanco telescope, Cerro Tololo Inter-American Observatory, NSF’s NOIRLab; the Bok telescope, Steward Observatory, University of Arizona; and the Mayall telescope, Kitt Peak National Observatory, NOIRLab. Pipeline processing and analyses of the data were supported by NOIRLab and the Lawrence Berkeley National Laboratory (LBNL). The Legacy Surveys project is honored to be permitted to conduct astronomical research on Iolkam Du’ag (Kitt Peak), a mountain with particular significance to the Tohono O’odham Nation.
NOIRLab is operated by the Association of Universities for Research in Astronomy (AURA) under a cooperative agreement with the National Science Foundation. LBNL is managed by the Regents of the University of California under contract to the U.S. Department of Energy.
This project used data obtained with the Dark Energy Camera (DECam), which was constructed by the Dark Energy Survey (DES) collaboration. Funding for the DES Projects has been provided by the U.S. Department of Energy, the U.S. National Science Foundation, the Ministry of Science and Education of Spain, the Science and Technology Facilities Council of the United Kingdom, the Higher Education Funding Council for England, the National Center for Supercomputing Applications at the University of Illinois at Urbana-Champaign, the Kavli Institute of Cosmological Physics at the University of Chicago, Center for Cosmology and Astro-Particle Physics at the Ohio State University, the Mitchell Institute for Fundamental Physics and Astronomy at Texas A\& M University, Financiadora de Estudos e Projetos, Fundacao Carlos Chagas Filho de Amparo, Financiadora de Estudos e Projetos, Fundacao Carlos Chagas Filho de Amparo a Pesquisa do Estado do Rio de Janeiro, Conselho Nacional de Desenvolvimento Cientifico e Tecnologico and the Ministerio da Ciencia, Tecnologia e Inovacao, the Deutsche Forschungsgemeinschaft and the Collaborating Institutions in the Dark Energy Survey. The Collaborating Institutions are Argonne National Laboratory, the University of California at Santa Cruz, the University of Cambridge, Centro de Investigaciones Energeticas, Medioambientales y Tecnologicas-Madrid, the University of Chicago, University College London, the DES-Brazil Consortium, the University of Edinburgh, the Eidgenossische Technische Hochschule (ETH) Zurich, Fermi National Accelerator Laboratory, the University of Illinois at Urbana-Champaign, the Institut de Ciencies de l’Espai (IEEC/CSIC), the Institut de Fisica d’Altes Energies, Lawrence Berkeley National Laboratory, the Ludwig Maximilians Universitat Munchen and the associated Excellence Cluster Universe, the University of Michigan, NSF’s NOIRLab, the University of Nottingham, the Ohio State University, the University of Pennsylvania, the University of Portsmouth, SLAC National Accelerator Laboratory, Stanford University, the University of Sussex, and Texas A\&M University.
BASS is a key project of the Telescope Access Program (TAP), which has been funded by the National Astronomical Observatories of China, the Chinese Academy of Sciences (the Strategic Priority Research Program “The Emergence of Cosmological Structures” Grant \# XDB09000000), and the Special Fund for Astronomy from the Ministry of Finance. The BASS is also supported by the External Cooperation Program of Chinese Academy of Sciences (Grant \# ~114A11KYSB20160057), and Chinese National Natural Science Foundation (Grant \#~12120101003, \#~11433005).
The Legacy Survey team makes use of data products from the Near-Earth Object Wide-field Infrared Survey Explorer (NEOWISE), which is a project of the Jet Propulsion Laboratory/California Institute of Technology. NEOWISE is funded by the National Aeronautics and Space Administration.
The Legacy Surveys imaging of the DESI footprint is supported by the Director, Office of Science, Office of High Energy Physics of the U.S. Department of Energy under Contract No. DE-AC02-05CH1123, by the National Energy Research Scientific Computing Center, a DOE Office of Science User Facility under the same contract; and by the U.S. National Science Foundation, Division of Astronomical Sciences under Contract No. AST-0950945 to NOAO.
\end{acknowledgements}

%
%

\bibliography{STSWG, Euclid, Q1}

%

\begin{appendix}
  \onecolumn 
  
\begin{landscape}
\section{\label{sec:apdx:cat}Complete catalogue of sources}
In this appendix we supply the complete catalogue of sources which we were able from TNS and the measurements which we were able to make. An enhanced version of this table is available in the supplementary material supplied with this work. 

\begin{longtable}{lccS[table-format=3.2]cccccc}
\caption{Complete catalogue of sources with Q1 coverage. $\Delta$MJD is discovery date minus the \Euclid observation date, where negative values denote where \Euclid observed after the detection. Measurements reported with errors are detections, those with `$>$' being upper limits. `Offset' is the separation between the TNS reported position and the location of the source in \IE (\YE, \IE, and \HE available in supplementary table). In notes the following convention is adopted; 1--AGN; 2--variable star; 3--nuclear transient; 4--bright host; 5--bad pixels; 6--point like; 7--nearby source.
}\label{tab:completeList}\\
\hhline{==========}
{Name} & {Coordinates} & Discovery  & {$\Delta$MJD}  & {\IE} & {\YE}  & {\JE} & {\HE} & {Offset} & {Notes}  \\
 &  & {Mag} &  &  & & & &$(\arcsec)$&\\
\hline
\endfirsthead
\caption{continued.}\\
\hhline{==========}
{Name} & {Coordinates} & {Discovery}  & {$\Delta$MJD}  & {\IE} & {\YE}  & {\JE} & {\HE} & {Offset} & {Notes}\\
 &  & {Mag} &  &  & & & &$(\arcsec)$&\\
\hline
\endhead
\hline
\endfoot
{AT 2021abmd} & \ra{17;45;39.878}, \ang{+64;46;43.51} & 17.44 & -1023.00 & $>19.15$ & $>17.10$ & $--$ & $--$ & 0.00 & 4 \\
{AT 2021abvx} & \ra{17;44;56.722}, \ang{+67;18;10.03} & 19.94 & -1007.61 & $>25.90$ & $>22.39$ & $>22.13$ & $>21.87$ & 0.09 & 4 \\
{AT 2021ahou} & \ra{17;41;15.31}, \ang{+66;47;17.7} & 19.72 & -955.00 & $19.7\pm0.009$ & $19.223\pm0.007$ & $18.716\pm0.007$ & $18.482\pm0.007$ & 0.05 & 1 \\
{AT 2021ufr} & \ra{17;51;33.37}, \ang{+65;14;54.1} & 19.00 & -1086.60 & $>24.35$ & $>20.64$ & $>20.07$ & $>19.78$ & 0.00 &  \\
{AT 2021vje} & \ra{18;02;51.326}, \ang{+66;05;42.09} & 19.35 & -1081.59 & $18.86\pm0.01$ & $18.369\pm0.006$ & $17.96\pm0.007$ & $17.676\pm0.005$ & 0.03 & 1 \\
{AT 2021xdk} & \ra{17;48;08.182}, \ang{+65;10;46.73} & 19.97 & -1056.01 & $>20.91$ & $>18.97$ & $>18.42$ & $>18.0$ & 0.00 & 3 \\
{AT 2021zha} & \ra{18;01;21.8}, \ang{+65;18;02.12} & 18.31 & -1030.08 & $>26.58$ & $>24.52$ & $>23.44$ & $>23.17$ & 0.00 &  \\
{AT 2022acr} & \ra{17;51;25.825}, \ang{+68;09;05.75} & 19.44 & -908.27 & $24.347\pm0.017$ & $24.408\pm0.027$ & $24.247\pm0.011$ & $>24.66$ & 0.02 & 6 \\
{AT 2022aenb} & \ra{3;30;26.329}, \ang{-26;29;54.36} & 19.89 & -684.19 & $25.925\pm0.05$ & $>24.082$ & $>23.943$ & $>23.667$ & 0.26 & 7 \\
{AT 2022bpn} & \ra{18;07;06.94}, \ang{+63;54;09.36} & 17.62 & -895.50 & $17.998\pm0.008$ & $18.049\pm0.005$ & $18.073\pm0.02$ & $17.718\pm0.004$ & 0.04 & 2 \\
{AT 2022bzw} & \ra{18;18;16.104}, \ang{+66;25;11.62} & 19.77 & -888.15 & $>24.46$ & $>21.80$ & $>21.23$ & $>21.07$ & 0.00 &  \\
{AT 2022cbq} & \ra{18;01;11.85}, \ang{+66;42;48.02} & 19.06 & -887.73 & $>20.51$ & $>17.783$ & $>17.35$ & $>17.13$ & 0.00 & 4 \\
{AT 2022cwf} & \ra{17;48;30.906}, \ang{+64;02;28.38} & 19.56 & -886.66 & $--$ & $--$ & $24.009\pm0.024$ & $24.169\pm0.03$ & -- &  \\
{AT 2022cxg} & \ra{18;19;32.797}, \ang{+66;18;29.62} & 18.49 & -878.29 & $>20.29$ & $>18.42$ & $>17.88$ & $>17.71$ & 0.00 & 4 \\
{AT 2022pxw} & \ra{4;07;21.072}, \ang{-47;05;08.36} & 19.39 & -774.20 & $>26.71$ & $>23.84$ & $>23.88$ & $>23.21$ & 0.00 &  \\
{AT 2022tqo} & \ra{3;35;19.101}, \ang{-27;37;04.8} & 19.23 & -702.38 & $>24.50$ & $>21.30$ & $>20.64$ & $>20.46$ & 0.00 & 4 \\
{AT 2022ugf} & \ra{17;44;33.899}, \ang{+65;32;31.28} & 19.75 & -690.76 & $--$ & $--$ & $>20.82$ & $>21.79$ & -- &  \\
{AT 2022xxw} & \ra{3;32;05.105}, \ang{-27;23;35.34} & 19.64 & -659.31 & $>23.45$ & $>20.797$ & $>20.23$ & $>20.06$ & 0.00 & 4 \\
{AT 2022yxt} & \ra{18;11;01.057}, \ang{+68;11;24.15} & 19.75 & -632.26 & $>22.97$ & $>19.296$ & $>19.03$ & $>18.55$ & 0.00 &  \\
{AT 2023aafo} & \ra{4;12;09.459}, \ang{-49;55;12.67} & 20.04 & -266.73 & $>25.79$ & $>22.329$ & $>21.78$ & $>21.53$ & 0.00 &  \\
{AT 2023abyp} & \ra{3;34;27.69}, \ang{-28;36;59.83} & 20.65 & -243.19 & $23.844\pm0.017$ & $23.447\pm0.028$ & $23.089\pm0.026$ & $22.982\pm0.021$ & 0.06 &  \\
{AT 2023adss} & \ra{3;32;43.921}, \ang{-27;47;16.14} & 29.35 & -313.56 & $>27.7$ & $>25.90$ & $>25.85$ & $>25.64$ & 0.00 &  \\
{AT 2023adst} & \ra{3;32;32.207}, \ang{-27;48;59.15} & 29.88 & -313.46 & $>27.29$ & $--$ & $--$ & $--$ & 0.00 &  \\
{AT 2023adsu} & \ra{3;32;33.75}, \ang{-27;46;44.41} & 30.41 & -313.54 & $>27.56$ & $>26.18$ & $>25.77$ & $>25.62$ & 0.00 &  \\
{AT 2023adsv} & \ra{3;32;39.457}, \ang{-27;50;19.67} & 28.04 & -313.54 & $>27.67$ & $--$ & $--$ & $--$ & 0.00 &  \\
{AT 2023adsw} & \ra{3;32;27.3}, \ang{-27;48;39.31} & 30.11 & -313.41 & $>28.13$ & $--$ & $--$ & $--$ & 0.00 &  \\
{AT 2023adsx} & \ra{3;32;40.801}, \ang{-27;46;05.98} & 30.15 & -313.56 & $>26.74$ & $--$ & $--$ & $--$ & 0.00 &  \\
{AT 2023adsy} & \ra{3;32;32.365}, \ang{-27;49;15.24} & 28.98 & -313.46 & $>27.97$ & $--$ & $--$ & $--$ & 0.00 &  \\
{AT 2023adsz} & \ra{3;32;34.955}, \ang{-27;48;36.68} & 29.52 & -313.54 & $>28.08$ & $--$ & $--$ & $--$ & 0.00 &  \\
{AT 2023adta} & \ra{3;32;32.468}, \ang{-27;48;52.26} & 28.54 & -313.46 & $>26.39$ & $--$ & $--$ & $--$ & 0.00 &  \\
{AT 2023adtb} & \ra{3;32;40.584}, \ang{-27;45;43.62} & 29.60 & -313.56 & $>27.27$ & $>26.10$ & $>25.79$ & $>25.65$ & 0.00 &  \\
{AT 2023adtc} & \ra{3;32;39.679}, \ang{-27;49;36.61} & 29.69 & -313.54 & $>27.94$ & $--$ & $--$ & $--$ & 0.00 &  \\
{AT 2023adtd} & \ra{3;32;31.707}, \ang{-27;47;48.51} & 28.00 & -313.46 & $>27.92$ & $>24.05$ & $>25.86$ & $>25.62$ & 0.00 &  \\
{AT 2023adte} & \ra{3;32;45.146}, \ang{-27;45;53.14} & 27.97 & -313.56 & $>26.04$ & $>25.51$ & $>25.06$ & $>24.35$ & 0.00 &  \\
{AT 2023adtf} & \ra{3;32;38.823}, \ang{-27;49;21.98} & 29.07 & -313.54 & $>27.81$ & $--$ & $--$ & $--$ & 0.00 &  \\
{AT 2023adtg} & \ra{3;32;46.37}, \ang{-27;44;55.79} & 28.69 & -313.56 & $>26.71$ & $>25.61$ & $>25.74$ & $>24.97$ & 0.00 &  \\
{AT 2023adth} & \ra{3;32;43.761}, \ang{-27;47;00.27} & 28.31 & -313.56 & $>26.94$ & $>25.33$ & $>25.69$ & $>25.63$ & 0.00 &  \\
{AT 2023adtj} & \ra{3;32;38.964}, \ang{-27;44;20.65} & 29.61 & -313.56 & $>27.57$ & $--$ & $--$ & $--$ & 0.00 &  \\
{AT 2023adtk} & \ra{3;32;35.852}, \ang{-27;47;19.04} & 28.29 & -313.54 & $>25.51$ & $>23.55$ & $--$ & $--$ & 0.00 &  \\
{AT 2023adtl} & \ra{3;32;45.105}, \ang{-27;45;34.51} & 27.40 & -313.56 & $>27.79$ & $>25.99$ & $>25.96$ & $>25.82$ & 0.00 &  \\
{AT 2023adtm} & \ra{3;32;40.225}, \ang{-27;49;49.49} & 27.07 & -313.54 & $>27.42$ & $--$ & $--$ & $--$ & 0.00 &  \\
{AT 2023adtn} & \ra{3;32;51.174}, \ang{-27;45;29.21} & 28.73 & -313.56 & $>26.42$ & $>24.83$ & $>24.22$ & $>24.42$ & 0.00 &  \\
{AT 2023adto} & \ra{3;32;29.518}, \ang{-27;49;16.08} & 27.22 & -313.41 & $>26.98$ & $--$ & $--$ & $--$ & 0.00 &  \\
{AT 2023adtp} & \ra{3;32;32.185}, \ang{-27;46;14.71} & 30.25 & -313.54 & $>28.33$ & $>25.16$ & $--$ & $--$ & 0.00 &  \\
{AT 2023adtq} & \ra{3;32;37.175}, \ang{-27;47;36.75} & 27.32 & -313.54 & $>27.41$ & $>24.82$ & $>25.77$ & $>25.59$ & 0.00 &  \\
{AT 2023adtr} & \ra{3;32;38.533}, \ang{-27;49;22.01} & 29.26 & -313.54 & $>26.12$ & $--$ & $--$ & $--$ & 0.00 &  \\
{AT 2023adts} & \ra{3;32;39.256}, \ang{-27;45;47.9} & 30.07 & -313.56 & $>27.77$ & $>25.38$ & $>25.89$ & $>25.67$ & 0.00 &  \\
{AT 2023adtt} & \ra{3;32;37.266}, \ang{-27;50;08.8} & 27.39 & -313.46 & $>26.16$ & $--$ & $--$ & $--$ & 0.00 &  \\
{AT 2023adtu} & \ra{3;32;32.362}, \ang{-27;47;20.55} & 26.57 & -313.54 & $>27.97$ & $>26.17$ & $>25.98$ & $>25.72$ & 0.00 &  \\
{AT 2023adtv} & \ra{3;32;37.607}, \ang{-27;48;38.1} & 29.98 & -313.54 & $>26.80$ & $--$ & $--$ & $--$ & 0.00 &  \\
{AT 2023adtw} & \ra{3;32;43.472}, \ang{-27;46;34.4} & 25.91 & -313.56 & $>26.79$ & $--$ & $--$ & $--$ & 0.00 &  \\
{AT 2023adtx} & \ra{3;32;49.039}, \ang{-27;45;19.38} & 27.46 & -313.56 & $>25.46$ & $>23.29$ & $>22.47$ & $>22.25$ & 0.00 &  \\
{AT 2023adty} & \ra{3;32;48.416}, \ang{-27;45;50.6} & 29.02 & -313.56 & $>25.94$ & $>23.63$ & $>23.24$ & $>23.07$ & 0.00 &  \\
{AT 2023adtz} & \ra{3;32;43.409}, \ang{-27;44;13.11} & 28.63 & -313.56 & $>27.56$ & $--$ & $--$ & $--$ & 0.00 &  \\
{AT 2023adua} & \ra{3;32;29.311}, \ang{-27;48;26.47} & 28.90 & -313.46 & $>26.01$ & $>25.16$ & $--$ & $--$ & 0.00 &  \\
{AT 2023adub} & \ra{3;32;42.281}, \ang{-27;47;46.47} & 28.27 & -313.56 & $>25.78$ & $>22.78$ & $--$ & $--$ & 0.00 &  \\
{AT 2023aqk} & \ra{4;14;59.71}, \ang{-49;07;56.42} & 17.76 & -592.35 & $17.879\pm0.016$ & $17.673\pm0.006$ & $17.491\pm0.004$ & $17.619\pm0.005$ & 0.04 & 1 \\
{AT 2023aqr} & \ra{17;59;35.815}, \ang{+68;19;31.76} & 19.79 & -541.31 & $>26.66$ & $>23.475$ & $>23.06$ & $>22.91$ & 0.00 &  \\
{AT 2023btx} & \ra{4;06;40.84}, \ang{-46;02;23.39} & 18.86 & -572.18 & $>23.70$ & $>20.902$ & $>20.5$ & $>20.29$ & 0.00 & 4 \\
{AT 2023cxi} & \ra{4;02;31.254}, \ang{-50;56;51.28} & 19.33 & -567.28 & $>27.68$ & $>25.186$ & $>24.83$ & $>24.50$ & 0.00 &  \\
{AT 2023dan} & \ra{17;44;53.811}, \ang{+64;02;20.4} & 18.72 & -493.55 & $--$ & $--$ & $>21.01$ & $>20.89$ & -- &  \\
{AT 2023ea} & \ra{3;29;20.213}, \ang{-27;36;20.26} & 19.61 & -575.98 & $>25.55$ & $>22.177$ & $>21.73$ & $>22.22$ & 0.00 &  \\
{AT 2023fhq} & \ra{3;25;00.98}, \ang{-28;29;30.19} & 18.82 & -489.14 & $>24.77$ & $>21.637$ & $>21.23$ & $>20.96$ & 0.00 &  \\
{AT 2023huo} & \ra{17;56;39.948}, \ang{+66;47;58.35} & 19.41 & -436.02 & $24.362\pm0.14$ & $>21$ & $0$ & $>20.3$ & 0.22 & 7 \\
{AT 2023mrk} & \ra{18;09;08.725}, \ang{+65;41;09.1} & 19.93 & -376.00 & $26.118\pm0.038$ & $25.984\pm0.069$ & $26.107\pm0.094$ & $25.842\pm0.072$ & 0.29 & 6 \\
{AT 2023ncz} & \ra{3;54;27.754}, \ang{-48;22;37.96} & 19.00 & -420.23 & $>25.27$ & $>23.926$ & $>23.52$ & $>23.34$ & 0.00 &  \\
{AT 2023pfg} & \ra{17;46;52.023}, \ang{+64;03;58.09} & 20.25 & -344.71 & $>24.276$ & $>20.642$ & $>19.981$ & $>19.9$ & 0.00 &  \\
{AT 2023qcd} & \ra{18;18;20.198}, \ang{+65;07;19.92} & 19.72 & -349.22 & $>26.286$ & $>23.47$ & $>23.824$ & $>23.824$ & 0.00 &  \\
{AT 2023qyv} & \ra{17;56;11.182}, \ang{+68;19;52.09} & 20.97 & -337.53 & $>25.055$ & $>24.194$ & $>23.237$ & $>23.237$ & 0.00 &  \\
{AT 2023spc} & \ra{3;33;03.761}, \ang{-27;20;37.52} & 19.93 & -330.23 & $>26.189$ & $>22.712$ & $>22.712$ & $>22.712$ & 0.00 &  \\
{AT 2023tjw} & \ra{3;57;03.825}, \ang{-47;52;48.64} & 19.54 & -354.58 & $>23.80$ & $>20.852$ & $>20.37$ & $>19.94$ & 0.00 & 4 \\
{AT 2023uke} & \ra{18;09;37.142}, \ang{+65;33;57.32} & 20.06 & -286.20 & $>26.943$ & $>23.063$ & $>22.08$ & $>22.08$ & 0.00 &  \\
{AT 2023vcs} & \ra{3;25;27.759}, \ang{-27;41;42.46} & 18.63 & -295.74 & $>22.163$ & $>19.253$ & $>18.312$ & $>18.339$ & 0.00 &  \\
{AT 2023veh} & \ra{3;38;36.934}, \ang{-28;27;46.66} & 19.29 & -297.51 & $>25.222$ & $>21.063$ & $>21.146$ & $>21.144$ & 0.00 &  \\
{AT 2023vnw} & \ra{3;32;15.166}, \ang{-27;51;26.92} & 28.34 & -294.41 & $>28.16$ & $>24.91$ & $--$ & $--$ & 0.00 &  \\
{AT 2023vnx} & \ra{3;32;22.751}, \ang{-27;51;55.58} & 28.76 & -294.41 & $>27.80$ & $>25.671$ & $--$ & $--$ & 0.00 &  \\
{AT 2023vny} & \ra{3;32;22.49}, \ang{-27;51;00.13} & 27.89 & -294.41 & $>27.78$ & $>25.446$ & $--$ & $--$ & 0.00 &  \\
{AT 2023vnz} & \ra{3;32;18.789}, \ang{-27;51;30.7} & 26.58 & -294.41 & $>27.09$ & $>23.507$ & $--$ & $--$ & 0.00 &  \\
{AT 2023voa} & \ra{3;32;14.467}, \ang{-27;51;27.13} & 28.91 & -294.41 & $>28.07$ & $>25.7$ & $--$ & $--$ & 0.00 &  \\
{AT 2023vob} & \ra{3;32;23.295}, \ang{-27;51;01.84} & 25.96 & -294.41 & $>25.47$ & $>23.66$ & $--$ & $--$ & 0.00 &  \\
{AT 2023voc} & \ra{3;32;19.143}, \ang{-27;52;14.51} & 29.48 & -294.41 & $>27.82$ & $>25.507$ & $--$ & $--$ & 0.00 &  \\
{AT 2023zvg} & \ra{4;20;07.188}, \ang{-49;01;05.71} & 19.14 & -294.01 & $>25.94$ & $>23.96$ & $>23.465$ & $>23.296$ & 0.00 &  \\
{AT 2024aaoo} & \ra{3;50;11.628}, \ang{-48;04;14.23} & 19.46 & 60.43 & $>27.902$ & $>25.109$ & $>24.741$ & $>24.424$ & 0.00 &  \\
{AT 2024abst} & \ra{17;52;57.117}, \ang{+64;40;56.16} & 19.63 & 123.92 & $>23.171$ & $>19.875$ & $>18.96$ & $>18.94$ & 0.00 &  \\
{AT 2024bkj} & \ra{17;45;08.568}, \ang{+67;41;30.51} & 18.08 & -167.12 & $23.04\pm0.2$ & $>19.83$ & $>19.452$ & $>19.282$ & 0.16 & 7 \\
{AT 2024clt} & \ra{3;34;10.033}, \ang{-28;24;45.86} & 18.92 & -175.51 & $>24.88$ & $>21.134$ & $>20.698$ & $>21.458$ & 0.00 &  \\
{AT 2024eht} & \ra{17;43;53.406}, \ang{+66;47;49.03} & 20.16 & -125.61 & $>22.94$ & $>19.598$ & $>19.039$ & $>18.838$ & 0.00 & 4 \\
{AT 2024eta} & \ra{17;52;50.084}, \ang{+65;13;52.89} & 19.73 & -120.64 & $>23.55$ & $>19.549$ & $>19.168$ & $--$ & 0.00 & 4 \\
{AT 2024fw} & \ra{3;29;29.771}, \ang{-28;42;14.63} & 20.29 & -214.82 & $24.259\pm0.035$ & $>21.665$ & $>22.481$ & $>22.253$ & 0.18 &  \\
{AT 2024gja} & \ra{4;15;40.098}, \ang{-49;47;50.7} & 19.57 & -148.77 & $24.108\pm0.026$ & $24.251\pm0.023$ & $>23.507$ & $>23.274$ & 0.14 &  \\
{AT 2024iml} & \ra{18;14;11.31}, \ang{+64;47;31.78} & 19.75 & -70.63 & $20.554\pm0.015$ & $20.705\pm0.006$ & $21.995\pm0.007$ & $21.62\pm0.005$ & 0.07 &  \\
{AT 2024kkp} & \ra{17;38;21.56}, \ang{+66;48;59.02} & 21.93 & -41.65 & $23.396\pm0.018$ & $22.3\pm0.019$ & $>21.596$ & $>21.334$ & 0.16 &  \\
{AT 2024mou} & \ra{3;35;45.57}, \ang{-27;57;09.11} & 17.24 & -50.25 & $19.414\pm0.009$ & $19.445\pm0.003$ & $20.919\pm0.007$ & $20.43\pm0.004$ & 0.04 &  \\
{AT 2024mzs} & \ra{3;34;57.65}, \ang{-29;04;11.21} & 18.70 & -41.75 & $18.685\pm0.01$ & $18.997\pm0.005$ & $18.852\pm0.004$ & $18.738\pm0.003$ & 0.06 & 1 \\
{AT 2024npq} & \ra{4;12;13.351}, \ang{-48;21;16.51} & 18.26 & -68.02 & $21.001\pm0.013$ & $21.482\pm0.004$ & $23.028\pm0.008$ & $22.441\pm0.004$ & 0.22 & 6 \\
{AT 2024pcm} & \ra{18;04;15.593}, \ang{+67;29;32.52} & 20.56 & -14.98 & $--$ & $--$ & $20.284\pm0.004$ & $--$ & -- &  \\
{AT 2024pnv} & \ra{3;28;29.23}, \ang{-28;45;13.46} & 18.87 & -22.75 & $18.904\pm0.008$ & $20.399\pm0.005$ & $21.173\pm0.006$ & $21.043\pm0.003$ & 0.06 &  \\
{AT 2024tdq} & \ra{18;12;12.61}, \ang{+65;09;46.53} & 22.61 & -16.81 & $>23.57$ & $--$ & $--$ & $--$ & 0.00 &  \\
{AT 2024tdr} & \ra{18;02;20.97}, \ang{+67;17;59.07} & 21.96 & -15.62 & $21.5\pm0.009$ & $22.621\pm0.007$ & $--$ & $>23.488$ & 0.09 &  \\
{AT 2024tds} & \ra{17;56;21.37}, \ang{+66;45;43.52} & 22.42 & -14.61 & $21.378\pm0.018$ & $20.235\pm0.024$ & $19.781\pm0.022$ & $20.547\pm0.056$ & 0.01 &  \\
{AT 2024tdv} & \ra{17;47;26.95}, \ang{+64;42;09.64} & 21.06 & -100.33 & $>23.74$ & $>20.804$ & $>20.3$ & $>20.095$ & 0.00 & 4 \\
{AT 2024tdw} & \ra{18;07;03.26}, \ang{+66;54;42.09} & 21.39 & -134.66 & $23.569\pm0.065$ & $>21.578$ & $>21.247$ & $>20.974$ & 0.03 &  \\
{AT 2024tdx} & \ra{17;55;19.25}, \ang{+65;27;51.53} & 22.06 & -135.10 & $--$ & $--$ & $--$ & $>21.348$ & -- & 2 \\
{AT 2024tdy} & \ra{17;45;56.4}, \ang{+65;28;56.8} & 21.65 & -134.95 & $--$ & $--$ & $>19.6$ & $>19.622$ & -- &  \\
{AT 2024tgm} & \ra{18;11;15.69}, \ang{+65;39;09.7} & 22.20 & -133.47 & $24.52\pm0.03$ & $>23.25$ & $>22.825$ & $>22.671$ & 0.14 &  \\
{AT 2024tgn} & \ra{17;51;06.4}, \ang{+67;24;02.41} & 22.45 & -132.01 & $26.187\pm0.048$ & $>26.372$ & $>26.021$ & $>26.224$ & 0.19 &  \\
{AT 2024tgo} & \ra{17;50;11.64}, \ang{+66;38;02.62} & 21.00 & -130.24 & $>19.42$ & $>18.352$ & $>18.064$ & $>18.149$ & 0.00 & 3 \\
{AT 2024tgw} & \ra{17;51;18.77}, \ang{+64;46;48.65} & 21.97 & -120.28 & $>20.07$ & $>17.975$ & $--$ & $--$ & 0.00 & 3 \\
{AT 2024tgx} & \ra{17;58;42.41}, \ang{+66;39;38.4} & 22.54 & -129.40 & $--$ & $--$ & $>19.764$ & $--$ & -- & 4 \\
{AT 2024thq} & \ra{17;44;30.12}, \ang{+66;16;51.23} & 23.09 & -99.24 & $>20.72$ & $>19.322$ & $>18.922$ & $>18.59$ & 0.00 & 4 \\
{AT 2024thv} & \ra{17;53;07.23}, \ang{+66;53;15.35} & 22.12 & -118.19 & $>25.3$ & $>23.674$ & $>23.225$ & $>23.418$ & 0.00 &  \\
{AT 2024tix} & \ra{17;49;25.25}, \ang{+66;04;50.03} & 22.28 & -110.42 & $25.23\pm0.018$ & $>24.977$ & $>24.747$ & $>24.284$ & 0.05 &  \\
{AT 2024tjh} & \ra{17;56;44.45}, \ang{+66;15;05.05} & 21.93 & -95.67 & $>25.92$ & $>25.154$ & $>24.942$ & $>24.885$ & 0.00 &  \\
{AT 2024tjp} & \ra{17;44;53.3}, \ang{+66;27;00.43} & 22.65 & -105.23 & $--$ & $22.933\pm0.007$ & $--$ & $--$ & -- &  \\
{AT 2024tkg} & \ra{18;03;17.61}, \ang{+66;23;54.18} & 22.36 & -93.89 & $24.062\pm0.011$ & $23.297\pm0.006$ & $23.021\pm0.006$ & $23.764\pm0.036$ & 0.08 & 6 \\
{AT 2024tkn} & \ra{17;44;17.7}, \ang{+65;45;10.57} & 23.06 & -102.02 & $24.95\pm0.021$ & $24.904\pm0.044$ & $24.882\pm0.025$ & $>24.767$ & 0.13 & 6 \\
{AT 2024tkt} & \ra{17;51;30.28}, \ang{+64;31;28.36} & 22.47 & -100.28 & $24.483\pm0.013$ & $23.946\pm0.019$ & $23.565\pm0.021$ & $23.689\pm0.02$ & 0.05 &  \\
{AT 2024tkx} & \ra{18;03;58.87}, \ang{+64;39;09.26} & 22.58 & -96.70 & $24.041\pm0.029$ & $22.519\pm0.019$ & $22.142\pm0.015$ & $22.017\pm0.015$ & 0.09 & 6 \\
{AT 2024tld} & \ra{17;50;18.84}, \ang{+64;50;57.46} & 22.14 & -87.34 & $24.691\pm0.014$ & $24.963\pm0.031$ & $24.537\pm0.023$ & $>25.284$ & 0.01 & 6 \\
{AT 2024tli} & \ra{18;01;30.47}, \ang{+67;17;36.56} & 21.64 & -84.41 & $23.047\pm0.015$ & $22.66\pm0.012$ & $22.435\pm0.01$ & $22.635\pm0.011$ & 0.06 &  \\
{AT 2024tmf} & \ra{17;54;54.34}, \ang{+64;38;45.41} & 22.54 & -80.58 & $>21.258$ & $>19.372$ & $>18.747$ & $>18.482$ & 0.00 & 4 \\
{AT 2024tmg} & \ra{17;59;44.47}, \ang{+66;30;47.14} & 22.48 & -75.67 & $22.881\pm0.013$ & $23.047\pm0.007$ & $23.095\pm0.007$ & $23.49\pm0.009$ & 0.18 &  \\
{AT 2024tmz} & \ra{18;03;12.96}, \ang{+66;25;19.52} & 22.07 & -79.89 & $24.178\pm0.013$ & $23.814\pm0.009$ & $23.781\pm0.009$ & $>23.728$ & 0.10 &  \\
{AT 2024tnw} & \ra{18;10;22.43}, \ang{+65;25;47.16} & 21.88 & -76.79 & $24.521\pm0.015$ & $24.885\pm0.018$ & $24.815\pm0.045$ & $>25.134$ & 0.09 &  \\
{AT 2024tnx} & \ra{18;09;37.33}, \ang{+65;06;24.44} & 22.42 & -76.88 & $23.353\pm0.079$ & $--$ & $--$ & $--$ & 0.02 & 4 \\
{AT 2024tny} & \ra{17;52;22.65}, \ang{+67;04;35.92} & 22.34 & -75.27 & $23.994\pm0.01$ & $23.67\pm0.012$ & $23.659\pm0.013$ & $24.183\pm0.023$ & 0.16 & 6 \\
{AT 2024tpf} & \ra{17;47;19.07}, \ang{+64;31;05.18} & 22.40 & -74.32 & $25.187\pm0.029$ & $>20.208$ & $>19.592$ & $>19.65$ & 0.04 &  \\
{AT 2024tqg} & \ra{18;10;50.54}, \ang{+65;11;44.98} & 22.01 & -62.68 & $23.149\pm0.011$ & $--$ & $--$ & $--$ & 0.01 &  \\
{AT 2024tql} & \ra{18;08;10.48}, \ang{+66;02;50.56} & 21.99 & -54.45 & $22.833\pm0.026$ & $21.871\pm0.026$ & $>21.286$ & $>20.793$ & 0.19 &  \\
{AT 2024tqm} & \ra{18;12;07.66}, \ang{+65;23;57.03} & 21.74 & -54.69 & $22.741\pm0.184$ & $19.937\pm0.008$ & $19.468\pm0.009$ & $19.237\pm0.008$ & 0.01 &  \\
{AT 2024trh} & \ra{17;50;46.32}, \ang{+67;22;22.95} & 20.96 & -49.22 & $21.094\pm0.009$ & $21.014\pm0.006$ & $20.913\pm0.005$ & $21.242\pm0.005$ & 0.06 &  \\
{AT 2024tsi} & \ra{17;59;42.81}, \ang{+65;29;09.82} & 22.23 & -33.59 & $21.615\pm0.01$ & $21.718\pm0.004$ & $21.881\pm0.005$ & $22.163\pm0.005$ & 0.13 & 6 \\
{AT 2024tsv} & \ra{18;03;48.81}, \ang{+64;38;32.92} & 22.01 & -20.85 & $22.393\pm0.012$ & $22.536\pm0.06$ & $22.481\pm0.07$ & $22.695\pm0.07$ & 0.06 & 6 \\
{AT 2024vwt} & \ra{17;54;43.534}, \ang{+64;52;44.56} & 20.26 & 64.89 & $>26.012$ & $--$ & $--$ & $--$ & 0.00 &  \\
{AT 2024wdv} & \ra{17;46;46.86}, \ang{+64;47;11.04} & 22.15 & -135.27 & $>24.159$ & $>21.638$ & $>21.033$ & $>20.757$ & 0.00 & 4 \\
{AT 2024xjo} & \ra{4;19;25.939}, \ang{-47;27;56.15} & 18.93 & 27.31 & $21.01$ & $21.376$ & $21.318$ & $22.015$ & 0.15 &  \\
{AT 2024yii} & \ra{17;34;46.621}, \ang{+65;33;58.03} & 19.98 & 92.08 & $>20.204$ & $>19.725$ & $>19.505$ & $>19.166$ & 0.12 & 3 \\
{AT 2024ymn} & \ra{3;32;26.131}, \ang{-29;38;31.31} & 21.13 & 59.00 & $>20.785$ & $>20.921$ & $>20.639$ & $>20.423$ & 0.00 & 6 \\
{AT 2024zdi} & \ra{17;53;20.19}, \ang{+65;08;17.04} & 22.90 & -16.55 & $>23.528$ & $--$ & $>23.673$ & $>23.654$ & 0.00 &  \\
{AT 2024zed} & \ra{17;48;43.7}, \ang{+64;52;26.74} & 22.46 & -134.28 & $24.494\pm0.026$ & $24.441\pm0.025$ & $>23.721$ & $>23.552$ & 0.16 &  \\
{AT 2024zee} & \ra{18;01;20.47}, \ang{+65;08;32.92} & 22.51 & -134.38 & $>24.808$ & $>22.24$ & $>22.24$ & $>23.041$ & 0.00 & 4 \\
{AT 2024zef} & \ra{18;03;56.6}, \ang{+67;45;02.16} & 22.13 & -133.38 & $25.944\pm0.033$ & $>24.428$ & $>24.276$ & $>24.042$ & 0.18 &  \\
{AT 2024zeg} & \ra{18;00;14.75}, \ang{+64;51;05.38} & 22.04 & -134.49 & $25.824\pm0.036$ & $>25.994$ & $>25.979$ & $>25.747$ & 0.16 &  \\
{AT 2024zfy} & \ra{18;14;38.86}, \ang{+65;39;57.49} & 21.95 & -21.72 & $22.612\pm0.036$ & $>21.079$ & $>20.823$ & $>20.803$ & 0.18 & 4 \\
{AT 2024zgs} & \ra{18;00;21.57}, \ang{+67;39;20.47} & 21.94 & -89.02 & $24.01\pm0.025$ & $>23.435$ & $>23.435$ & $>22.696$ & 0.10 &  \\
{AT 2024zih} & \ra{17;55;29.94}, \ang{+65;48;45.83} & 22.12 & -53.99 & $24.98\pm0.05$ & $24.079\pm0.065$ & $24.563\pm0.051$ & $24.386\pm0.065$ & 0.05 &  \\
{AT 2024zin} & \ra{17;46;37.48}, \ang{+65;12;38.47} & 22.65 & -50.36 & $22.221\pm0.05$ & $>20.371$ & $>19.82$ & $>19.787$ & 0.09 &  \\
{AT 2024zis} & \ra{18;10;29.32}, \ang{+66;40;56.77} & 22.40 & -43.89 & $22.845\pm0.025$ & $22.454\pm0.2$ & $22.598\pm0.2$ & $22.562\pm0.2$ & 0.04 &  \\
{AT 2024zjg} & \ra{17;58;06.28}, \ang{+64;49;26.79} & 22.61 & -18.64 & $23.179\pm0.01$ & $--$ & $--$ & $--$ & 0.07 &  \\
{AT 2024zjh} & \ra{17;55;30.28}, \ang{+64;44;41.86} & 22.69 & -18.55 & $23.684\pm0.012$ & $24.091\pm0.009$ & $24.242\pm0.013$ & $>25.139$ & 0.11 &  \\
{SN 2021aele} & \ra{3;35;32.564}, \ang{-28;30;51.52} & 19.17 & -999.19 & $>22.941$ & $>20.893$ & $>20.534$ & $>20.408$ & 0.00 & 3 \\
{SN 2022ooo} & \ra{18;13;56.288}, \ang{+64;35;03.99} & 18.75 & -745.41 & $>25.293$ & $>23.04$ & $>22.686$ & $>22.357$ & 0.00 &  \\
{SN 2022sje} & \ra{18;22;05.32}, \ang{+66;37;28.16} & 18.15 & -690.45 & $>26.485$ & $>24.209$ & $>23.738$ & $>23.626$ & 0.00 &  \\
{SN 2022wen} & \ra{17;47;14.639}, \ang{+64;15;50.57} & 19.94 & -660.90 & $>21.594$ & $18.662$ & $>18.132$ & $>17.741$ & 0.00 &  \\
{SN 2023aew} & \ra{17;40;51.37}, \ang{+66;12;22.75} & 18.05 & -541.69 & $26.214\pm0.06$ & $25.579\pm0.03$ & $>25.662$ & $25.074\pm0.019$ & 0.16 &  \\
{SN 2023psq} & \ra{17;54;19.329}, \ang{+64;13;32.18} & 18.54 & -335.85 & $>25.442$ & $>23.013$ & $>22.55$ & $>22.5$ & 0.00 & 5 \\
{SN 2023uqu} & \ra{17;47;47.83}, \ang{+64;20;57.24} & 20.15 & -285.94 & $23.811\pm0.011$ & $24.105\pm0.008$ & $>24.77$ & $24.617\pm0.015$ & 0.05 &  \\
{SN 2023vcn} & \ra{3;35;17.472}, \ang{-27;45;03.81} & 19.61 & -295.12 & $25.194\pm0.037$ & $>23.135$ & $>22.842$ & $>22.81$ & 0.18 &  \\
{SN 2024abla} & \ra{3;50;18.947}, \ang{-47;46;57.96} & 18.38 & 73.24 & $>26.28$ & $>23.708$ & $>23.447$ & $>23.438$ & 0.00 &  \\
{SN 2024pvw} & \ra{18;19;44.103}, \ang{+66;00;47.48} & 19.23 & 3.05 & $20.623\pm0.01$ & $20.859\pm0.005$ & $21.066\pm0.007$ & $21.633\pm0.009$ & 0.18 &  \\
\hline
\end{longtable}
\end{landscape}

\section{Target \IE cutouts}\label{apx:cutouts}
In this appendix we show the \IE cutouts for each of the the TNS reported targets we followed up in this work. This is primarily for reference and to demonstrate visually the capablity of \Euclid to detect these objects.
\begin{figure*}[ht!]
    \centering
    \includegraphics[width=\textwidth]{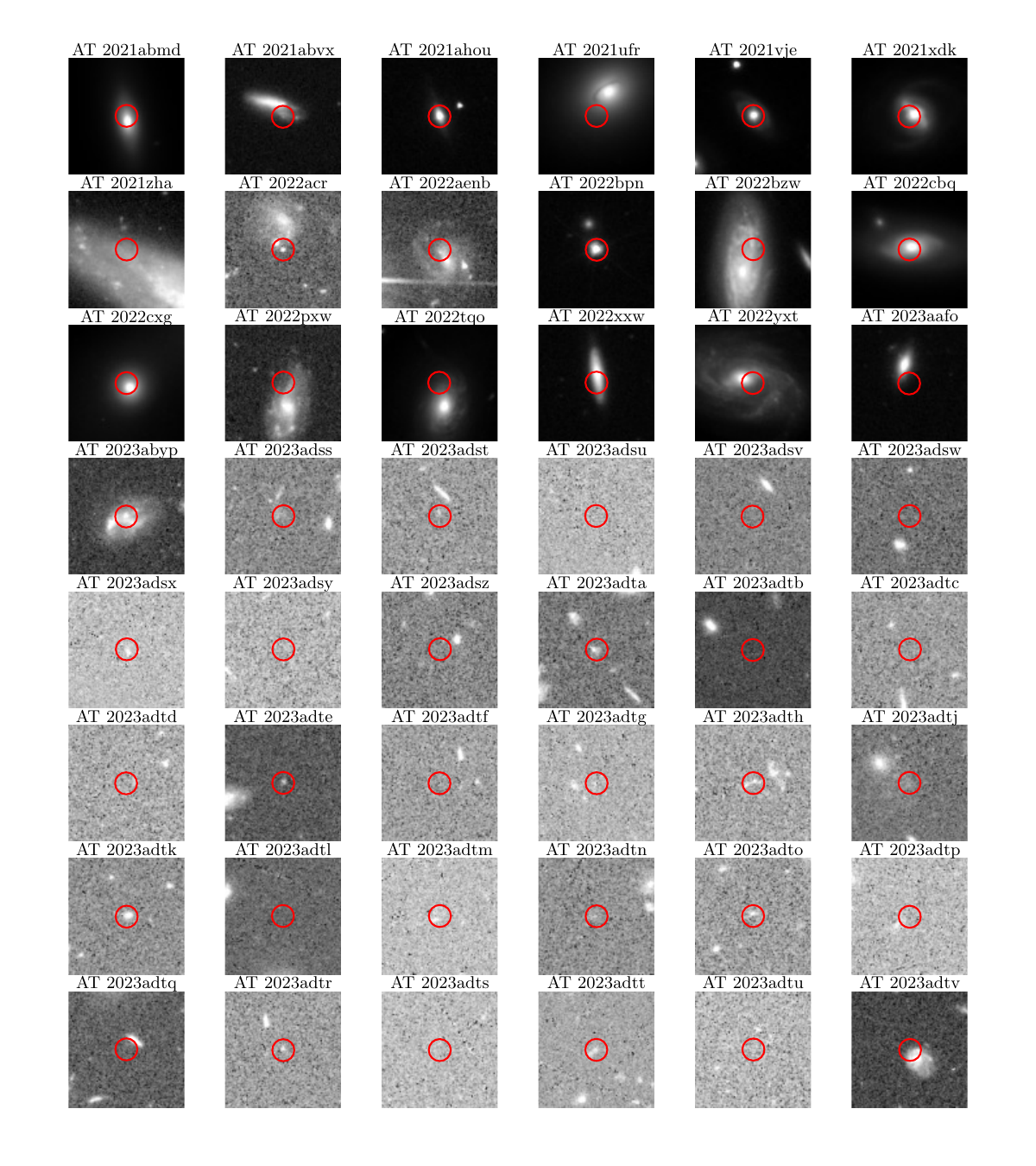}
    \caption{$12\arcsec\times12\arcsec$ cutouts of the targets as seen in \Euclid \IE mosaic images. The red circles are centred on the position of the transient as reported in TNS. A total of 156 cutouts are shown, corresponding to the total number of cutouts for which it was possible to recover an \IE cutout, including those where the photometric procedure failed (see Sect.~\ref{sec:selection} for a discussion of this).}\label{fig:cutouts_1}
\end{figure*}

\begin{figure*}[ht!]
    \centering
    \includegraphics[width=\textwidth]{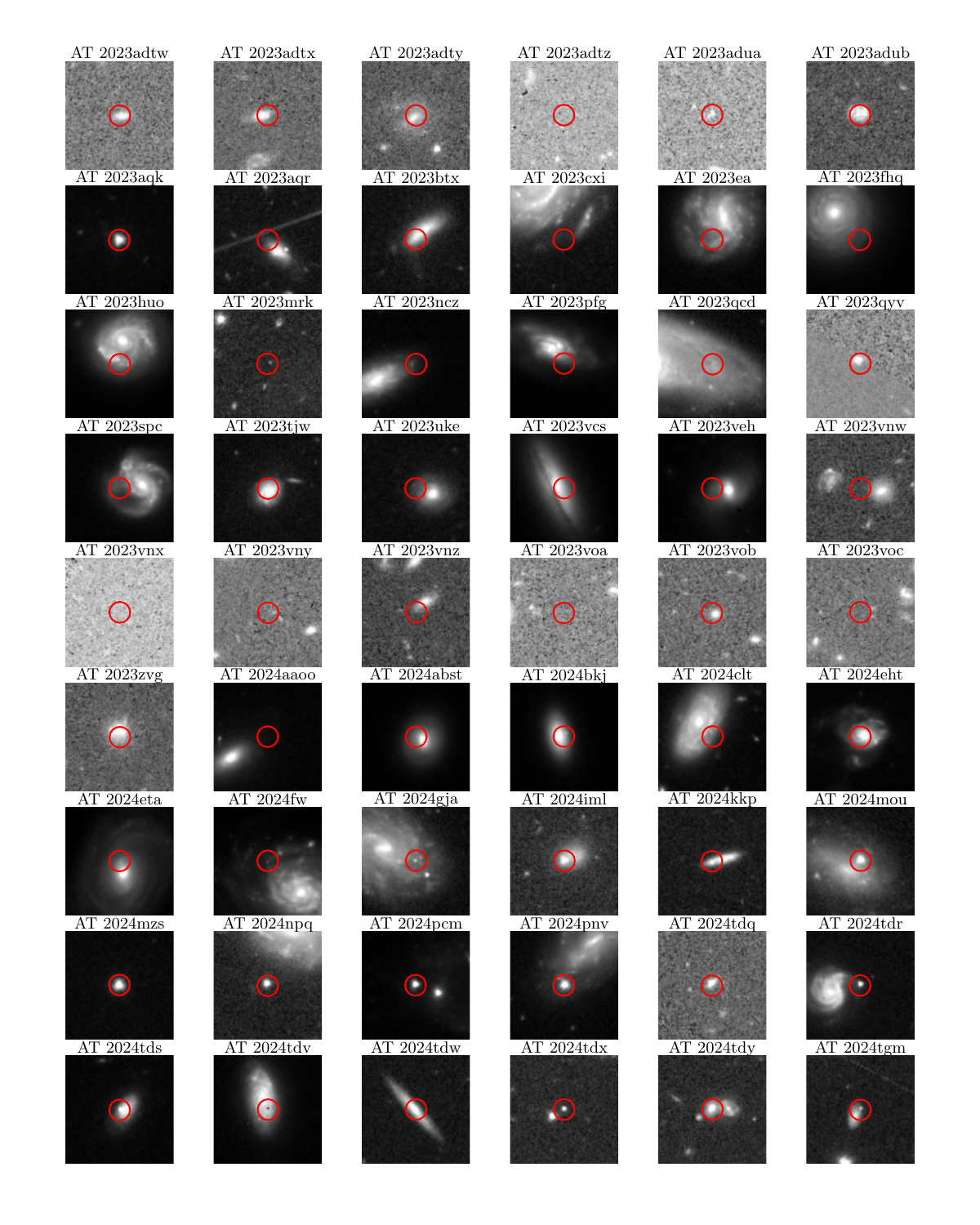}
    \caption{As Fig.~\ref{fig:cutouts_1}}\label{fig:cutouts_2}
\end{figure*}

\begin{figure*}[ht!]
    \centering
    \includegraphics[width=\textwidth]{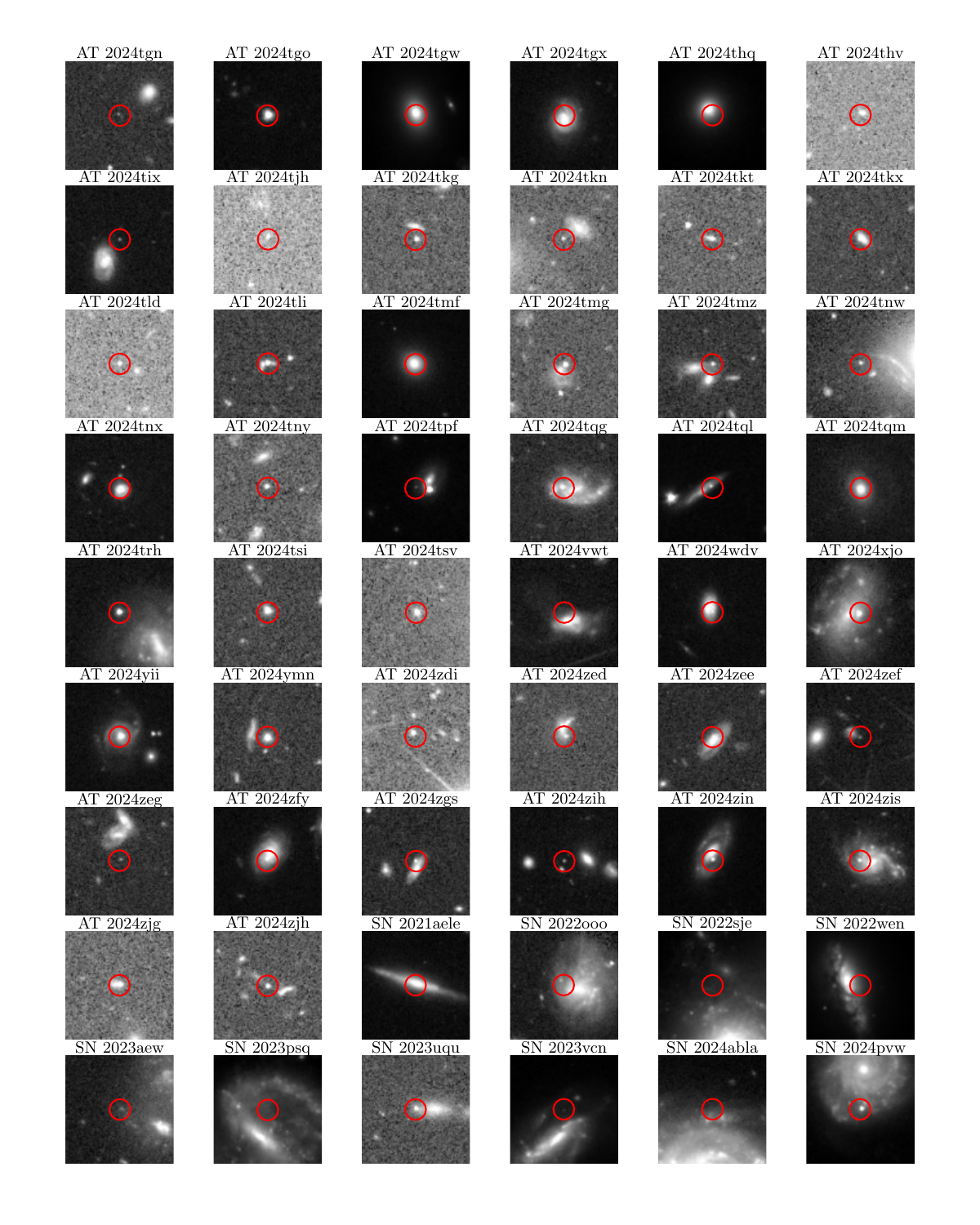}
    \caption{As Fig.~\ref{fig:cutouts_1}}\label{fig:cutouts_3}
\end{figure*}

\newpage 

\section{Light curves}

\begin{figure*}[h]
    \centering
    \includegraphics[width=0.45\linewidth]{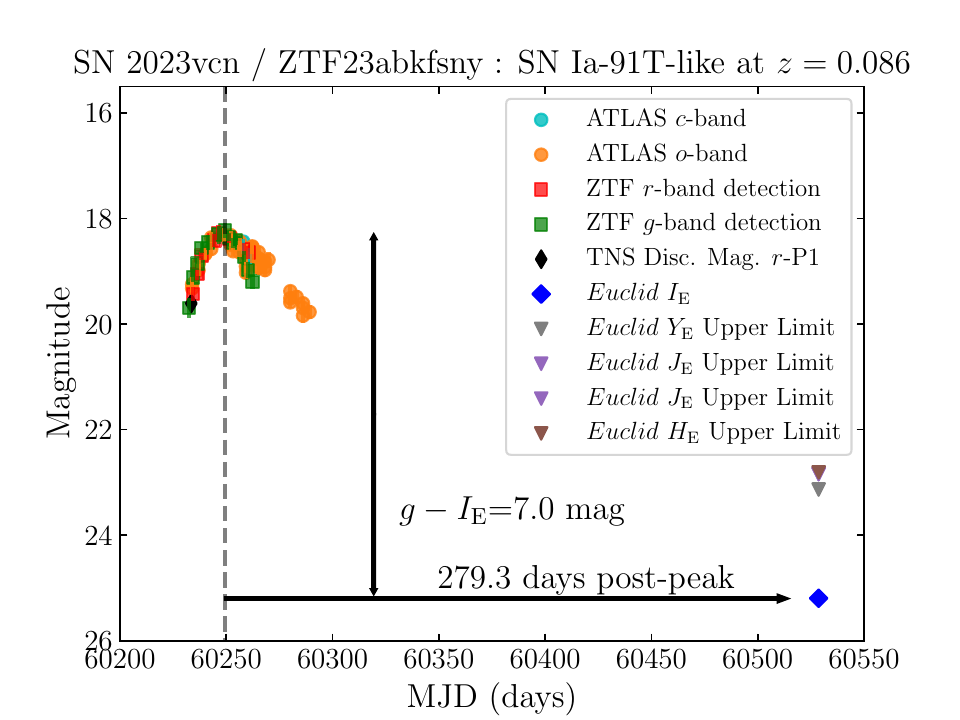}
    \includegraphics[width=0.45\linewidth]{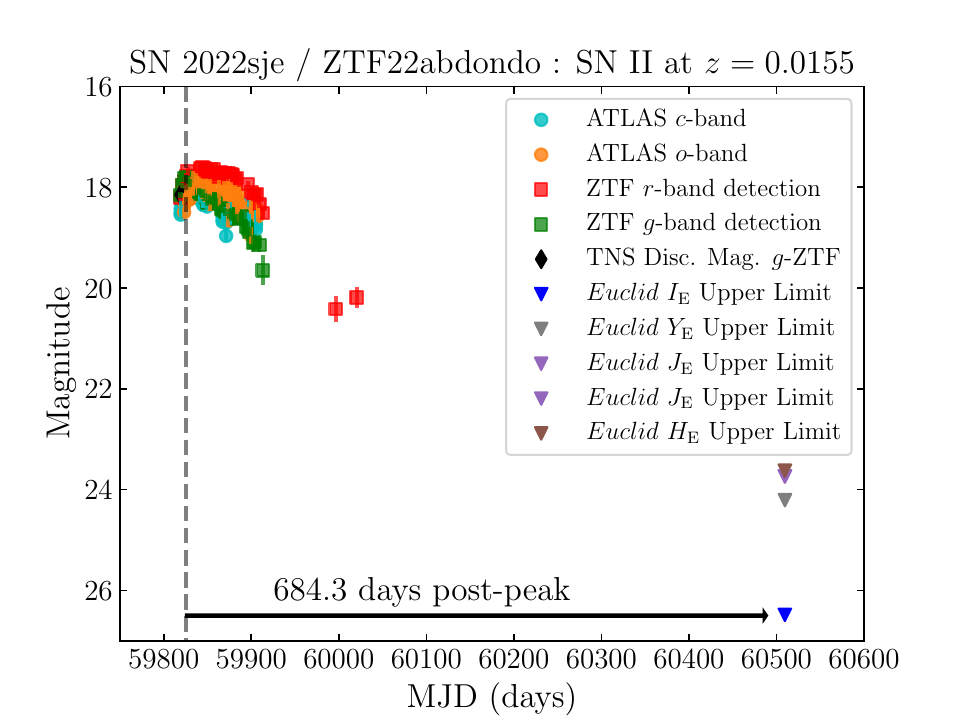}
    \includegraphics[width=0.45\linewidth]{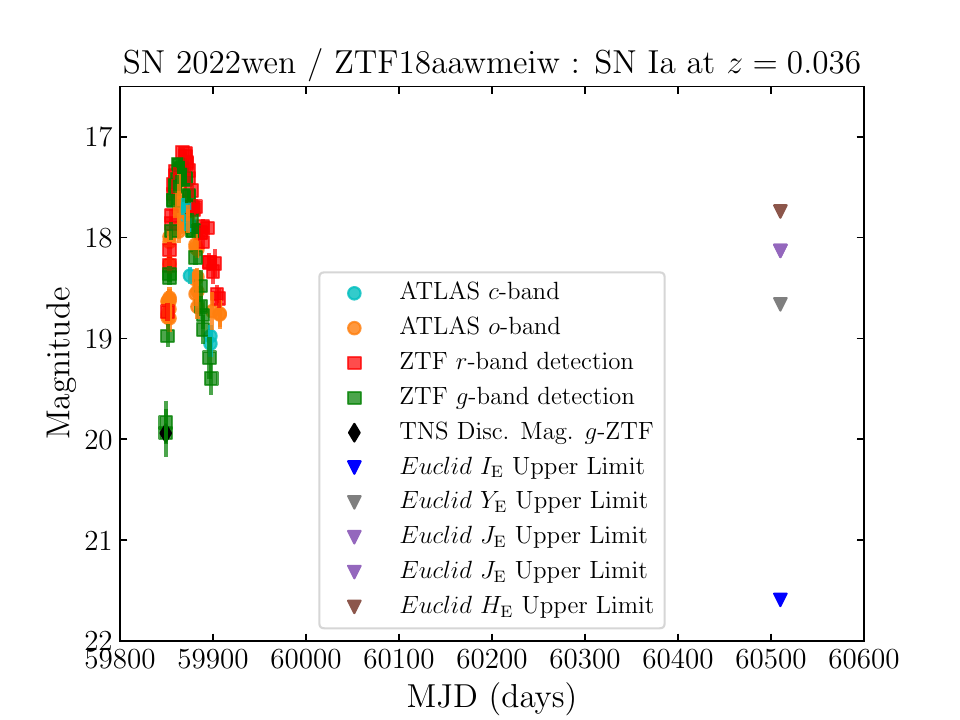}
    \includegraphics[width=0.45\linewidth]{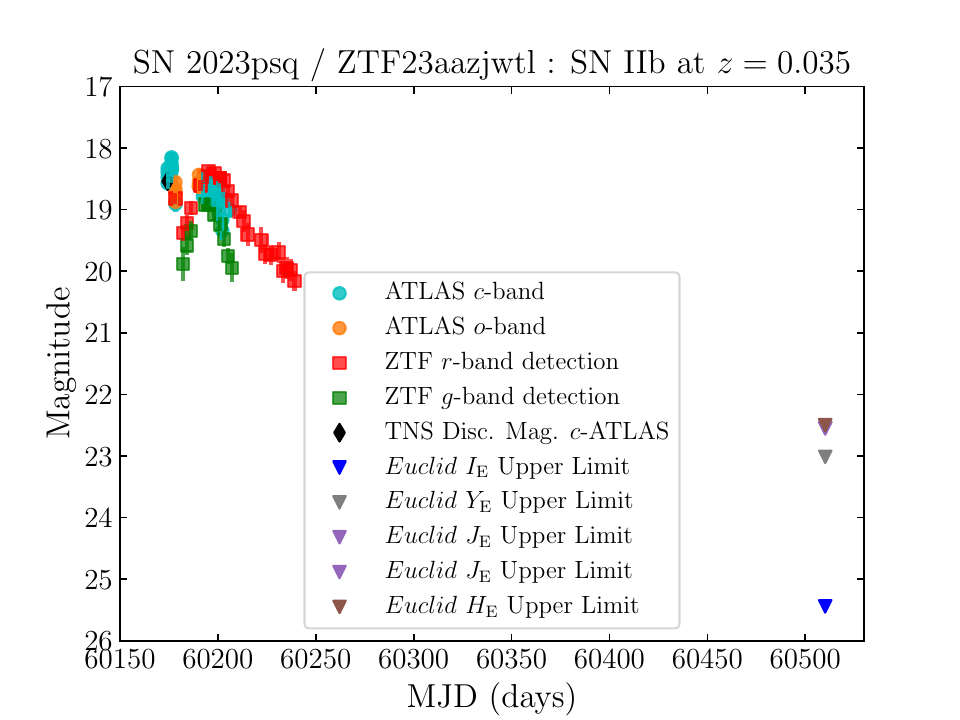}
    \includegraphics[width=0.45\linewidth]{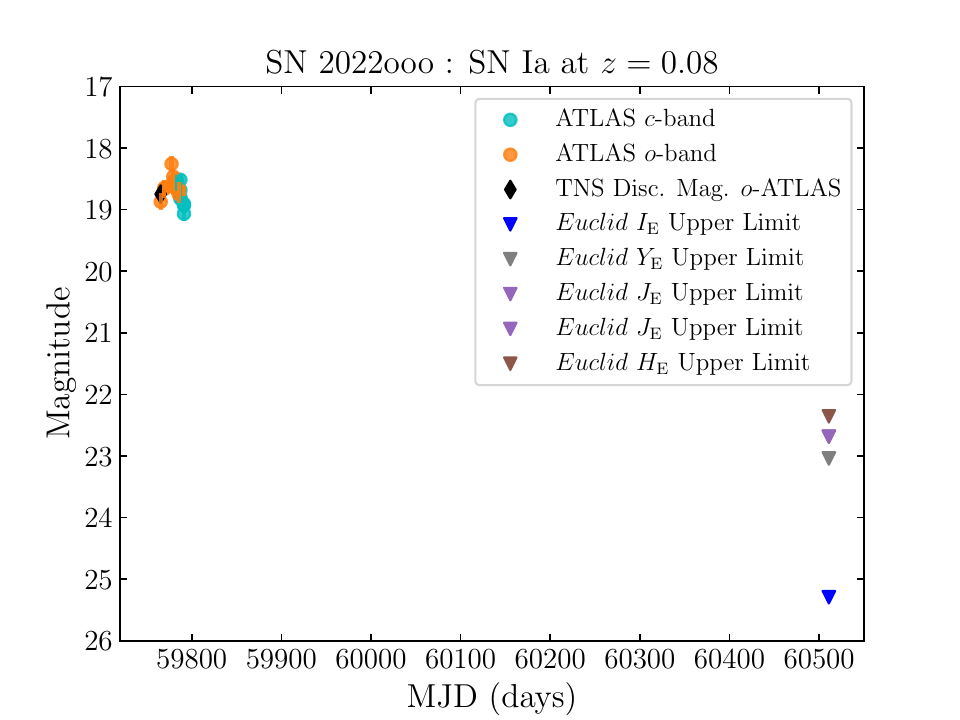}
    \includegraphics[width=0.45\linewidth]{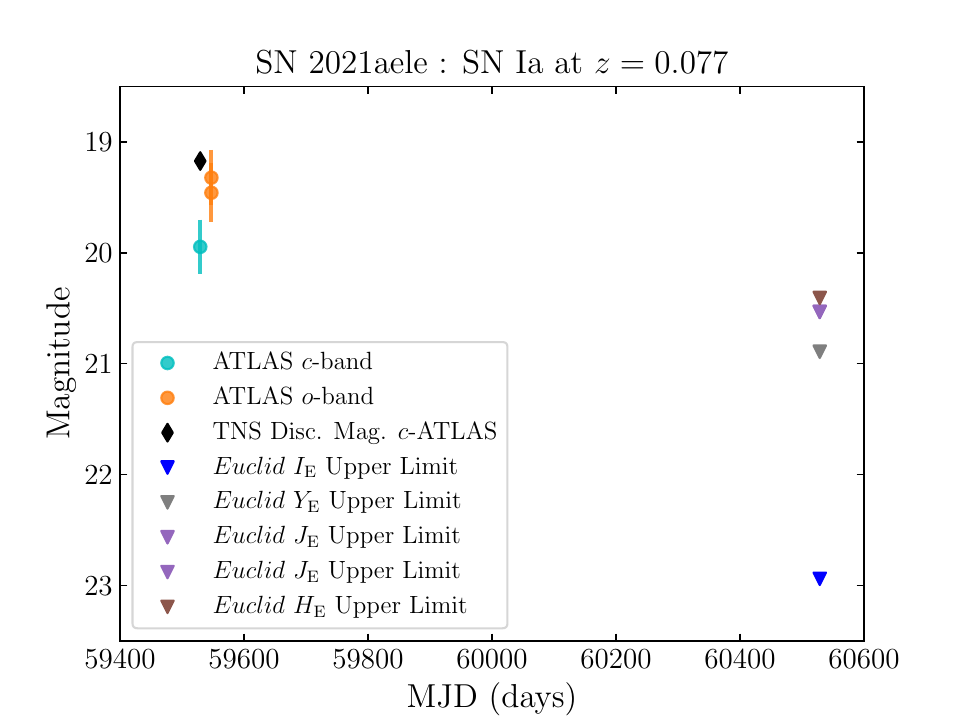}
    \caption{Light curves plots of classified SNe including estimates of the epochs of the \Euclid detections and variations in magnitudes from peak, for some of them.}
    \label{fig:sne_lcs}
\end{figure*}

\begin{figure*}[h]
    \centering
    \includegraphics[width=0.45\linewidth]{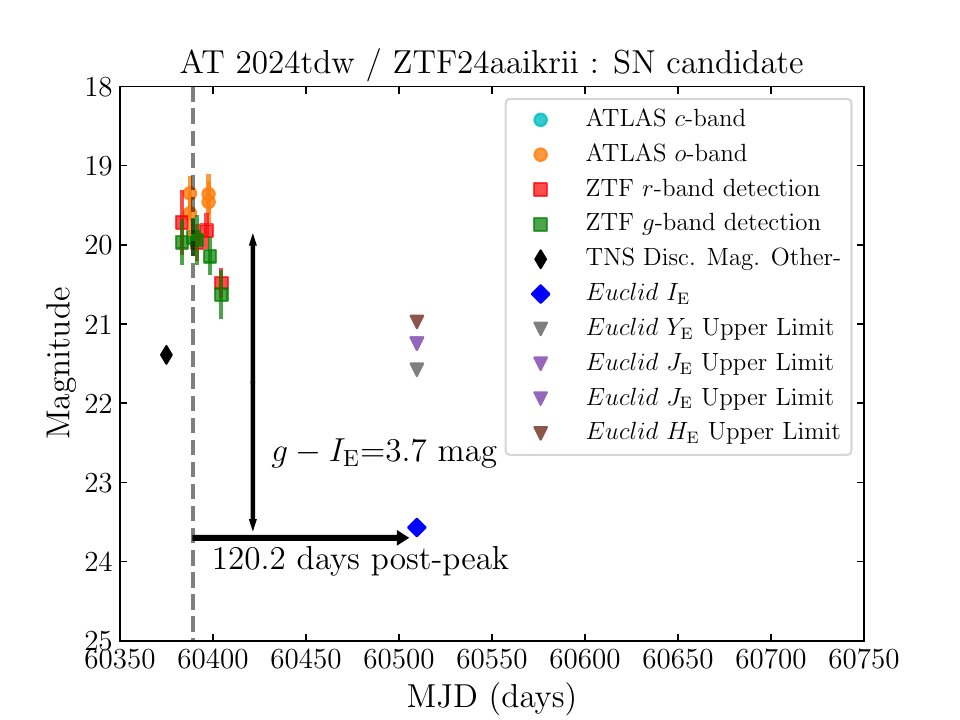}
    \includegraphics[width=0.45\linewidth]{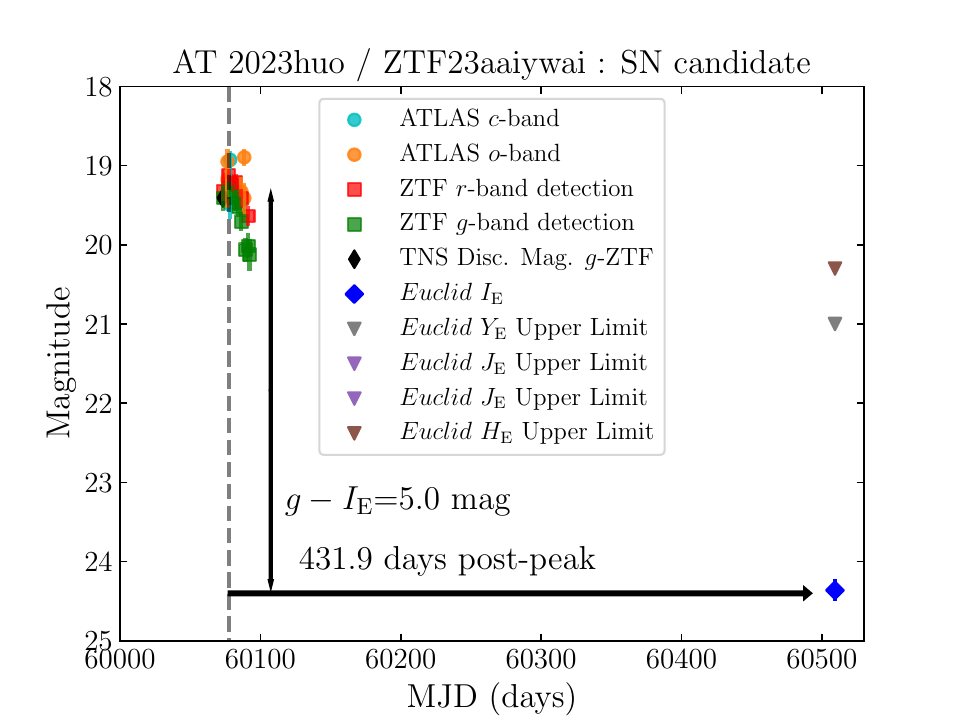}
    \includegraphics[width=0.45\linewidth]{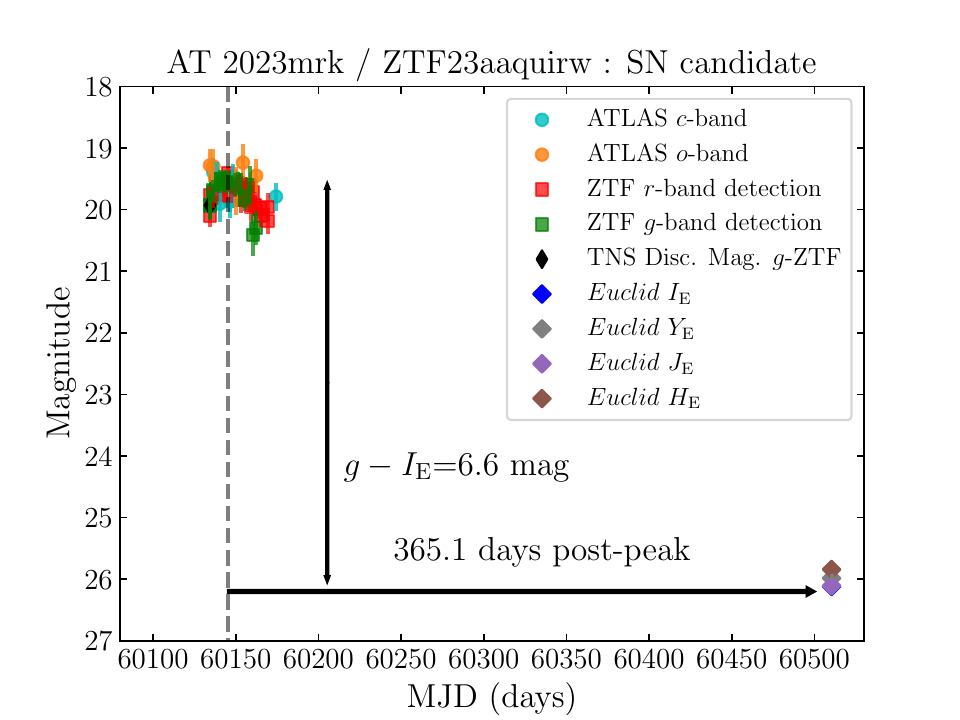}
   \caption{Light curves plots of some unclassified transients including \Euclid photometry or upper limits.}
    \label{fig:at_lcs}
\end{figure*}

\begin{figure*}
    \centering
    \includegraphics[width=0.45\linewidth]{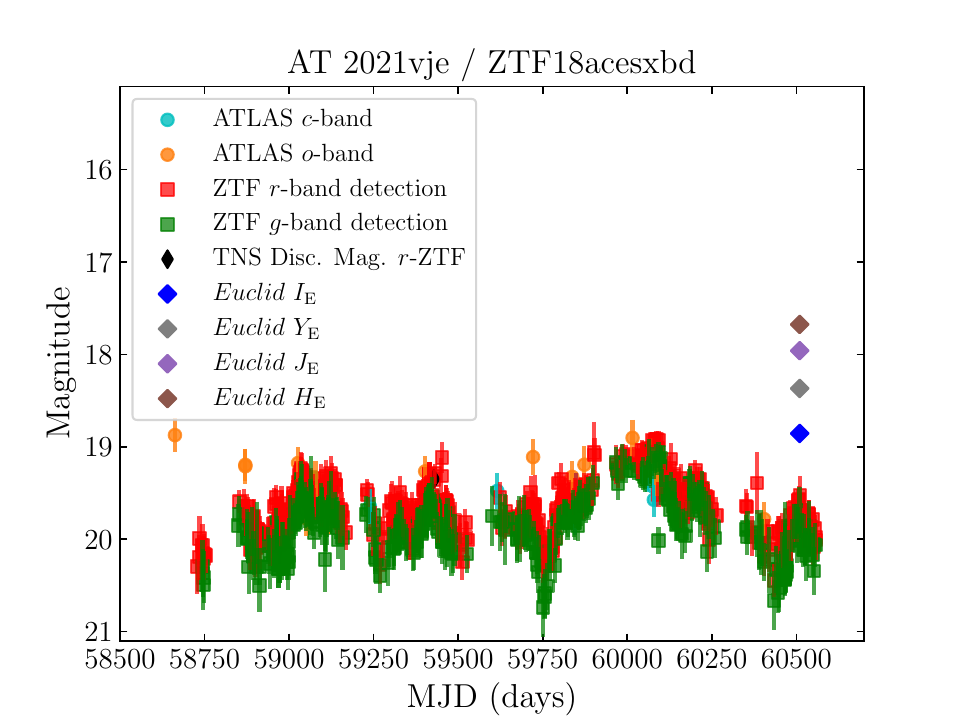}
    \includegraphics[width=0.45\linewidth]{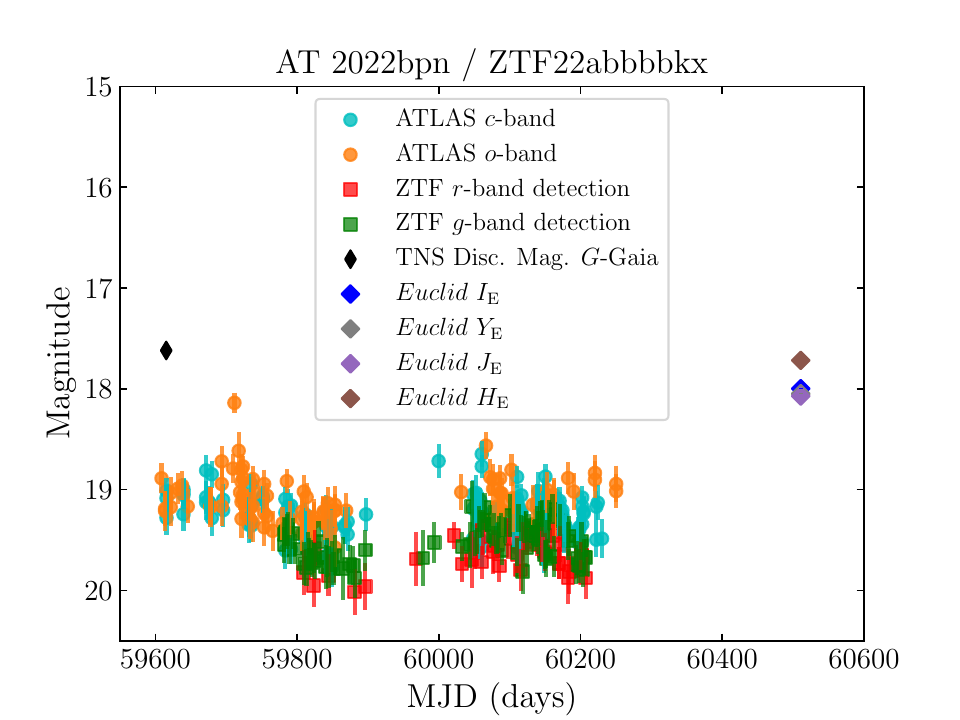}
    \includegraphics[width=0.45\linewidth]{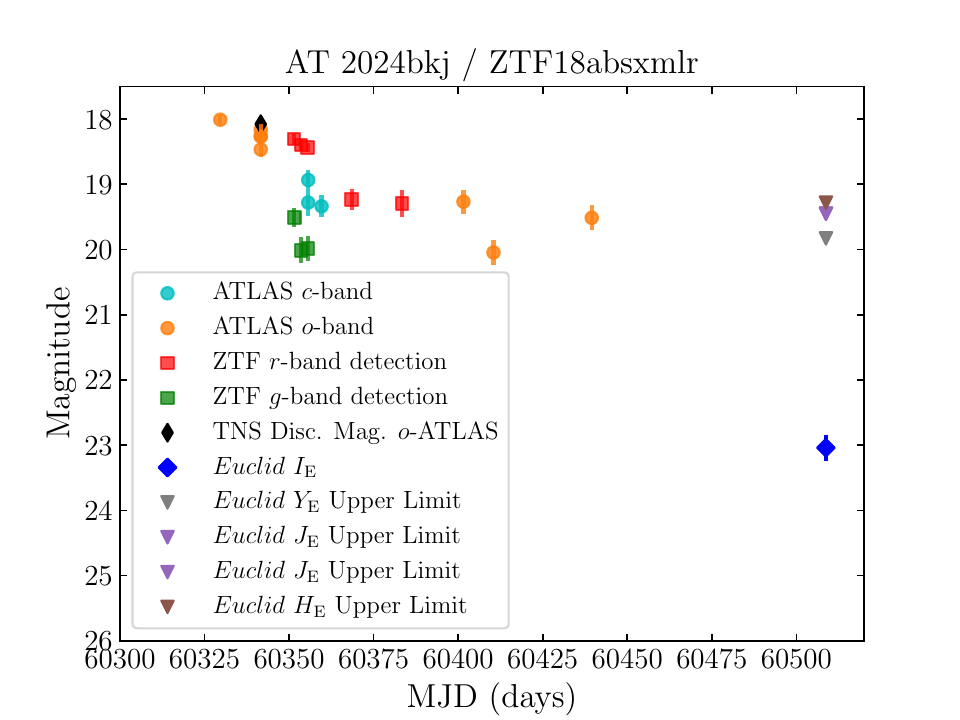}
   \caption{Light curves plots of some unclassified transients likely produced by AGN activity (AT\,2021vje), by stellar activity (AT\,2022bpn), or of unclear nature (AT\,2024bkj).}
    \label{fig:other_lcs}
\end{figure*}

\begin{figure*}
    \centering
    \includegraphics[width=0.45\linewidth]{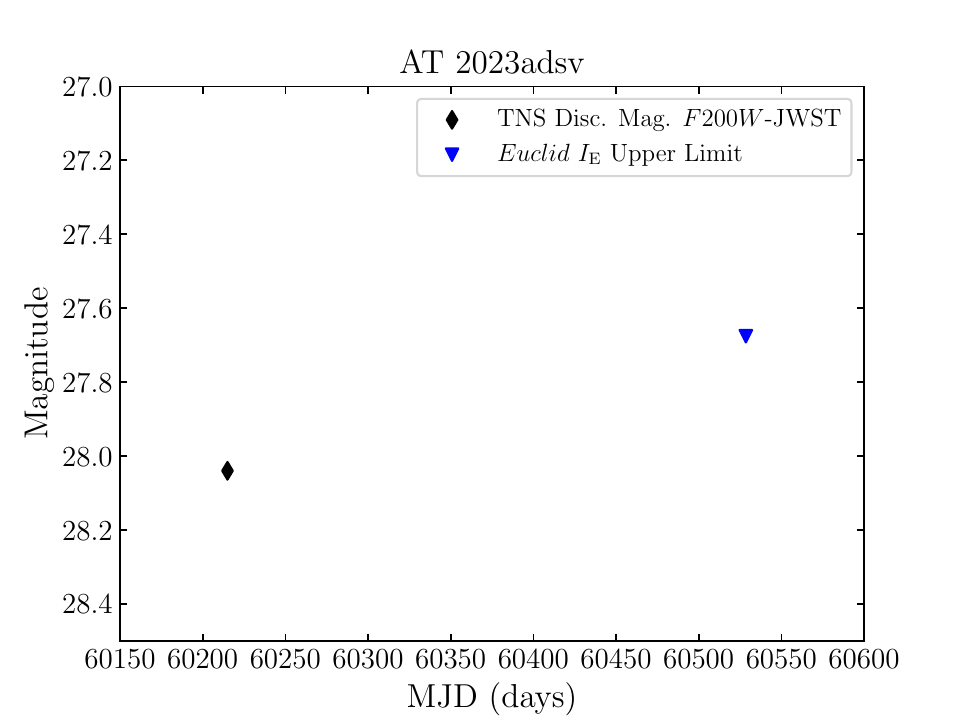}
    \includegraphics[width=0.45\linewidth]{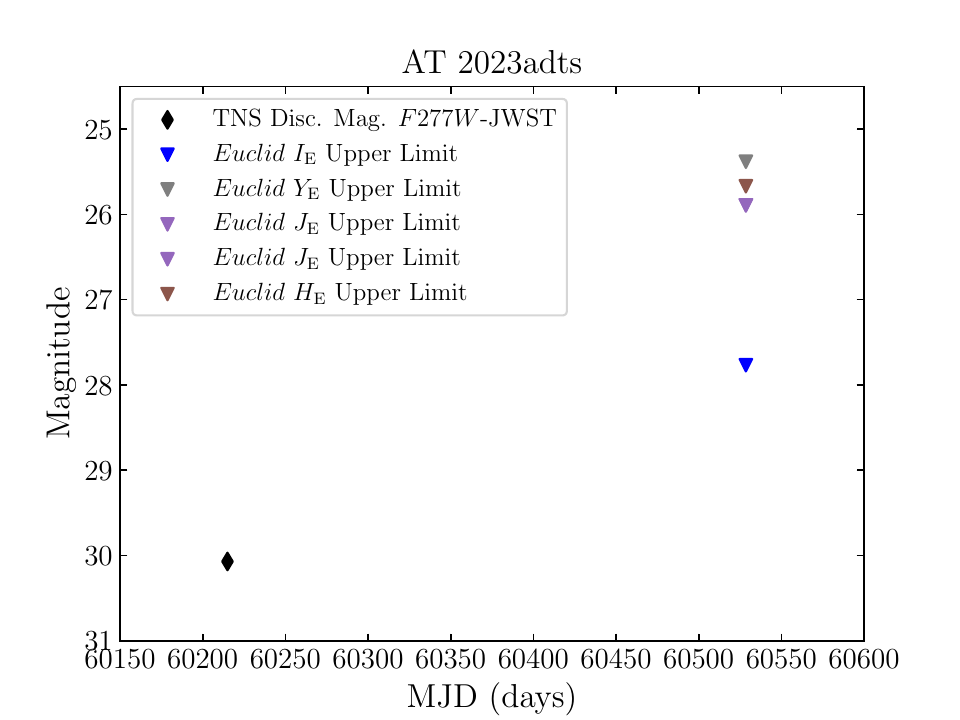}
    \includegraphics[width=0.45\linewidth]{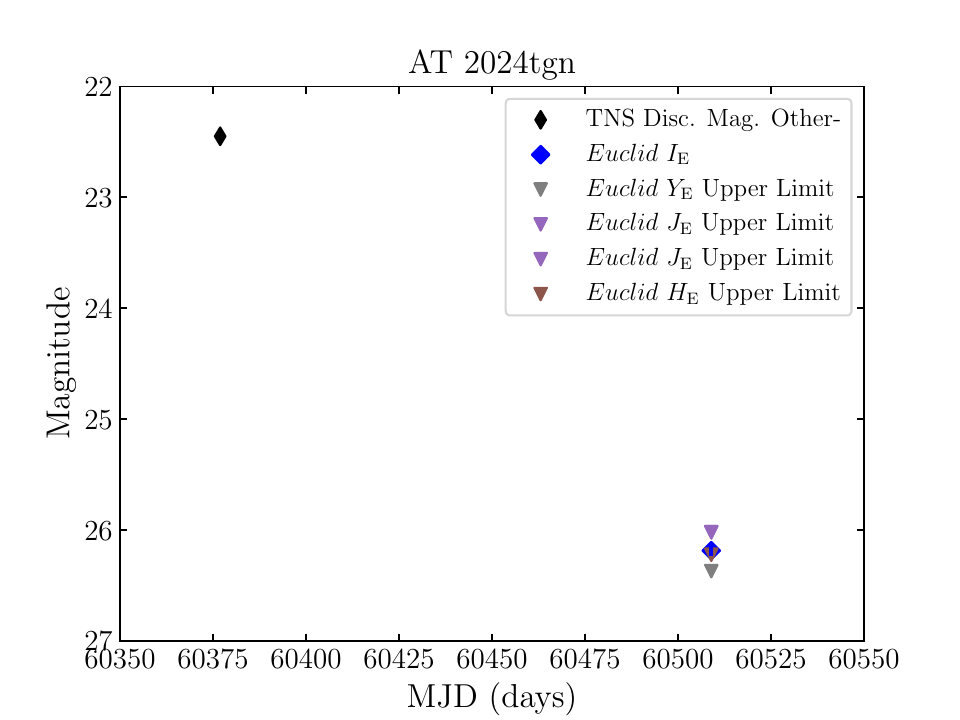}
   \caption{Light curves plots of some poorly sampled unclassified transients discovered in the JADES (AT\,2023adsv and AT\,2023adts) and WFST (AT\,2024tgn) surveys and including \Euclid photometry or upper limits.}
    \label{fig:deep_lcs}
\end{figure*}

In this appendix we show additional light curves of several transients. In Fig.~\ref{fig:sne_lcs} we show the light curves of the classified SNe listed in \autoref{tab:completeList} that are not discussed in Sect.~\ref{notable_transients}. In Fig.~\ref{fig:at_lcs} we show additional light curves of some of the unclassified transients listed in \autoref{tab:completeList}, showing \Euclid photometry detections. In Fig.~\ref{fig:other_lcs} we show light curves likely produced by AGN activity (AT\,2021vje), by stellar activity (AT\,2022bpn), or of unclear nature (AT\,2024bkj). Finally, in Fig.~\ref{fig:deep_lcs} we show some examples of poorly-sampled light curves discovered in the JADES (AT\,2023adsv and AT\,2023adts) and WFST (AT\,2024tgn) surveys.

\end{appendix}

\end{document}